\documentstyle[11pt,titlepage]{article}

\newcommand{\ym}{{\scriptscriptstyle (\cy \cm)}}
\newcommand{\pl}{{\scriptscriptstyle (+)}}
\newcommand{\mn}{{\scriptscriptstyle (-)}}
\newcommand{\uo}{{\scriptscriptstyle (1)}}
\newcommand{\gb}{{\scriptscriptstyle (GB)}}

\newcommand{\im}{{\rm Im}}

\newcommand{\sect}[1]{\setcounter{equation}{0}\section{#1}}

\newcommand{\be}{\begin{equation}}
\newcommand{\eeq}{\end{equation}}

\newcommand{\bea}{\begin{eqnarray}}
\newcommand{\ena}{\end{eqnarray}}
\newcommand{\nn}{\nonumber}

\newcommand{\ba}{\begin{array}}
\newcommand{\ea}{\end{array}}

\newcommand{\br}{{\bar{r}}}

\newcommand{\bD}{{\bar{D}}}
\newcommand{\bF}{{\bar{F}}}
\newcommand{\bM}{{\bar{M}}}
\newcommand{\bS}{{\bar{S}}}
\newcommand{\bT}{{\bar{T}}}
\newcommand{\bW}{{\bar{W}}}
\newcommand{\bX}{{\bar{X}}}

\newcommand{\al}{{\alpha}}
\newcommand{\bt}{{\beta}}
\newcommand{\gm}{{\gamma}}
\newcommand{\dt}{{\delta}}
\newcommand{\eps}{{\epsilon}}
\newcommand{\vep}{{\varepsilon}}

\newcommand{\la}{{\lambda}}
\newcommand{\si}{{\sigma}}
\renewcommand{\th}{\theta}
\newcommand{\vp}{\varphi}

\newcommand{\Si}{\Sigma}
\newcommand{\Dt}{\Delta}
\newcommand{\La}{\Lambda}

\newcommand{\Om}{{\Omega}}

\newcommand{\thb}{\bar{\theta}}
\newcommand{\vpb}{\bar{\vp}}
\newcommand{\sib}{{\bar{\sigma}}}
\newcommand{\lab}{{\bar{\lambda}}}

\newcommand{\psib}{{\bar{\psi}}}
\newcommand{\chib}{{\bar{\chi}}}
\newcommand{\phib}{{\bar{\phi}}}

\newcommand{\Lab}{\bar{\Lambda}}

\newcommand{\dg}{{\dot{\gamma}}}
\newcommand{\da}{{\dot{\alpha}}}
\newcommand{\dv}{{\dot{\varphi}}}
\newcommand{\db}{{\dot{\beta}}}
\newcommand{\dd}{{\dot{\delta}}}
\newcommand{\dep}{{\dot{\epsilon}}}

\newcommand{\dmu}{{\dot{\mu}}}

\newcommand{\hcq}{{\hat{\cal Q}}}
\newcommand{\hB}{\hat{B}}
\newcommand{\hW}{\hat{W}}

\newcommand{\Omt}{\tilde{\Omega}}
\newcommand{\hh}{\tilde{h}}
\newcommand{\Qt}{\tilde{Q}}

\newcommand{\undal}{{\underline{\alpha}}}
\newcommand{\undel}{\underline{\delta}}
\newcommand{\undbt}{{\underline{\beta}}}
\newcommand{\undgm}{{\underline{\gamma}}}
\newcommand{\unddt}{{\underline{\delta}}}

\newcommand{\undph}{{\underline{\varphi}}}
\newcommand{\undSi}{{\underline{\Sigma}}}
\newcommand{\undOm}{{\underline{\Omega}}}

\newcommand{\undB}{{\underline{B}}}

\newcommand{\ca}{{\cal A}}

\newcommand{\cd}{{\cal D}}
\newcommand{\ce}{{\cal E}}
\newcommand{\cf}{{\cal F}}
\newcommand{\ch}{{\cal H}}

\newcommand{\cl}{{\cal L}}
\newcommand{\cm}{{\cal M}}
\newcommand{\cq}{{\cal Q}}
\newcommand{\car}{{\cal R}}
\newcommand{\cs}{{\cal S}}
\newcommand{\ct}{{\cal T}}

\newcommand{\cu}{{\cal U}}

\newcommand{\cw}{{\cal W}}
\newcommand{\cy}{{\cal Y}}

\newcommand{\crt}{\tilde{\cal{R}}}

\newcommand{\cdb}{{\bar{\cal{D}}}}
\newcommand{\cwb}{{\bar{\cal{W}}}}

\newcommand{\csi}{\cs^I}
\newcommand{\cti}{\ct^I}

\newcommand{\rd}{{R^\dagger}}

\def\lp{\left(}
\def\rp{\right)}

\def\l[{\left[}
\def\r]{\right]}

\newcommand{\ie}{{\em i.e. }}
\newcommand{\pp}{{\lp \cd^2 - 8 \rd \rp}}
\newcommand{\qq}{{\lp \cdb^2 - 8 R \rp}}

\newcommand{\hs}{\hspace}
\newcommand{\vs}{\vspace}
\newcommand{\cem}{\hspace{1cm}}

\newcommand{\ka}{{K\"{a}hler\ }}

\newcommand{\f}[2]{{\textstyle\frac{#1}{#2}}}

\newcommand{\Xb}{{\bar{X}}}

\newcommand{\llsym}[1]{\stackrel{\scriptstyle#1}{\mbox{\huge $\smile$}}}
\newcommand{\lsym}[1]{\stackrel{\scriptstyle#1}{\mbox{$\smile$}}}
\newcommand{\sym}[1]{\stackrel{\scriptstyle#1}{\mbox{\tiny $\smile$}}}
\newcommand{\ci}[1]{\raise5pt\hbox{$\scriptstyle#1$}} 

\def\2#1{\mbox{#1}}

\def\12{\f{1}{2}}
\def\i2{\f{i}{2}}
\def\13{\f{1}{3}}
\def\s2{\sqrt{2}}

\newcommand{\prt}{\partial}
\newcommand{\tr}{{\rm tr}}

\newcommand{\chb}{\raise2.5pt\hbox{$\chib$}}
\begin{document}


\font\nine=cmbx10 at 9pt
\font\ten=cmbx10 at 10pt
\font\eleven=cmbx10 at 11pt
\font\twelve=cmbx10 at 12pt
\font\thirteen=cmbx10 at 13pt
\font\fifteen=cmbx10 at 15pt

\begin{titlepage}

\begin{center}

\renewcommand{\thefootnote}{\fnsymbol{footnote}}

{\twelve Centre de Physique Th\'eorique\footnote{Unit\'e Propre de
Recherche 7061}, CNRS Luminy, Case 907}

{\twelve F-13288 Marseille -- Cedex 9}

\vspace{1.4 cm}

{\LARGE {\bf The superspace geometry of gravitational \\
             Chern-Simons forms and their couplings \\[3mm]
             to linear multiplets : a review}}

\vspace{1 cm}

{\large {\bf {Georges GIRARDI}}}\footnote{{\em Laboratoire d'Annecy le Vieux de Physique
Th\'eorique} - LAPTH - (URA 14-36 du CNRS, associ\'ee
\`a l'Universit\'e de Savoie) {\em Chemin de Bellevue, BP 110,
F-74941 Annecy-le-Vieux Cedex, France}}  {\large {\bf {and Richard
GRIMM}}}\footnote{{\em Centre de Physique Th\'eorique, CNRS Luminy,
Case 907, F-13288 Marseille --Cedex 9}}

\vspace{2,3 cm}

{\bf Abstract}
\end{center}
The superspace geometry of Chern-Simons forms is shown to be closely related
to that of the 3-form multiplet. This observation allows
to simplify considerably the geometric structure of supersymmetric
Chern-Simons forms and their coupling to linear multiplets.
The analysis is carried through in $U_K(1)$ superspace, relevant at the same
time for supergravity-matter couplings and for chirally extended
supergravity.

\vfill


\bigskip

\noindent January 1998

\noindent {\large L}{\large A}{\large P}{\large T}{\large H}-673/98

\noindent CPT-98/P.3615

\bigskip

\noindent anonymous ftp : ftp.cpt.univ-mrs.fr

\noindent web : www.cpt.univ-mrs.fr

\setcounter{footnote}{0}
\renewcommand{\thefootnote}{\arabic{footnote}}

\end{titlepage}


\newpage

\thispagestyle{empty}
$\mbox{}$ {}

\newpage

\setcounter{page}{1}
\tableofcontents


\sect{Introduction}

\indent

\subsection{Context and possible motivations}

\indent

The purpose of this article is to provide a concise introductory
review of the superspace geometry relevant for
$N=1$ supersymmetric theories in four dimensions which go
beyond the by now commonly used standard formulation of the general
supergravity-matter system \cite{CFG82}, \cite{CFG83},
\cite{Bag83}. {\em Beyond} is meant here in the sense that, in
addition to the general couplings of chiral matter multiplets and
of Yang-Mills multiplets to supergravity, couplings of linear
supermultiplets in the presence of Chern-Simons forms will be
included. Relying on the mechanisms used successfully for the
implementation of Chern-Simons forms of the Yang-Mills type in
supersymmetric theories \cite{Gri85}, \cite{BGGM87b}, an attempt is
made to clarify, as concisely as possible, the geometrical
structures and the special features occurring in the corresponding
descriptions of gravitational Chern-Simons forms.

As the most popular motivation for this work we recall that
couplings of antisymmetric tensor gauge fields to supersymmetric
theories, and the appearance of Chern-Simons forms, are important
ingredients in the construction of low energy effective
approximations of some underlying fundamental superstring theory.

As usual in this kind of approach it is prohibitively complicated, if
not impossible, to explicitly derive the effective from the exact theory.
There are, however, criteria which allow nevertheless to obtain nontrivial
information on the form of the low energy theory, one of them being
the requirement of absence of anomalies in the fundamental theory.

In the case of superstring theory, this kind of reasoning leads,
among other things, to the coupling of the antisymmetric tensor
gauge field together with Yang-Mills and gravitational Chern-Simons
forms via the so-called Green-Schwarz mechanism \cite{GS84} in the
ten-dimensional effective theory. At this point, however,
supersymmetry of the mechanism is far from evident, not only due to
technical complications but also at a more fundamental level due to
the lack of complete understanding of higher-dimensional and/or
extended supersymmetries.

In attempts of relating such kinds of theories to effective four-dimensional
ones with $N=1$ supersymmetry the remnants of anomaly cancellation mechanisms
should show up in one way or another.
As, in particular, one requires these couplings to appear
in a supersymmetric way, a more profound understanding of
the general structure of supersymmetric theories in themselves
is important, irrespective of the motivation put forward
in relation with superstring theory. Stated differently, one might turn the
argument around and study the general form of $N=1$ four-dimensional
supersymmetric theories as a framework into which any of the candidates
of such low energy approximations should fit. It is actually this point
of view which will be adopted in our investigations.

Hence turning to $N=1$ supersymmetry in four dimensions, we recall
that an antisymmetric tensor appears in the so-called linear
multiplet, together with a real scalar and a Majorana spinor. The
lesson to be learnt is then to couple this multiplet to the general
supergravity-matter system, which, together with its intrinsic
\ka invariance is by now rather well understood in geometric
terms \cite{BGGM87a}, \cite{BGG90}, \cite{Gri90b}.

The structure of chiral \ka transformations, inherent in any
coupling of supergravity with matter has led in \cite{BGGM87a} to a
unified geometric description of such theories: chiral \ka
transformations appear together with Lorentz transformations in the
structure group of superspace.

It is well-known from the standard superspace formulations
\cite{Zum79a}, \cite{WeB83}, that the {\em spin connection} in
supergravity theories is expressed as a function of the vierbein
field, its derivatives and of quadratic Rarita-Schwinger field
terms. Likewise, in the new approach, the {\em \ka connection}
\cite{Bag85} (not to be confused with the Christoffel connection on
the \ka manifold itself) is given in terms of the bosonic matter
component fields, their space-time derivatives and their fermionic
supersymmetric partners (for this reason it is sometimes referred
to as {\em composite connection}). Both these geometric objects
appear naturally in the framework of so-called {\em \ka superspace
geometry}. Moreover, this geometric formulation gives rise to a
unified dynamical description: the supersymmetric action of the
kinetic terms of the complete supergravity-matter system (with
canonically normalized Einstein term) is given by one single term
in superspace, namely the superdeterminant of the frame of \ka
superspace.

Given this powerful and elegant formulation it is natural to search
for a generalization which allows to accommodate the
couplings of linear multiplets as well.
In addition, such a construction should be able to embody
the additional structures arising from the
supersymmetric inclusion of Chern-Simons terms in the
field strength of the antisymmetric tensor.

A promising way to implement more general couplings of any number
of linear multiplets to the supergravity-matter system consists in
generalizing the notion of \ka superfield potential: in addition of
being exclusively a function of the chiral and antichiral matter
superfields it is allowed to depend on the linear superfields as
well. This approach has given already a number of interesting
results in particular cases:
\begin{itemize}
\item The coupling of a single linear multiplet to supergravity
(without matter) came under the disguise of the so-called
$16-16$ supergravity \cite{GGMW84b}, \cite{HU86}, \cite{Sie86}, \cite{ADO86}.
\item This coupling was subsequently amended to include
Chern-Simons forms of supersymmetric Yang-Mills
theory \cite{Gri85}.
\item Based on the geometric description in superspace,
a particularly interesting special coupling \cite{Wit85}
of one linear multiplet with Chern-Simons form of Yang-Mills type
to the complete supergravity-matter system has been worked out in full
detail \cite{BGGM87b}.
Moreover the extension of this construction to the case of general couplings of
an arbitrary number of linear multiplets has been indicated.
\item The superspace geometry relevant for
gravitational Chern-Simons forms has been developped and a
superspace action has been proposed \cite{GG87}, \cite{GG87b}.
\end{itemize}

In a slightly different language, making use of the duality between
linear and chiral multiplet formulations, component field
expressions arising from couplings of gravitational Chern-Simons
forms in "higher derivative supergravity theories" have been
reported in a series of publications of S. Ferrara et al.
\cite{CFG*85}, \cite{CFG*86}, \cite{CFV87}, \cite{FV87},
\cite{CFG*88}, clearly demonstrating the complexity of the subject
and pointing out the appearance of new unorthodox structures
(related to higher derivative couplings of matter fields and the
r\^ole of previously auxiliary firlds which may become
dynamical in some sense), questions which clearly deserve
further study.

Recall also that the issue of linear multiplet
couplings and supersymmetric Chern-Simons forms
is relevant in the context of string loop corrections to effective
gauge coupling functions \cite{DKL91}, \cite{Lou91}, \cite{CLO95} and
its relations
to \ka sigma model anomaly cancellation mechanisms \cite{CO91}, \cite{CO92a},
\cite{CO92b}, \cite{DFKZ92}.

From the point of view of general properties of supersymmetric
theories, as alluded to above, the situation may be interpreted as
follows. In distinction to the traditional approaches, \ie
without linear multiplets and Chern-Simons forms, where the effective
gauge coupling functions are restricted to be the sum of a
holomorphic and an anti-holomorphic function of the scalar matter
superfields, the explicit calculations of string loop corrections
yielded non-holomorphic gauge coupling functions. This apparent
contradiction can be explained in the framework of theories which
include linear superfield and Chern-Simons couplings \cite{DFKZ92},
\cite{BGG91}, \cite{ABGG92}, already contained in the general
formulation proposed in \cite{BGGM87b}.

Again, in this kind of investigations, the Yang-Mills case is comparatively
well understood, whereas in the corresponding mechanisms involving
supersymmetric gravitational Chern-Simons forms many questions are
still awaiting a satisfactory answer.

The present paper is intended to contribute to a clarification and
a better understanding of the structure of supersymmetric theories
describing couplings of linear multiplets and Chern-Simons forms, in particular
gravitational ones.

The basic idea of the approach presented here is the use of methods
of superspace geometry, that is to proceed as far as possible in
terms of superfields in order to encode compactly the embarrassing
complications of explicit component field expressions. Done in an
appropriate way this allows to analyse concisely the principal
features of a supersymmetric theory, in particular when it comes to
the formulation of the invariant action used to describe the
supersymmetric dynamics. Only after having completely set the
stage in geometrical terms, the transition to
the description in terms of component
fields ({\em viz.} invariant action, supersymmetry transformation
laws, etc.) is performed, using standard textbook methods and
without any further ambiguities.

The basic strategy pursued in the present approach
will be to generalize the description which works very
well in the Yang-Mills case, to the gravitational problems.
Without pretending that this approach is the only possible one, we feel
nevertheless that for the time being it is the only realistic viable one.

One of the purposes of this paper is to show that the gravitational
case differs from the Yang-Mills case not only in
being technically more involved, but also in certain conceptual respects.
Loosely speaking this may be assigned to the fact that contrary to
Yang-Mills given as superspace geometry in the supergravity-matter
"background", the gravitational Chern-Simons forms are to be included
concisely in the supergravity geometry itself.

\indent

\subsection{Linear multiplet without supergravity}

\indent

The linear supermultiplet is the supersymmetric
extension of the antisymmetric tensor gauge potential
\be b_{mn} \ = \ - b_{nm}. \eeq

Historically the antisymmetric
tensor was studied already some time ago by V. I. Ogievetsky and
I. V. Polubarinov \cite{OP67}, later on it appeared in
the context of string theory
in the work of Kalb and Ramond \cite{KR74}.
The linear multiplet \cite{FZW74}, \cite{Sie79} is the prototype
supermultiplet which contains an antisymmetric tensor gauge field,
but there are other ones, in
$N=1$ four dimensional supergravity, the new-minimal multiplet,
as well as in extended and higher dimensional supersymmetry.
It was its ten-dimensional incarnation
which was used by Green and Schwarz in their anomaly-cancellation
mechanism \cite{GS84}.
In four dimensional effective theories this mechanism is expected to result,
among other things,
in couplings of a modified linear multiplet to the supergravity-matter
system. {\em Modified} means here that the effects of Chern-Simons forms of
Yang-Mills and gravitational types should be taken into account.

To begin with, we
present in this section the basic features of a linear multiplet with
Yang-Mills Chern-Simons forms in global supersymmetry.
This discussion is intended to provide a first impression of the
geometric methods used in the more general and complicated case
when the traditional supergravity-matter couplings and
gravitational Chern-Simons forms will be taken into account.

\indent

Consider first the non-supersymmetric case,
\ie the simple case of the antisymmetric tensor gauge
potential $b_{mn}$ in four dimensions with gauge transformations generated by a
four vector $\rho_m$ such that
\be
b_{mn} \ \mapsto \ b_{mn} + \prt_m \rho_n - \prt_n \rho_m,
\eeq
and with invariant field strength given as
\be
h_{0 \, lmn} \ = \ \prt_l b_{mn} + \prt_m b_{nl} + \prt_n b_{lm}.
\eeq
The subscript $0$ denotes here the absence of Chern-Simons forms.
As a consequence of its definition the field strength satisfies
the Bianchi identity
\be
\vep^{klmn} \prt_k h_{0 \, lmn}\ = \ 0.
\eeq
The invariant kinetic action is given as
\be
\cl \ = \ \f12 \, \hh_0^m \hh_{0 \, m},
\label{hh}
\eeq
with $\hh_0^k=\f{1}{3!} \vep^{klmn} h_{0 \, lmn}$ denoting
the dual of the field strength tensor.

\indent

Consider next the case where a Chern-Simons term
for a Yang-Mills potential $a_m$ is added such that
\be
h_{lmn} \ = \ h_{0 \, lmn}+ k \, Q_{lmn}.
\eeq
Here $k$ is a constant which helps keeping track of the terms induced by the
inclusion of the Chern-Simons combination
\be
Q_{lmn} \ = \ - \tr \lp a_{[l} \prt_m a_{n]} - \f{2i}{3} a_{[l} a_m a_{n]} \rp,
\label{cs}
\eeq
with $[lmn]=lmn + mnl + nlm - mln - lnm - nml$.
The gauge transformations of the Chern-Simons term
are compensated by assigning suitably
adjusted Yang-Mills gauge transformations to the antisymmetric
tensor, thus rendering the modified field strength invariant.
The presence of the Chern-Simons term modifies the Bianchi identity
as well, it reads now
\be
\vep^{klmn} \prt_k h_{lmn} \ = \ -\f{3}{2} k \, \vep^{klmn} \tr(f_{kl}f_{mn}).
\eeq

A dynamical theory may then be obtained from the invariant action
\be
\cl \ = \ \f12 \, \hh^m \hh_m -\f{1}{4} \, \tr(f^{mn}f_{mn}),
\label{hf}
\eeq
with $\hh^k=\f{1}{3!} \vep^{klmn} h_{lmn}$,
and Yang-Mills field strength
\be
f_{mn} \ = \ \prt_m a_n - \prt_n a_m - i \, [a_m,a_n].
\eeq
This action describes the dynamics of Yang-Mills potentials $a_m(x)$ and
an antisymmetric tensor gauge potential $b_{mn}$ with effective
$k$-dependent couplings induced through the Chern-Simons form.

\indent

This theory is dual to another one where the antisymmetric
tensor is replaced by a real scalar $a(x)$ in the following sense:
one starts from a first order action describing a
vector $X^m(x)$, a scalar $a(x)$ and the
Yang-Mills gauge potential $a_m(x)$,
\be
\cl \ = \ ( X^m - k \, \Qt^m ) \prt_m a
         + \f12 \, X^m X_m -\f{1}{4} \, \tr(f^{mn}f_{mn}),
\label{ford}
\eeq
where the gauge Chern-Simons form is included as
\be
\Qt^k \ = \ \f{1}{3!} \vep^{klmn} Q_{lmn}
      \ = \ - \vep^{klmn} \, \tr \lp a_l \prt_m a_n -\f{2i}{3} a_l a_m a_n \rp.
\eeq
Variation of the first order action with respect to the
scalar field $a$ gives rise to the equation of motion
\be
\prt_m (X^m -k\, \Qt^m) \ = \ 0,
\eeq
which is solved in terms of an antisymmetric tensor such that
\be
X^k - k \, \Qt^k \ = \ \f12 \vep^{klmn} \prt_l b_{mn}.
\eeq

Substituting back shows that the first term
in (\ref{ford}) becomes a total derivative and
one ends up with the previous action (\ref{hf})
with $\hh^m=X^m$, describing an
antisymmetric tensor gauge field coupled to a gauge Chern-Simons form.

On the other hand, varying the first order action with respect to $X^m$ yields
\be
X_m \ = \ - \prt_m a.
\eeq
In this case, substitution of the equation of motion, together with the
divergence equation for the Chern-Simons form, \ie
\be
\prt_k \Qt^k \ = \ - \f{1}{4} \vep^{klmn} \tr \lp f_{kl} f_{mn} \rp .
\eeq
gives rise to a theory describing a
real scalar field with an axion coupling term:
\be
\cl \ = \ - \f12 \prt^m a(x) \, \prt_m a(x)
          - \f{1}{4} \tr(f^{mn}f_{mn})
          - \f{k}{4} \, a(x) \, \vep^{klmn} \tr(f_{kl} f_{mn}).
\label{af}
\eeq

It is in this sense the two actions (\ref{hf}) and (\ref{af}) derived
here from the first order one (\ref{ford}) are dual to each other.
They describe the dynamics of an antisymmetric
tensor gauge field and of a real scalar, respectively, with special types of
Yang-Mills couplings. Observe that the kinetic term of the
Yang-Mills sector is not modified in this procedure.

\indent

We come now to the discussion of the globally supersymmetric case. The linear
supermultiplet consists of an antisymmetric tensor,
a real scalar and a Majorana spinor.
In superfield language it is described by a superfield $L_0$,
subject to the constraints\footnote{with the
usual notations $D^2=D^\al D_\al$ and
$\bD^2= D_\da D^\da$}
\be
D^2 L_0 \ = \ 0, \cem \bD^2 L_0 \ = \ 0.
\label{lincon}
\eeq
Again, the subscript $0$ means that we do not include, for the
moment, Chern-Simons forms. The linear superfield $L_0$ contains
the antisymmetric tensor only through its field strength $h_{0 \,
lmn}$. Indeed, the superfield $L_0$ is the supersymmetric analogue of
$h_{0 \, lmn}$ (it describes the multiplet of field strengths) and
the constraints (\ref{lincon}) are the supersymmetric version of
the Bianchi identities. The particular form of these constraints
implies that terms quadratic in $\th$ resp. in $\thb$ are
irrelevant (they are not independent component fields), it is for
this reason that $L_0$ has been called a {\em linear superfield}
\cite{FZW74}.

Instead of writing down explicitly the power series expansion in
$\th$, $\thb$ of the superfields and to identify the component
fields as the respective coefficient functions (keeping in mind the
constraint equations!), we shall use here suitable projections to
lowest superfield components for the identification of component
fields. This is reminiscent of the geometric superspace description
and convenient for keeping track of constraints and deriving
supersymmetry transformations (in particular later on in the case
of local supersymmetry \ie coupling to supergravity).

To begin with we identify the real scalar $L(x)$
of the linear multiplet as the lowest component
\be
L_0|_{\th=\thb=0} \ = \ L_0(x).
\eeq
The spinor derivatives of superfields are again superfields and we
define the Weyl components $(\La_\al (x), \Lab^\da (x))$ of the
Majorana spinor of the linear multiplet as
\be
D_\al L_0|_{\th=\thb=0} \ = \ \La_\al (x), \cem
D^\da L_0|_{\th=\thb=0} \ = \ \Lab^\da (x).
\eeq
The antisymmetric tensor appears in $L_0$ via its field strength
identified as
\be
\l[ D_\al, D_\da \r] L_0|_{\th=\thb=0} \ = \ - \f{1}{3} \si_{k \al \da}
                                         \, \vep^{klmn} h_{0 \, lmn},
\eeq
thus completing the identification of the independent component fields contained
in $L_0$.
The canonical supersymmetric kinetic action for the linear multiplet is then
given by the square of the linear superfield integrated over superspace.
In more explicit terms and in the language of projections
to lowest superfield components it is obtained from
\be
\cl \ = \ -\f{1}{32} \lp D^2 \bD^2 + \bD^2 D^2 \rp (L_0)^2|_{\th=\thb=0}.
\eeq
Evaluated in terms of component fields, it reads simply
\be
\cl \ = \ \f12 \hh_0^m \, \hh_{0 \, m}  -  \f12 \prt^m L_0 \, \prt_m L_0
                   -\f{i}{2} \si^m_{\al \da} ( \La^\al \prt_m \Lab^\da
                                        +\Lab^\da \prt_m \La^\al ),
\eeq
generalizing the purely bosonic action (\ref{hh}) given above and showing
that there is no auxiliary field in the linear multiplet.

\indent

We proceed now to introduce Chern-Simons forms in the
supersymmetric case, in other words to construct the
supersymmetric version of (\ref{hf}). As a prerequisite
we recall first some basic properties of the Yang-Mills
gauge multiplet.
It consists of the gauge potentials $a_m(x)$,
the gauginos $\la(x), \lab(x)$, which are Majorana spinors
and the auxiliary scalars $D(x)$. All of these
component fields are Lie-algebra valued.
They are identified in the gaugino superfields
$\cw^\al, \cw_\da$, which are Lie-algebra valued as well,
subject to the chirality conditions
\be
\cd_\al \cw^\da \ = \ 0, \cem \cd^\da \cw_\al \ = \ 0,
\label{ymc1}
\eeq
and to the additional constraints
\be
\cd^\al \cw_\al \ = \ \cd_\da \cw^\da.
\label{ymc2}
\eeq
The spinor derivatives occuring here are defined to be covariant
with respect to Yang-Mills transformations. Again, these constraint
equations have a geometric interpretation as Bianchi identities in
superspace.

To be more precise, the gaugino component fields are defined as
the lowest components of the gaugino superfields themselves,
\be
\cw_\al|_{\th=\thb=0} \ = \ -i \la_\al, \cem
\cw^\da|_{\th=\thb=0} \ = \ i \lab^\da,
\eeq
whereas the usual Yang-Mills field strengths $f_{mn}$ and the auxiliary
fields $D(x)$ occur at the linear level in the superfield
expansion, :
\bea
\cd_\bt \cw_\al |_{\th=\thb=0} &=& - i (\si^{mn} \eps)_{\bt \al} \, f_{mn}
                                  - \eps_{\bt \al} \, D(x), \nn \\
\cd_\db \cw_\da |_{\th=\thb=0} &=& - i (\eps \sib^{mn})_{\db \da} \, f_{mn}
                                  + \eps_{\db \da} \, D(x).
\ena

We come now to the supersymmetric description of the corresponding
Chern-Simons forms, that is the supersymmetric extension of
(\ref{cs}). As discussed in detail in appendix {\bf A}, it is described
in terms of the Chern-Simons superfield $\Om$, which has the
properties
\bea
\tr(\cw^\al \cw_\al) &=& \f{1}{2} \bD^2 \Om, \label{om1}
\\
\tr(\cw_\da \cw^\da) &=& \f{1}{2} D^2 \Om, \label{om2}
\ena
in accordance with the constraint equations (\ref{ymc1}) and (\ref{ymc2}):
the appearance of the differential operators $D^2$ and $\bD^2$
is due to the chirality constraint whereas the additional constraint
(\ref{ymc2}) is responsible for the fact that one and the same
superfield $\Om$ appears in both equations.
The component field Chern-Simons form (\ref{cs}) is then identified
in the lowest superfield component
\be
[D_\al , D_\da] \, \Om|_{\th=\thb=0} \ = \
- \f{1}{3} \si_{k \al \da} \, \vep^{klmn} Q_{lmn}
         - 4 \, \tr(\la_\al \lab_\da),
\eeq
with $Q_{lmn}$ given in eq. (\ref{cs}).

Since the terms on the left-hand sides in (\ref{om1}) and
(\ref{om2}) are gauge invariant, it is clear that a gauge
transformation adds a linear superfield to  $\Om$ (the explicit
construction is given in appendix {\bf A}). As a consequence, and in
analogy with the non-supersymmetric case discussed before, the
linear superfield $L_0$ can be assigned Yang-Mills transformations
such that the combination
\be   L \ = \ L_0 + k \, \Om,     \eeq
is gauge invariant. However, this superfield $L$ satisfies now
the modified linearity conditions
\bea
\bD^2 L &=& 2k \, \tr(\cw^\al \cw_\al),
\label{modlin1}
\\
D^2 L &=& 2k \, \tr(\cw_\da \cw^\da),
\label{modlin2}
\ena
Again, these equations together with
\be
[D_\al , D_\da] L \ = \ - \f{1}{3} \si_{d \al \da} \, \vep^{dcba} H_{cba}
                               - 4k \, \tr(\cw_\al \cw_\da),
\eeq
have an interpretation as Bianchi identities in superspace geometry.
The last one shows how the usual field strength of the
antisymmetric tensor together with the component field
Chern-Simons form appears in the superfield expansion of $L$:
\be
[D_\al , D_\da] L|_{\th=\thb=0} \ = \
\si_{k \al \da} \, \vep^{klmn} \lp \prt_n b_{ml} + \f{k}{3} Q_{nml} \rp
 - 4k \, \tr(\cw_\al \cw_\da).
\eeq

The invariant action for this supersymmetric system is given
as the lowest component of the superfield
\be
\cl \ = \ -\f{1}{32} \lp D^2 \bD^2 + \bD^2 D^2 \rp L^2
         -\f{1}{16} D^2 \tr(\cw^2)
         -\f{1}{16} \bD^2 \tr(\cwb^2).
\label{sushf}
\eeq
This action describes the supersymmetric version of the
purely bosonic action (\ref{hf}).
Its explicit component field gestalt will be displayed and
commented on in a short while.

As is well known \cite{LR83}, the notion of duality as described
above in the non-supersymmetric case, can be extended to
supersymmetric theories as well. This is most conveniently done in
the language of superfields. The supersymmetric version of the
first order action (\ref{ford}) is given as
\be
\cl \ = \ -\f{1}{32} \lp D^2 \bD^2 + \bD^2 D^2 \rp
                   \lp X^2 + \s2 (X - k \Om)(S + \bS) \rp
         -\f{1}{16} D^2 \tr(\cw^2) -\f{1}{16} \bD^2 \tr(\cwb^2).
\label{susford}
\eeq
 Here, $X$ is a real but otherwise unconstrained superfield,
whereas $S$ and $\bS$ are chiral,
\be
D_\al \bS \ = \ 0, \cem \bD^\da S \ = \ 0.
\eeq
Of course, the chiral multiplets are going to play the part of the scalar
field $a(x)$ in the previous non-supersymmetric discussion.

Varying the first order action with respect to the superfield $S$,
or, more correctly with respect to its unconstrained prepotential
$\Si$, defined as $S=\bD^2 \Si$, the solution of the chirality
constraint, shows immediately (upon integration by parts using
spinor derivatives) that the superfield $X$ must satisfy the
modified linearity condition. It is therefore identified with $L$
and we recover the action (\ref{sushf}) above.

On the other hand, varying the first order action (\ref{susford}) with respect
to $X$ yields the superfield equation of motion
\be X \ = \ -\f{1}{\s2} (S + \bS). \eeq
Substituting for $X$ in (\ref{susford}) and neglecting terms $S^2$
and $\bS^2$ which are trivial upon superspace integration, we
arrive at
\be
\cl \ = \ \f{1}{32} \lp D^2 \bD^2 + \bD^2 D^2 \rp
                   \lp \bS S + k \s2 \, \Om \, (S + \bS) \rp
         -\f{1}{16} D^2 \tr(\cw^2) -\f{1}{16} \bD^2 \tr(\cwb^2).
\eeq
It is already obvious to recognize the usual superfield kinetic
term for the chiral multiplet and the Yang-Mills kinetic terms, it
remains to have a closer look at the terms containing the
Chern-Simons superfield. Taking into account the chirality
properties for $S$ and $\bS$ and the derivative relations
(\ref{om1}) and (\ref{om2}) for the Chern-Simons superfields we
obtain
\bea
\cl &=& \f{1}{32} \lp D^2 \bD^2 + \bD^2 D^2 \rp \bS S
             -\f{1}{16} D^2 tr(\cw^2) -\f{1}{16} \bD^2 \tr(\cwb^2) \nn \\[2mm]
       & &   +\f{k \s2}{8} D^2 \lp S \, \tr(\cw^2) \rp
             +\f{k \s2}{8} \bD^2 \lp \bS \, \tr(\cwb^2) \rp.
\label{susaf}
\ena
This action is now the supersymmetric version of the action
(\ref{af}).

\indent

We display now the component field expresssions
for the two dual versions (\ref{sushf}) and
(\ref{susaf}) of the supersymmetric construction.
In the antisymmetric tensor version, the complete invariant component
field action deriving from (\ref{sushf}) is given as
\bea
\cl &=& \f12 \hh^m \hh_m - \f12 \prt^m L \ \prt_m L
                  -\f{i}{2} \si^m_{\al \da} \lp \La^\al \prt_m \Lab^\da
                           + \Lab^\da \prt_m \La^\al \rp \nn \\[2mm]
& & + (1+2kL) \, \tr \left[ - \f{1}{4} f^{mn} f_{mn}
                   - \f{i}{2} \si^m_{\al \da} \lp \la^\al \cd_m \lab^\da
            + \lab^\da \cd_m \la^\al \rp
  + \f{1}{2} \widehat{D} \, \widehat{D} \right] \nn \\[2mm]
& & -k \, \hh^m \, \tr(\la \si_m \lab)
    -k \, \La \, \si^{mn} \, \tr(\la f_{mn})
    -k \, \Lab \, \sib^{mn} \, \tr(\lab f_{mn}) \nn \\[2mm]
& & - \f{k^2}{4}(1+2kL)^{-1} \lp \La^2 \, \tr \la^2 + \Lab^2 \, \tr \lab^2
             - 2 \La \si^m \Lab \ \tr(\la \si_m \lab) \rp \nn \\[2mm]
& & - \f{k^2}{2} \lp \tr \la^2 \ \tr \lab^2
                  - \tr (\la \si^m \lab) \ \tr (\la \si_m \lab) \rp.
\label{comphf}
\ena
This is the supersymmetric version of (\ref{hf}).
The redefined auxiliary field
\be
\widehat{D} \ = \ D + \f{ik}{1+2kL} (\La \la - \Lab \lab),
\eeq
has trivial equation of motion.

On the other hand, in order to display the component field
Lagrangian in the chiral superfield version, we recall
the definition of the component field content of the chiral
superfields
\be
S|_{\th=\thb=0} \ = \ S(x), \cem D_\al S|_{\th=\thb=0} \ = \ \s2 \, \chi_\al(x),
\cem D^2 S|_{\th=\thb=0} \ = \ -4F(x),
\eeq
and
\be
\bS|_{\th=\thb=0} \ = \ \bS(x), \cem
\bD^\da \bS|_{\th=\thb=0} \ = \ \s2 \, \chib^\da (x), \cem
\bD^2 \bS|_{\th=\thb=0} \ = \ -4\bF(x).
\eeq

The component field action in the dual formulation, derived from
the superfield action (\ref{susaf}) takes then the form
\bea
\cl &=& - \prt^m \bS \, \prt_m S
        - \f{i}{2} \si^m_{\al \da} \lp \chi^\al \prt_m \chib^\da
                                       +\chib^\da \prt_m \chi^\al \rp
+ \widehat{F} \widehat{\bF} \nn \\[2mm]
& & + \lp 1-k\s2 \, (S+\bS) \rp \tr \left[ -\f{1}{4} f^{mn}f_{mn}
            -\f{i}{2} \si^m_{\al \da} \lp \la^\al \cd_m \lab^\da
                     + \lab^\da \cd_m \la^\al \rp
  +\f12 \widehat{D} \, \widehat{D} \right] \nn \\[2mm]
& & -\f{k}{4i\s2}(S-\bS) \left[ \vep^{klmn} \tr(f_{kl} f_{mn})
                        + 4 \prt_m \tr (\la \si^m \lab) \right] \nn \\[2mm]
& & +k \, \chi \si^{mn} \, \tr (\la f_{mn})
    +k \, \chib \sib^{mn} \, \tr (\lab f_{mn})
        -\f{k^2}{8} \, \tr \la^2 \, \tr \lab^2 \nn \\[2mm]
& & -\f{k^2}{4} \lp 1-k\s2 \, (S+\bS) \rp^{-1}
 \lp \chi^2 \, \tr \la^2 + \chib^2 \, \tr \lab^2
             -2 (\chi \si^m \chib) \, \tr (\la \si_m \lab) \rp.
\label{compaf}
\ena
This is the supersymmetric version of (\ref{af}).
Again, we have introduced the diagonalized combinations
for the auxiliary fields
\be
\widehat{F} \ = \ F + \f{k\s2}{4} \, \tr \lab^2, \cem
\widehat{\bF} \ = \ \bF + \f{k\s2}{4} \, \tr \la^2,
\eeq
and
\be
\widehat{D} \ = \
D - \f{ik}{1-k\s2 \, (S+\bS)} \, (\chi \la - \chib \lab).
\eeq

\indent

The two supersymmetric actions (\ref{comphf}) and (\ref{compaf})
are dual to each other, in the precise sense of the construction
performed above. In both cases the presence of the Chern-Simons
form induces $k$-dependent effective couplings, in
particular quadrilinear spinor couplings. Also, one recognizes
easily the axion term in the second version already encountered
in the purely bosonic case discussed before.

A striking difference to the non-supersymmetric case, however, is
the appearance of a $k$-dependent gauge coupling function,
multiplying the Yang-Mills kinetic terms. This shows that
supersymmetrization of (\ref{hf}) and (\ref{af}) results not only
in supplementary fermionic terms, but induces also genuinely new
purely bosonic terms.

\indent

\subsection{3-form multiplet}

\indent

Before turning to supergravity and to our main subject, gravitational
Chern-Simons forms, let us close this introduction with some
remarks on the 3-form gauge supermultiplet. This is, besides
the chiral and linear multiplet, yet another supermultiplet
describing helicity $(0,1/2)$. It consists of a three-index
antisymmetric gauge potential $C_{lmn}(x)$,
a complex scalar $T(x)$, a Majorana spinor with Weyl
components $\eta_\al(x)$, $\eta^\da(x)$ and a real scalar
auxiliary field $H(x)$. In superfield language
\cite{Gat81}, \cite{BPGG96} it
is described by a chiral superfield
\be
\bD^\da T \ = \ 0, \cem D_\al \bT \ = \ 0,
\eeq
which is subject to the additional constraint
\be
D^2 T - \bD^2 T \ = \ \f{8i}{3} \vep^{klmn} \Si_{klmn},
\eeq
with
\be
\Si_{klmn} \ = \ \prt_k C_{lmn} - \prt_l C_{mnk}
               + \prt_m C_{nkl} - \prt_n C_{klm},
\eeq
the field strength tensor of the three-index gauge potential
superfield. It is invariant under the transformation
\be
C_{lmn} \ \mapsto \ C_{lmn} + \prt_l \xi_{mn}
                    + \prt_m \xi_{nl} + \prt_n \xi_{lm},
\eeq
where the gauge parameter $\xi_{mn} = - \xi_{nm}$ is a 2-form.

An explicit realization of this multiplet structure is provided by
the composite superfield $\tr(\cw^2)$ and its complex conjugate
$\tr(\cwb^2)$. As the gaugino superfield appearing here is chiral
(\ref{ymc1}), these composites are chiral, resp. antichiral as well,
\be
\bD^\da \, \tr(\cw^2) \ = \ 0, \cem D_\al \, \tr(\cwb^2) \ = \ 0.
\eeq
On the other hand the gaugino superfields are subject to
an additional constraint (\ref{ymc2}), which translates into
an additional equation for the composites as well, namely
\be
D^2 \tr(\cw^2) - \bD^2 \tr(\cwb^2) \ = \ i \vep^{klmn} \tr(f_{kl}
f_{mn}),
\eeq
where the topological density
\be
\vep^{klmn} \tr(f_{kl}f_{mn}) \ = \ -\f{2}{3} \vep^{klmn} \prt_k Q_{lmn},
\eeq
plays now the r\^ole of the field-strength and the Chern-Simons form
(which, under Yang-Mills transformations changes indeed by the
derivative of a 2-form) the r\^ole of the 3-form gauge
potential. In other words, supersymmetric Chern-Simons forms fit
perfectly in the framework of the 3-form multiplet. It is this
analogy which will be exploited in this paper for the description
of supersymmetric Chern-Simons forms, in particular in the
gravitational case.


\sect{The basic superspace structures}

\indent

\subsection{Outline}

\indent

The purpose of this paper is to discuss the properties of locally supersymmetric
theories which contain gravitational Chern-Simons forms coupled via the
fieldstrengths of antisymmetric tensor gauge fields to the standard
supergravity-matter system.

So far, in spite of a number of efforts, no satisfactory answer has been
given to this problem. This is, in part, due to the
formidable technical complexity of such a theory.
Not only does one have to understand the structure of the multiplets
involved, but one should also be able to identify the component
fields and to derive the complete structure of their supersymmetry
transformation laws, not to forget the construction of invariant actions
describing supersymmetric dynamics.

In order to cope with the technical complexities, we propose to
employ methods of superspace geometry.
One of the advantages in using this approach is that a great deal of the
investigations can be carried out at a purely
geometrical level. This is in particular true for the structure of the
supersymmetry transformations, but also for issues like \ka transformations
which arise as a consequence of supersymmetry in supergravity-matter
coupling.

Moreover, when it comes to supersymmetric dynamics, superspace
provides methods to determine invariant
actions and to discuss their properties in a concise way.
Finally, component field results, in particular complete
supersymmetric component field actions with all their embarrassing wealth
of couplings can be derived.

We begin, in subsection {\bf 2.2}, with the description of the
superspace structure relevant for the {\em supergravity-matter system},
namely the general coupling of chiral superfields to supergravity
and supersymmetric Yang-Mills theory. In this formulation the \ka
structure is properly taken into account ab initio:
\ka transformations appear in the structure group
of superspace as field dependent chiral $U_K(1)$ transformations.
This kind of superspace geometry is called $U_K(1)$ superspace.

Moreover, the kinetic terms for the supersymmetric
sigma-model appear through a $D$-term construction for a superfield \ka potential,
with chiral superfields taking the r\^ole of the complex coordinates.

In subsection {\bf 2.3} we present a short reminder of the
generic method for constructing actions, invariant at the same time
with respect to supersymmetry and $U_K(1)$ transformations.

This then provides the geometrical background for the description
of the linear multiplet and of the various types of Chern-Simons
forms. The bare linear multiplet (\ie in the absence of
Chern-Simons forms) arises from the superspace
geometry of the 2-form gauge potential. In subsection {\bf 2.4}
this superspace geometry will be presented in $U_K(1)$ superspace,
featuring the corresponding {\em linearity conditions}.

In subsection {\bf 2.5} we first display the salient properties
of supersymmetric Yang-Mills theory in $U_K(1)$ superspace,
in particular the corresponding Chern-Simons forms.
In combination with the linear superfield geometry this gives then rise to the
coupling of Chern-Simons form and antisymmetric tensor gauge field,
summarized, at the superfield level, in terms of the
so-called {\em modified linearity conditions}.
In the course of this construction we also point out the close relation
between Chern-Simons forms and the superspace geometry of a 3-form
gauge potential.

\indent

\subsection{$U_K(1)$ superspace}

\indent

Supergravity is a generalization of general relativity. Since
supersymmetry, by definition, brings in fermionic degrees of
freedom, the relevant formulation of Einstein gravity is in terms
of vierbein field (local Lorentz frames) and spin connection. As is
well known, the supersymmetric extension consists then in adding
the Rarita-Schwinger field as the supersymmetry partner of the
vierbein together with certain auxiliary (usually non-propagating)
fields which serve to establish an off-shell realization of the
local supersymmetry algebra.

Supergravity may be viewed as the gauged theory of supersymmetry:
the anticommuting parameters of supersymmetry transformations
become space-time dependent and it is the Rarita-Schwinger field,
which, under supersymmetry transformations, acquires an
inhomogeneous term proportional to the (covariant) space-time
derivative of the local supersymmetry parameter.

In superspace one generalizes the notions of local frame and spin connection
in extending them to the anticommuting
directions of superspace equipped with a full-fledged graded differential
geometry.

In some more detail, the usual frame, viewed as a differential form over
space-time,
\be e^a \ = \ dx^m e_m{}^a(x), \eeq
is extended to a differential form over superspace,
\be E^A \ = \ dz^M E_M{}^A (z), \eeq
where transition from lower case to upper case indices signifies the passage
from ordinary to superspace geometry, based on coordinates
$z^M = (x^m,\th^\mu,\thb_\dmu)$. Accordingly,
\be E^A \ = \ \left(E^a,E^\al,E_\da \right), \eeq
has vectorial and spinor indices, the latter in Weyl-spinor
notation. In this general set-up the usual vierbein and
Rarita-Schwinger fields are identified as lowest superfield
components, \ie
\be
e_m{}^a(x) \ = \ E_m{}^a(x,0,0) \ = \ E_m{}^a|,
\eeq
and
\be
\f12 \psi_m{}^\al(x) \ = \ E_m{}^\al|,
\cem \f12 \psib_{m \, \da}(x) \ = \ E_{m \, \da}|.
\eeq

The symmetries in this superspace description are general
supercoordinate transformations, unifying the usual general
coordinate transformations and the local supersymmetry
transformations in their vector and spinor parts, respectively, and
local Lorentz transformations, in turn acting through vector and
spinor representations on $E^a$ and $E^\al, E_\da$.

As to the first, covariance is achieved through the use of differential form
and inverse frame language in superspace, whereas covariance with respect to
local Lorentz transformations is ensured by covariant derivatives using the
spin connection.

These are the symmetries of pure supergravity. If one wants to include
supersymmetric matter, described in terms of chiral superfields and their
complex conjugates, a new additional geometric structure shows up:
the chiral matter superfields in the general supergravity-matter system
are coordinates of a {\em superfield \ka manifold}. This property arises from
the requirement of supersymmetry of the dynamical theory, as pointed out
by Zumino already for the supersymmetric sigma model without supergravity.
When coupled to supergravity the sigma model action was recognized to be
\ka invariant provided the spinor fields (including those of supersymmetric
Yang-Mills theory) transform under well-prescribed chiral phase transformations
whose parameters are given as the imaginary part of the field dependent
\ka transformations.

It is precisely the \ka superspace formulation which clarifies this situation.
In particular it puts the structure of the chiral \ka transformations
on a sound geometrical basis, without reference to the dynamical construction.
The key mechanism of this formulation is to include the chiral \ka phase
transformations into the structure group of superspace, on the same footing as
the local Lorentz transformations.

The \ka potential appears then quite naturally as a prepotential
for these chiral transformations. Moreover, its $D$-term provides
an action invariant at the same time (and for the same reasons)
under supersymmetry and under (superfield) \ka transformations.

This formulation of the supergravity/matter/Yang-Mills system is presented in
full detail in \cite{BGG90}. At present, in this section we will consider a
generic chiral $U_K(1)$ and identify its prepotential $K$ with the \ka
potential only afterwards. More generally, as explained in \cite{BGGM87b},
\cite{BGG91},
\cite{ABGG92}, it may be allowed to depend on linear superfields as well, in
which case we will refer to it as {\em kinetic prepotential}.

Coming back to the basic object of the superspace formulation,
namely the frame $E^A$ in superspace, this means that in addition
to the aforementioned general supercoordinate and Lorentz
transformations we assign chiral transformations (in terms of a chiral
superfield $F$ and its complex conjugate) such that
\footnote{Given the explicit form of these chiral transformations, the issue of
gauged $R$-transformation comes immediately to ones mind.}
\be
E^A \ \mapsto \ E^A \, \exp \l[ -\f{i}{2} w(E^A) \ \im \, F \r],
\eeq
with chiral weights $w(E^A)$ defined as
\be
w(E^a) \ = \ 0, \cem w(E^\al) \ = \ 1, \cem w(E_\da) \ = \ -1.
\eeq

In view of all this, superspace torsion is defined as
\be
T^A \ = \ dE^A + E^B {\phi_B}^A + w(E^A) E^A A,
\eeq
that is, just the covariant exterior derivative of the frame in superspace.
The first two terms on the right are standard, in particular, the spin
connection is a one form in superspace,
\be
{\phi_B}^A  \ = \ dz^M {\phi_{MB}}^A (z),
\eeq
taking values in the Lie-algebra of the Lorentz group such that its
spinor components are given in terms of the vector ones as
\be
{\phi_\bt}^\al \ = \ - \f{1}{2} {(\si^{ba})_\bt}^\al
\phi_{ba}, \cem
{\phi^{\db}}_{\da} \ = \ - \f{1}{2} {(\sib^{ba})^{\db}}_{\da}
\phi_{ba}.
\eeq
The abelian gauge potential
\be
A \ = \ dz^M A_M(z),
\eeq
is new: it serves to
covariantize the chiral $U_K(1)$ transformations,
\be
A \ \mapsto \ A + \f{i}{2} \, d \, \im \, F.
\eeq

We do not intend here to give a complete and detailed
review of this geometrical structure.
For this we refer to our earlier work.
Here we try to concentrate on the crucial points which will be of relevance
later on in the discussion of the structure of linear superfield geometry and
of Chern-Simons forms in superspace. Recall nevertheless the definitions
of the fieldstrengths
\bea
{R_B}^A &=& d {\phi_B}^A \ + \ {\phi_B}^C {\phi_C}^A,
\\
F &=& dA.
\ena

Torsion, curvature and $U_K(1)$ fieldstrength are 2-forms in superspace,
their expansion in the covariant frame basis being defined as
\bea
T^A &=& \f{1}{2} E^B E^C {T_{CB}}^A,
\\
{R_B}^A &=& \f{1}{2} E^C E^D {R_{DC \ B}}^A,
\\
F &=& \f{1}{2} E^C E^D F_{DC}.
\ena

Recall that superspace torsion is subject to covariant constraints which
imply that all the coefficients of torsion, curvature and $U_K(1)$
fieldstrength are given in terms of the few
covariant supergravity superfields
\be
R, \cem \rd, \cem G_a, \cem W_{\lsym{\gm \bt \al}}, \cem
W_{\lsym{{\dg} {\db}\da}},
\eeq
and their covariant derivatives. As all these basic superfields are identified
in the torsion coefficients, with chiral weights
\be
w(T_{CB}{}^A) \ = \ w(E^A) - w(E^B) - w(E^C),
\eeq
their chiral weights are fixed to be
\bea
w(R) \ = \ 2, \cem w(\rd) &=& -2, \cem w(G_a) \ = \ 0, \nonumber \\
w(W_{\lsym{\gm \bt \al}}) \ = \ 1, \cem & &
w(W_{\lsym{{\dg} {\db}\da}}) \ = \ -1.
\ena

To be more explicit, the nonvanishing components of superspace torsion are
\be
{T_\gm}^{{\db} a} \ = \ -2i {(\si^a \eps)_\gm}^{\db},
\eeq
\be
T_{\gm b \da} \ = \ -i \si_{b \gm \da} \rd, \cem
T^{\dg \ \ \al}_{\ \ b} \ = \ -i {\sib_b}^{\dg \al} R,
\eeq
\be
{T_{\gm b}}^\al \ = \ \f{i}{2} {(\si_c \sib_b)_\gm}^\al G^c, \cem
{T^\dg}_{b \da} \ = \ -\f{i}{2} {({\sib_c \si_b)}^{\dg}}_\da G^c,
\eeq
and ${T_{cb}}^\al$ and ${T_{cb}}_\da$, the covariant
Rarita-Schwinger fieldstrength superfields. The superfields
$W_{\lsym{\gm \bt \al}}$ and $W_{\lsym{\dg \db\da}}$ are called
Weyl spinor superfields, because they occur in the decomposition of
these Rarita-Schwinger superfields in very much the same way as the
usual Weyl tensor occurs in the decomposition of the covariant
curvature tensor. For a detailed account of basic superspace geometry see
\cite{WeB83}, \cite{BGG90}.

Consistency of the superspace Bianchi identities with the special
form of the torsion components displayed so far implies the
chirality conditions:
\be
{\cd}_\al \rd \ = \ 0, \cem {\cd}_{\da} R \ = \ 0,
\eeq
\be
{\cd}_\al \bar{W}_{\lsym{{\dg} {\db} \da}} \ = \ 0 , \cem
{\cd}_{\da}W_{\lsym{\gm \bt \al}} \ = \ 0.
\eeq

Leaving aside, for the moment, the description  of the explicit
form of the curvatures we turn to the fieldstrengths of the chiral
$U_K(1)$ sector, which are described in terms of the superfields
$X_\al$ and $\Xb^\da$, defined as spinor derivatives of the basic
superfields $R$, $\rd$ and $G_a$ as follows:
\bea
X_\al &=&\cd_\al R -\cd^\da G_{\al \da},
\\
{\Xb}^{\ \da} &=&\cd^\da \rd + \cd_\al G^{\al \da}.
\ena
Again, consistency with the Bianchi identities implies chirality, \ie
\be
\cd_\al {\Xb}^\da \ = \ 0, \cem \cd^\da X_\al \ = \ 0,
\eeq
as well as the condition
\be
\cd^\al X_\al - \cd_\da {\Xb}^\da  \ = \  0.
\eeq

As  mentioned earlier, the coefficients of the $U_K(1)$ gauge
potential $A$,
\be
A \ = \ E^A A_A \ = \ E^a A_a + E^\al A_\al + E_\da A^\da,
\eeq
are given in terms of the $U_K(1)$ prepotential superfield $K$ as
\bea
A_\al & = &+\f{1}{4}{E_\al}^M \prt_M K
\label{Aal}
\\
A^\da & = &-\f{1}{4}E^{\da M} \prt_M K
\label{Ada}
\\
A_{\al \da}-\f{3i}{2} G_{\al \da} & = & \f{i}{2}
(\cd_\al A_\da + \cd_\da A_\al).
\label{Av}
\ena

In the last equation the $G_{\al \da}$ term appears due to the special choice
$F_{\al \da} = -3 G_{\al \da}$ of conventional constraint
for the $U_K(1)$ fieldstrength.
Using this explicit form of $A$ yields then
\bea
X_\al & = & -\f{1}{8} \qq \cd_\al K,
\\
\Xb^\da & = & -\f{1}{8} \pp \cd^\da K.
\ena
As a consequence one has then
\be \cd^\al \! X_\al \ = \ \cd_\da \Xb^\da, \eeq
for the $D$-term pertaining to the $U_K(1)$ factor.
As long as $K$ is an independent superfield, the lowest component of the
superfield $\cd^\al \! X_\al$ is an independent component, as usually in
supersymmetric gauge theory. On the other hand if one allows $K$ to be a
function of chiral and linear superfields, this $D$-term yields the
corresponding kinetic terms after successive applications of the spinorial
covariant derivatives to $K$ and due to the chirality and linearity
conditions.

The superfield $\cd^\al \! X_\al$ is related to the basic supergravity
superfields such that
\be
\cd^2 R + \cdb^2 \rd \ = \ -\f{2}{3} \, \car -\f{2}{3} \, \cd^\al \! X_\al
         + 4 \, G^a G_a + 32 \, \rd R,
\eeq
where $\car$ is the curvature scalar (see appendix {\bf B} for notational
details).
This relation is at the heart of the construction of the supersymmetric
component field action, as will become clear in the following subsection.
On the other hand, the orthogonal combination
\be \cd^2 R + \cdb^2 \rd \ = \ 4i \, \cd_a G^a, \eeq
has an intriguing resemblance to the 3-form superspace, as reviewed in
appendix {\bf A}.

\indent

\subsection{Construction of generic invariant actions}

\indent

An important topic which can already be addressed here is the question
of constructing actions invariant under supersymmetry transformations.
As in the geometrical formulation outlined so far, supersymmetry
transformations occur in the general supercoordinate transformations,
the method of constructing invariant actions proceeds along the same
lines as in usual general relativity, namely with the help of invariant
densities. In general relativity the basic object is the determinant
$e(x)$ of the vierbein field $e_m{}^a(x)$.
It was pointed out by Wess and Zumino
that in supergravity this should be generalized to the superdeterminant
$E(x,\th,\thb)$ of $E_M{}^A(x,\th,\thb)$, the frame in superspace, as
introduced above, together with the the usual concept of superspace
integration.

It is not hard to see that the construction of
Wess and Zumino, originally performed in traditional superspace, holds
in $U_K(1)$ superspace as well.
This is due to the fact that the geometric framework concerning general
superspace coordinate transformations remains the same in $U_K(1)$
superspace, what changes is the structure group - but the determinant
of the frame is invariant under the new chiral phase transformations.

Already in the original formulation it is quite intriguing that
the complete action for supergravity is given by the
superspace volume element
\[ \int d^4 \! x \; d^4 \th  \, E. \]

Even more amazing, when matter fields are included through $U_K(1)$
superspace, this action describes the kinetic terms for the
complete supergravity-matter system, the differences to the
previous case arising of course from the different geometric
structures to be taken into account in evaluating the corresponding
component field expressions. In this sense one might speak of a
unified description of gravity and matter fields, the complete
action arising from one single distinguished geometrical object -
the $U_K(1)$ superspace volume element.

In both cases the component field Lagrangians deduced from these
superspace expressions contain the Einstein
curvature scalar term
\[ -\f12 \, e \, \car \]
with the usual canonical normalization.
The use of $U_K(1)$ superspace avoids the cumbersome component field
rescalings of the original constructions of Cremmer et. al..

In the construction of the supersymmetric superpotential term
or the kinetic action for supersymmetric Yang-Mills theory one employs
the so-called chiral volume elements of superspace,
\[
\int d^4 \th \, \frac{E}{R} \, r \ \sim \ \int d^2 \th \, \ce r,
\cem \mbox{and} \cem
\int d^4 \th \, \frac{E}{\rd} \, \br \ \sim \ \int d^2 \thb \, \bar{\ce} \br,
\]
where $R$ and $\rd$ are the chiral supergravity superfields appearing in
the torsion as explained above and $r$ and $\br$ are generic chiral
superfields of weights $w(r)=+2$ and $w(\br)=-2$, respectively, which should
be specified according to the kind of invariant action one intends to
construct. Observe that in particular the choice $r=R$ gives back the kinetic
supergravity actions discussed above.

In other words, this chiral density construction is the
generalization to local supersymmetry of what is called the
$F$-term construction in rigid supersymmetry, applied to the
generic chiral superfield $r$. The explicit algorithm consists then
simply in writing out the superymmetric completion of the component
field expression of the $F$-term $\cd^2 r|$ according to
\be
\cl_{gen} \ = \
- \f{1}{4} e \lp \cd^2 - 24 \rd \rp r|
+ \f{i}{2} e (\psib_m \sib^m)^\al \cd_\al r|
- e (\psib_m \sib^{mn} \psib_n) r| \ + \ h.c. \ .
\label{2.36}
\eeq
Here, the lowest components of the supergravity superfields $R$ and
$\rd$ are defined as
\be R| \ = \ -\f{1}{6} M(x), \cem \rd| \ = \ -\f{1}{6} \bM(x). \eeq
Hence, in this prescription for the construction of a
supersymmetric component field action one has to choose appropriately
some chiral superfield $r$ with $w(r)=2$, work out the projections to
component fields of $r$, $\cd_\al r$ and $\cd^2 r$ and substitute in the
equation for $\cl_{gen}$. The supergravity-matter action (with properly
normalized curvature scalar term) is then obtained in taking the superfields
$R$ and $\rd$, thus justifying the remark at the end of the previous
subsection, whereas the Yang-Mills action is obtained from the
superfield $\cw^\al \cw_\al$ and its complex conjugate.

\indent

\subsection{Linear multiplet geometry and supergravity}

\indent

The linear multiplet has a geometrical interpretation as a 2-form gauge
potential in superspace geometry. Since we wish to construct theories where
the linear multiplet is coupled to the supergravity-matter system, we will
formulate this 2-form geometry in the background of $U_K(1)$
superspace. The basic object is the 2-form gauge potential defined as
\[ B \ = \ \f12 dz^M dz^N \, B_{NM}. \]
It is subject to gauge transformations of parameters $\xi=dz^M \xi_M$ which are
themselves one forms in superspace:
\[ B \ \mapsto \ B + d \xi, \]
or, in more detail,
\[ B_{NM} \ \mapsto \ B_{NM} + \prt_N \xi_M - (-)^{d(N) d(M)} \prt_M \xi_N. \]
The invariant fieldstrength is a 3-form, defined as
\[ H \ = \ dB. \]
As a 3-form in superspace, $H$ is given as
\[ H \ = \ \frac{1}{3!} E^A E^B E^C H_{CBA}, \]
with $E^A$ the frame of $U_K(1)$ superspace.
As a consequence of $dd=0$ one obtains the {\em Bianchi identities}
\[ dH \ = \ 0. \]
Fully developed this reads
\be
\f{1}{4!} E^A E^B E^C E^D \lp 4 \, \cd_D H_{CBA}
                + 6 \, T_{DC}{}^F H_{FBA} \rp \ = \ 0.
\eeq

The linear superfield is recovered from this general
structure in imposing covariant constraints on the fieldstrength
coefficients $H_{CBA}$, a rather common procedure in the
superspace formulation of supersymmetric theories. The constraints
to be chosen here are
\be
H_{\undgm \undbt \undal} \ = \ 0, \cem
H_{\gm \bt a} \ = \ 0, \cem H_{\dg \db a} \ = \ 0,
\eeq
where as usual underlined indices serve to denote both dotted and
undotted ones, $\undal=(\al,\da)$. The consequences of these constraints
on the other coefficients are obtained either by explicitly solving
the constraints in terms of (unconstrained) prepotentials or else
by working through the covariant Bianchi identities.
As a result one finds that all the fieldstrength
components of the 2-form are expressed in terms of one superfield
$L$ which is identified in
\be
H_{\gm}{}^\db{}_a \ = \ -2i \, (\si_b \eps)_\gm{}^\db \ L.
\eeq
Furthermore one obtains
\bea
H_{\gm ba} &=& 2 (\si_{ba}{)_\gm}^\vp \cd_\vp L, \\
H^\dg_{\ \ ba} &=& 2 (\sib_{ba}{)^\dg}_\dv \cd^\dv L.
\ena

Compatibility of the constraints imposed above with the
structure of the Bia\-nchi iden\-ti\-ties then implies the
linearity conditions
\bea
\qq L &=& 0,
\\
\pp L &=& 0,
\ena
for a linear superfield in interaction with the supergravity-matter
system. Finally, the vector component $H_{cba}$ appears at the
level
\be
\lp [\cd_\al , \cd_\da] - 4 \si^a_{\al \da} G_a \rp L
\ = \  - \f{1}{3} \si_{d \, \al \da} \, \vep^{dcba} H_{cba}.
\eeq
In terms of component fields this means that $H_{cba}$ is
identified in the $\th \thb$ component (in the language where
superfield expansion is defined through successive application of
covariant spinor derivatives).

Let us close this subsection with a few remarks concerning the
definition of the component fields of the linear multiplet
as obtained from this superspace formulation, in particular the
identification of the antisymmetric tensor gauge potential:
\be
B|| \ = \ b \ = \ \f{1}{2} dx^m dx^n b_{nm}(x).
\eeq
The so-called double bar construction \cite{BBG87}, used here, projects at the
same time superspace differentials on their purely vector parts,
$dz^M \mapsto dx^m$, and the corresponding superfield coefficients
on their lowest superfield components at $\th = \thb = 0$. For the
covariant fieldstrength, this gives :
\be
H|| \ = \ h \ = \ db \ = \ \f{1}{3!} dx^l dx^m dx^n h_{nml},
\eeq
or
\be
h_{nml} \ = \ \prt_n b_{ml} + \prt_m b_{ln} + \prt_l b_{nm}.
\eeq
We also shall frequently make use of the dual, defined as
\be
\hh^k \ = \ \f{1}{3!} \vep^{klmn} h_{lmn}
      \ = \ \f{1}{2} \vep^{klmn} \prt_l b_{mn}.
\eeq

On the other hand, if the double bar projection is applied to the
expansion of $H$ in terms of the covariant frame, as given above,
we have to use the projection
\be
E^a|| \ = \ e^a(x) \ = \ dx^m e_m{}^a(x),
\eeq
for the vector part and
\be
E^\al|| \ = \ e^\al(x) \ = \ \f12 dx^m \psi_m{}^\al(x), \cem
E_\da|| \ = \ e_\da(x) \ = \ \f12 dx^m \psib_{m \da}(x),
\eeq
for the spinor ones. Using then the decomposition
\be
H|| \ = \ \f{1}{3!} e^a e^b e^c H_{cba}|
        + \f{1}{2} e^a e^b e^\gm H_{\gm ba}|
        + \f{1}{2} e^a e^b e_\dg H^\dg{}_{\ \ ba}|
        + e^a e^\bt e_\dg H^\dg{}_{\ \bt a}|,
\eeq
one derives in a straightforward way the expression
\bea
\lefteqn{-\f{1}{3} \si_{d \al \da} \vep^{dcba} H_{cba}| \ = \
\si_{k \al \da} \vep^{klmn} \prt_n b_{ml}} \nn \\
& & +i L \, \si_{k \al \da} \vep^{klmn} (\psi_n \si_m \psib_l)
    +2iL \, \si_{k \al \da} \lp \psi_m \si^{mk} \La
                                  -\psib_m \sib^{mk} \Lab \rp,
\ena
where we have used the definitions
\be
L| \ = \ L(x),
\eeq
and
\be
D_\al L| \ = \ \La_\al (x), \cem
D^\da L| \ = \ \Lab^\da (x).
\eeq

This short excursion was made to show how the superspace
construction provides in a rather straightforward and compact way
the basic building blocks which will be used later on in the
evaluation of supersymmetry invariant actions. In the example
worked out here the supercovariant component fieldstrength
$H_{cba}|$ exhibits terms linear and quadratic in the
Rarita-Schwinger field when coupled to supergravity.

\indent

\subsection{Super Chern-Simons forms: the Yang-Mills case}

\indent

We define
the Yang-Mills gauge potential as a Lie algebra valued one-form in
the background of $U_K(1)$ superspace, \ie
\be
\ca \ = \ E^A \ca_A^{(r)} T_{(r)} \ = \ \ca^{(r)} T_{(r)}.
\eeq
Latin indices in parentheses are used here to denote
the basis of the Lie algebra, the
commutation relations of the generators $T_{(r)}$ being defined as
\be
\left[ T_{(r)},T_{(s)} \right] \ = \ i{c_{(r)(s)}}^{(t)}
T_{(t)}.
\eeq
Gauge transformations are parametrized by group elements $g$
in the usual way except that now the parameters of the gauge
transformations are promoted from real functions to real superfields.
\be
\ca \ \mapsto \ {}^g \ca \ = \ g^{-1} \ca g - g^{-1} d g.
\eeq
The covariant fieldstrength
\be
\cf \ = \ d\ca+ \ca\ca,
\eeq
is a 2-form in superspace defined as
\be
\cf \ = \ \f{1}{2} E^A E^B \cf_{BA},
\eeq
with coefficients
\be
\cf_{BA} \ = \ \cd_B \ca_A - (-)^{ab} \cd_A \ca_B - (\ca_B,\ca_A)
                +{T_{BA}}^C \ca_C.
\eeq
Note the appearance of the supergravity torsion terms.
The derivatives occuring here covariantize
Lorentz and $U_K(1)$ transformations, following the usual
prescriptions, with chiral weights $w(\ca_A)=-w(E^A)$.
We use the notation
\be
(\ca_B, \ca_A) \ = \ \ca_B \ca_A - (-)^{ba} \ca_A \ca_B,
\eeq
for the graded commutation relations.
Of course, the fieldstrength is Lie algebra valued as well,
\be
\cf \ = \ \cf^{(r)}T_{(r)},
\eeq
and it is sometimes useful to display it in the form
\be
\cf^{(r)} \ = \ d\ca^{(r)}
                  + \f{i}{2}\ca^{(p)}\ca^{(q)}{c_{(p)(q)}}^{(r)}.
\eeq

Based on these definitions one can then go ahead with the construction
of the superspace analogue of Chern-Simons forms \cite{GG87}.
In the present context we restrict ourselves to
the case of the Chern-Simons 3-form.
Following the notation of \cite{MM86} we define
\be
\cq^{\ym} \ = \ \tr \lp \ca d \ca + \f{2}{3} \ca \ca \ca \rp,
\eeq
which, as a 3-form in superspace, has the decomposition
\be
\cq^{\ym} \ = \ \f{1}{3!} dz^K dz^L dz^M \cq^{\ym}_{MLK}
        \ = \ \f{1}{3!} E^A E^B E^C \cq^{\ym}_{CBA}.
\eeq
Clearly, the exterior derivative of this superspace
Chern-Simons form yields the fieldstrength squared term
\be
d \cq^{\ym} \ = \  \tr \lp \cf \cf \rp.
\eeq

The coupling to the antisymmetric tensor multiplet is then obtained
by incorporating this Chern-Simons forms in the fieldstrength of
2-form gauge potential as follows:
\be
H^{\ym} \ = \ dB + k\, \cq^{\ym}.
\eeq
The superscript ${\ym}$ indicates the presence of the Chern-Simons
form in the definition of the fieldstrength. Since the Chern-Simons
3-form $\cq^{\ym}$ changes under gauge transformations of the
Yang-Mills connection $\ca$ with the exterior derivative of a
2-form,
\be
\cq^{\ym} \ \mapsto \ {}^g \cq^{\ym} \ = \ \cq^{\ym} + d \Dt(g),
\eeq
covariance of $H^{\ym}$ can be achieved in assigning an inhomogeneous
compensating gauge transformation
\be
B \ \mapsto \ {}^g B \ = \ B - \Dt(g),
\eeq
to the 2-form gauge potential. Finally, the addition of the Chern-Simons
forms gives rise to the modified Bianchi identities
\be
d H^{\ym} \ = \ k \, \tr \lp \cf \cf \rp.
\eeq

We discuss now the restrictions on the covariant fieldstrengths
$H^{\ym}$ and $\cf$. As we have pointed out in the preceding
subsection, the linear multiplet corresponds to a 2-form geometry
with constraints on the fieldstrength. On the other hand it is well
known that in supersymmetric Yang-Mills theory the fieldstrength
$\cf$ is constrained as well. As a consequence, a question of
compatibility arises when these two superspace structures are
combined in the way we propose here.

The answer to this question is that the coupling of Chern-Simons
forms to the antisymmetric tensor multiplet is indeed consistent.
The most immediate way to see this is to investigate explicitly the structure
of the modified Bianchi identities in the presence of the constraints.

To this end let us first recall that supersymmetric Yang-Mills theory is defined
by the covariant constraints
\be
\cf^{\da\db} \ = \ 0, \cem \cf_{\bt \al} \ = \ 0, \cem
{\cf_\bt}^\da \ = \ 0,
\eeq
which can be understood as compatibility conditions for the
covariant chirality constraints on the matter superfields and, the
third one, as a covariant redefinition of the vector component
$\ca_a$ of the superspace Yang-Mills connection.

These constraints severely restrict the form of the remaining
components of the Yang-Mills fieldstrength, as can be seen from
their explicit solution or by a simple analysis of the Bianchi
identities. In any case one finds
\bea
\cf_{\bt a} &=&+i{(\si_a\eps)_\bt}^\db \cwb_\db,
\\
{\cf^\db}_a &=&-i{(\sib_a \eps)^\db}_\bt \cw^\bt,
\\
\cf_{ba}&=&\f{1}{2}(\eps\si_{ba})^{\bt\al}
\cd_\al\cw_\bt +\f{1}{2}(\sib_{ba} \eps)^{\db\da}
\cd_\da \cwb_\db.
\ena

All the superspace fieldstrength components are given in terms of
the covariant Yang-Mills superfields
\be
\cwb^\da \ = \ \cwb^{(r) \da}T_{(r)}, \cem
\cw_\al \ = \ \cw^{(r)}_\al T_{(r)},
\eeq
which, with respect to $U_K(1)$ superspace, have chiral weights
\be
w(\cwb^\da) \ = \ -1, \cem
w(\cw_\al) \ = \ +1.
\eeq
Moreover, as a consequence of the constraints
the Bianchi identities boil down to the equations
\be
\cd_\al\cwb^\da \ = \ 0, \cem \cd^\da \cw_\al \ = \ 0,
\label{ymcon1}
\eeq
\be
\cd^\al \cw_\al \ = \ \cd_\da \cwb^\da.
\label{ymcon2}
\eeq
We also define the $D$-term superfield ${\bf D}^{(r)}$ as
\be
{\bf D}^{(r)} \ = \ -\f{1}{2}\cd^\al\cw_\al^{(r)},
\eeq
which, by construction, has vanishing chiral weight,
\be
w({\bf D}^{(r)}) \ = \ 0.
\eeq

Observe that in solving the Yang-Mills Bianchi identities the complete
structure of $U_K(1)$ superspace as presented earlier has been
taken into account. Derivatives are covariant with respect to Lorentz,
chiral $U_K(1)$ and Yang-Mills gauge transformations.

We now turn back to the modified Bianchi identities for the fieldstrength
of the 2-form gauge potential in the presence of Yang-Mills
Chern-Simons forms. Assuming for $H^{\ym}$ the same constraints as
in the preceding subsection for $H$ on the one hand and taking into
account the special properties arising from the Yang-Mills constraints
in the fieldstrength squared terms on the other hand one arrives,
after some algebra, at the result that the general modified Bianchi
identities are simply replaced by the {\em modified linearity conditions}
\bea
\pp L^{\ym} &=& 2k \, \tr \lp \cwb_\da \cwb^\da \rp, \\
\qq L^{\ym} &=& 2k \, \tr \lp \cw^\al \cw_\al \rp,
\ena
written in $U_K(1)$ superspace, together with the relation
\be
\lp \l[ \cd_\al, \cd_\da \r] - 4 \si^a_{\al \da} G_a \rp L^{\ym} \ = \
         -\f{1}{3} \si_{d \al \da} \vep^{dcba} H^{\ym}_{cba}
         -4k \, \tr \lp \cw_\al \cwb_\da \rp,
\eeq
which identifies the fieldstrength tensor $H^{\ym}_{cba}$
in the superfield expansion of
\be L^{\ym} \ = \ L + k \, \Om^{\ym}. \eeq

The Chern-Simons superfield $\Om^{\ym}$ will be discussed in detail shortly.
The compatibility of the two superspace structures involved in this
construction has an explanation in the language of the
superspace geometry of the so-called 3-form gauge potential.
To see this in some more detail, we denote
\be \Si^{\ym} \ = \ \tr \lp \cf \cf \rp,
\eeq
   the fieldstrength squared term. From the explicit decomposition
\be
\Si^{\ym} \ = \ \f{1}{4!} E^A E^B E^C E^D {\Si^{\ym}}_{DCBA} \ = \
          \f{1}{4!} E^A E^B E^C E^D \, 6 \, \cf_{DC} \, \cf_{BA},
\eeq
and from the constraints on $\cf$ it is immediate to deduce that
\be
{\Si^{\ym}}_{\unddt \, \undgm \, \undal \, A} \ = \ 0.
\eeq
These are just the constraints which characterize the fieldstrength
of the 3-form gauge potential.

Let us therefore open here a parenthesis and digress shortly on the
features of the corresponding superspace formulation.
In the generic
case we have a 3-form gauge potential $B^3$ with covariant fieldstrength
$\Si = d B^3$ subject to precisely this set of constraints\footnote{in
the generic case the superscript ${\ym}$ is omitted}.
In appendix {\bf A} we point out in some detail how the explicit
solution of these constraints can be described in terms of one
single real scalar superfield $\Om$. This means that, up to pure
gauge contributions, all the coefficients $B^3_{CBA}$ of the
3-form gauge potential $B^3$ are expressible in terms of the
prepotential $\Om$.

On the other hand, at the level of the covariant fieldstrengths, this implies
restrictions on the other coefficients of $\Si$. As usual in
constrained superspace geometry the explicit structure of the
fieldstrength components may be obtained
from the Bianchi identities, in this case $d \Si=0$.
It turns out that they are completely
determined by superfields $S$ and $T$ subject to chirality conditions
\be \cd_\al S \ = \ 0, \cem \cd^\da T \ = \ 0, \eeq
and appearing as follows in the coefficients of the 4-form fieldstrength:
\bea
\Si_{\dt \gm \, ba} &=& \f12 (\si_{ba}\eps)_{\dt \gm} \, S, \\
\Si^{\dd \dg}{\,}_{ba} &=& \f12 (\sib_{ba}\eps)^{\dd \dg} \, T.
\ena
As another consequence of the constraints one finds that
\be
\Si_\dt{}^\dg{\,}_{ba} \ = \ T_\dt{}^{\dg \, c} \, \Si_{cba},
\eeq
where $\Si_{cba}$ is totally antisymmetric in its three indices.
But this means that it can be absorbed in a redefinition of the
coefficient $B^3_{cba}$ of 3-form gauge potential $B^3$.
This is easily deduced from the explicit expression
\be
\f{1}{4!} E^A E^B E^C E^D \, \Si_{DCBA} \ = \
\f{1}{4!} E^A E^B E^C E^D \lp 4 \, \cd_D B^3_{CBA}
                             + 6 \, T_{DC}{}^F B^3_{FBA} \rp,
\eeq
where for the fieldstrength coefficient we are interested in one has
\be
\Si_\dt{}^\dg{\,}_{ba} \ = \ T_\dt{}^{\dg \, c} B^3_{cba}
+ \mbox{ derivative and other torsion terms} \ .
\eeq
This shows that the modified 3-form gauge potential
\be
\undB^3_{cba} \ = \ B^3_{cba} - \Si_{cba},
\eeq
corresponds to the modified fieldstrength coefficient
\be
\undSi_\dt{}^\dg{\,}_{ba} \ = \ 0.
\eeq
Since this equation is obtained from a covariant and linear redefinition of
the gauge potential, it is sometimes referred to as {\em conventional constraint}.

Taking, from now on, into account this modification, the remaining
coefficients, at canonical dimensions 3/2 and 2, i.e.
$\undSi_{\unddt \, cba}$ and $\undSi_{dcba}$, respectively, are
given in terms of spinor derivatives of the basic superfields $S$
and $T$. To be more precise, at dimension 3/2 one obtains
\bea
\undSi_{\dt \, cba} &=&
           -\f{1}{16} \, \si^d_{\dt \dd} \, \vep_{dcba} \cd^\dd S, \\
\undSi^\dd {}_{cba} &=&
           +\f{1}{16} \, \sib^{d \, \dd \dt} \, \vep_{dcba} \cd_\dt T,
\ena
and the Bianchi identity at dimension two takes the simple form
\be
\lp \cd^2 - 24 \rd \rp T - \lp \cdb^2 - 24 R \rp S \ = \
         \f{8i}{3} \vep^{dcba} \undSi_{dcba}.
\eeq

This equation should be understood as another condition which
serves to further restrict the chiral superfields $S$ and $T$,
thus describing the supermultiplet of 3-form gauge potential
in $U_K(1)$ superspace.

Correspondingly, from the explicit solution of the constraints
one finds that
$S$ and $T$ are given as the chiral projections of $U_K(1)$
superspace geometry acting on one and the same prepotential $\Om$:
\bea S &=& -4 \pp \Om, \\ T &=& -4 \qq \Om. \ena

To discuss the relevance of this geometric structure for the discussion
of supersymmetric Chern-Simons forms,
we come back to the explicit expression of
the superspace 4-form as fieldstrength squared term and interprete
it as the fieldstrength of the 3-form gauge potential identified
in turn with the Chern-Simons 3-form,
\be
\Si^{\ym} \ = \ \tr \lp \cf \cf \rp \ = \ d \cq^{\ym}.
\eeq

First of all, it is straightforward to convince oneself that in this case
the previously generic superfields $S$ and $T$ are given as
\be
S^{\ym} \ = \ -8 \, \tr \lp \cwb_\da \cwb^\da \rp, \cem
T^{\ym} \ = \ -8 \, \tr \lp \cw^\al \cw_\al \rp.
\eeq

On the other hand, the explicit solution of the constraints shows
the existence of a so-called Chern-Simons superfield $\Om$
such that
\bea
\tr \lp \cw_\da \cw^\da \rp &=& \f12 \pp \Om^{\ym}, \\
\tr \lp \cw^\al \cw_\al \rp &=& \f12 \qq \Om^{\ym}.
\ena

By definition, the Chern-Simons superfield is given in terms of the
(pre)potentials which define supersymmetric Yang-Mills theory.
In order to present an explicit expression for $\Om^{\ym}$ one has to take into
account the solution of the constraints in terms of prepotentials.

We have tried to make clear in this section that the superspace geometry
of the 3-form gauge potential provides a generic framework for the
discussion of Chern-Simons forms in superspace. Established
in full detail for the Yang-Mills case, this property will be
advantageously exploited in the more complicated gravitational case.


\sect{Gravitational Chern-Simons forms in superspace}

\indent

\subsection{Some general considerations}

\indent

We come now to the description of gravitational Chern-Simons forms.
The discussion
will proceed in two steps: in this present section we display the geometric
structure in the framework of superspace geometry, whereas in the next section
we will discuss a relatively simple example
including gravitational Chern-Simons forms
via the antisymmetric tensor coupling \`a la Green and Schwarz.

In the geometrical description we will exploit what we have learned
in the case of supersymmetric Yang-Mills theory. There, the geometric structure
of Chern-Simons forms in superspace is quite well
understood and invariant actions for the antisymmetric tensor with Chern-Simons
form in its fieldstrength, coupled to the general supergravity-matter system,
can be obtained by means of the standard chiral density construction.

It is natural to ask whether the techniques which work quite well
in the Yang-Mills case can be generalized to include gravitational
Chern-Simons forms. We will see in the following that this is
indeed true to a large extent, although substantially more involved
technically. But we shall also see that novel features on the
conceptual level appear, in particular when it comes to the
construction of supersymmetric dynamics.

In order to cope with this new situation we shall investigate the structure
of Chern-Simons forms in superspace in a more systematic way, based on
the observation that the structure of Chern-Simons forms
fits remarkably well into the superspace geometry of
the  3-form gauge potential.

As a starting point we take a number
of 2-form gauge potentials $B^I$ numbered by $I=1,...,n$ and certain types
of Chern-Simons forms $\cq^\Dt$ with constant couplings\footnote{
One should distinguish carefully indices $I$ counting linear superfields
from indices $A=a, \al, \da$ denoting superspace.} $k^I_\Dt$.
The corresponding fieldstrengths are then defined as
\be
H^I \ = \ dB^I + k^I_\Dt \, \cq^\Dt.
\label{eq:7.1}
\eeq
In practice we will include here
Yang-Mills, gravitational (two chiralities) and $U_K(1)$ Chern-Simons forms
with $\Dt$ taking values
\be
\Dt \ \in \
\left\{ \, (+) \, , \, (-) \, , \, (1) \, , \, (\cy \cm) \, \right\}.
\label{eq:7.2}
\eeq
We denote the derivative of the Chern-Simons form
\be
d \cq^\Dt \ = \ \Psi^\Dt.
\label{eq:7.3}
\eeq
In more explicit terms
\bea
\Psi^{\ym} \ = \ \tr (\cf \cf), && \cem \Psi^\uo \ = \ FF,
\label{eq:7.4}\\
\Psi^\pl \ = \ {R_\bt} \raisebox{.3ex}{${}^\al$} {R_\al}^\bt, && \cem
\Psi^\mn \ = \ {R^\db}_\da {R^\da}_\db.
\label{eq:7.5}
\ena
The Bianchi identity for the 2-form gauge field is then given as
\be
d H^I \ = \  k^I_\Dt \, \Psi^\Dt.
\label{3.6}
\eeq
This fixes our notations. The first, and crucial, nontrivial
point in this approach
is the observation that $\Psi^\Dt$ allows for the decomposition
\be
\Psi^\Dt \ = \ \Si^\Dt + d M^\Dt,
\label{eq:7.6}
\eeq
such that
\begin{itemize}
\item the coefficients of
the 3-form $M^\Dt$, as well as those of the 4-form
$\Si^\Dt$, are covariant expressions in terms of the corresponding
fieldstrength, torsion and curvature superfields.
\item the tensorial structure of the coefficients
of the 4-form $\Si^\Dt$ corresponds exactly to that of the
constraints in the 3-form geometry.
\end{itemize}

Of course, this decomposition must be explicitly established in every
particular case. Before doing so we recall however a number of generic
features valid in all cases.

First of all, upon substitution into the Bianchi identity for $H^I$,
one arrives at
\be
d \ch^I \ = \ k^I_\Dt \Si^\Dt,
\label{3.8}
\eeq
where we use the definition
\be \ch^I \ = \ H^I - k_\Dt^I M^\Dt \ = \ d B^I +  k_\Dt^I (\cq^\Dt - M^\Dt).
\label{3.9}
\eeq
It is in this form that the analogy with the 3-form gauge
potential shows up. Defined as a differential 4-form in
superspace,
\be
\Si^\Dt \ = \ \frac{1}{4!} \, E^A E^B E^C E^D \, {\Si^\Dt}_{DCBA},
\label{eq:7.8}
\eeq
is subject to Bianchi identities
\be d \Si^\Dt \ = \ 0.
\label{eq:7.9}
\eeq

We shall show below, that the coefficients of $\Si^\Dt$ can be
determined such that
\be
{\Si^\Dt}_{\undel \undgm \undbt A} \ = \ 0,
\label{eq:7.10}
\eeq
(where $\undal=\al, \da$ and $A=a, \al, \da$).
This is the tensorial
structure of the constraints of the 3-form geometry.
 We can then exploit our knowledge of the supersymmetric 3-form
gauge potential to gain more insight into the structure of
curvature-squared terms without needing to know all the details of
the explicit form of the decomposition (which may be rather
complicated and which are given below).

The Bianchi identities, given the property (\ref{eq:7.10}),
determine the tensorial structure of the remaining fieldstrength
components and give rise to certain relations involving covariant
spinor derivatives. These general features of the 4-form
$\Si^\Dt$ do not depend on the particular properties of the type of
Chern-Simons forms under consideration.

Let us briefly recall the outcome of the analysis of the Bianchi
identities. The components of $\Si^\Dt$ are completely described in
terms of two superfields $S^\Dt$ and $T^\Dt$, appearing as follows
in the tensor decomposition:
\bea
{\Si^\Dt}_{\dt \gm \ ba} & = &
\f{1}{2} (\si_{ba} \eps)_{\dt \gm} \, S^\Dt,
\label{eq:7.11} \\
\Si^{\Dt \, \dd \dg}{}_{\ ba} & = &
\f{1}{2} (\bar{\si}_{ba} \eps)^{\dd {\dg}} \, T^\Dt.
\label{eq:7.12}
\ena
By a special choice of conventional constraints, it is possible to impose
\be
{{{\Si^\Dt}_\dt}^{\dg}}_{\ ba} \ = \ 0.
\label{eq:7.13}
\eeq
As to the superfields $S^\Dt$ and $T^\Dt$ the Bianchi identities reduce
to the chirality conditions
\be
\cd_\al S^\Dt \ = \ 0, \cem \cd^\da T^\Dt \ = \ 0.
\label{eq:7.14}
\eeq
The remaining components at higher (canonical) dimensions are then
\bea
\Si^\Dt{}_{\ \dt \ cba} &=& - \f{1}{16} \, \si^d_{\dt \dd} \, \vep_{dcba}
                          \, \cd^\dd \! S^\Dt,
\label{eq:7.15} \\
\Si^{\Dt \ \dd \ }{}_{cba} &=& \  + \f{1}{16} \bar{\si}^{d \, \dd \dt}
                          \, \vep_{dcba} \, \cd_\dt T^\Dt,
\label{eq:7.16}
\ena
and, finally,
\be
{\Si^\Dt}_{dcba} \ = \ \vep_{dcba} \, {\bf \Si}^\Dt.
\label{eq:7.17}
\eeq
The boldscript scalar superfields ${\bf \Si}^\Dt$ appearing at this
level are given as the second order spinor derivatives of the basic
superfields:
\be
2i {\bf \Si}^\Dt \ = \ - \f{1}{32} \pp T^\Dt + \f{1}{32} \qq S^\Dt.
\label{eq:7.18}
\eeq

In conclusion, we have seen that all the coefficients of the
superspace 4-form $\Si^\Dt$, subject to the constraints
(\ref{eq:7.10}), are given in terms of the superfields  $S^\Dt$ and
$T^\Dt$ and their spinor derivatives. As to the analysis of the
curvature-squared terms, this shows that it is sufficient to
identify the superfields  $S^\Dt$ and $T^\Dt$ in terms of the
underlying geometry (Yang-Mills, supergravity or $U_K(1)$) for full
knowledge of the corresponding superspace 4-form $\Si^\Dt$.

This decomposition is particularly useful in the determination
of the modified linearity conditions. To this end we observe first of
all that the redefined fieldstrengths $\ch^I$ are subject to
the same constraints as before, \ie without Chern-Simons forms:
\be
{\ch^I}_{\undgm \undbt \undal} \ = \ 0, \cem
{\ch^I}_{\gm \bt \, a} \ = \ 0, \cem {\ch^I}_{\dg \db \, a} \ = \ 0.
\eeq
In other words, whereas the redefined quantities $\ch^I$ have a very
simple form, the original $H^I$ can be quite complicated \cite{GG87}.
The linear superfield is then identified in
\be
{\ch^I}_{\gm}{}^\db{}_a \ = \ -2i (\si_a \eps)_\gm{}^\db \ \cl^I.
\label{3.22}
\eeq
Furthermore, one still has
\bea
{\ch^I}_{\gm ba} &=& 2 (\si_{ba}{)_\gm}^\vp \, \cd_\vp \cl^I, \\
{\ch^I}^\dg_{\ \ ba} &=& 2 (\sib_{ba}{)^\dg}_\dv \, \cd^\dv \cl^I.
\ena
The Bianchi identities boil then down to the modified linearity
conditions
\bea
\pp \cl^I &=& -\f{1}{4} \csi, \cem \cem \csi \ = \ k^I_\Dt S^\Dt, \\
\qq \cl^I &=& -\f{1}{4} \cti, \cem \cem \cti \ = \ k^I_\Dt T^\Dt.
\ena

Note that we allow in general $\csi$ and $\cti$ to be linear combinations of
terms pertaining to Yang-Mills, gravitational or $U_K(1)$ Chern-Simons forms.

Finally, the vector component ${\ch^I}_{cba}$ appears in the same
way as before in the $\th \thb$-component of $\cl^I$:
\be
\lp [\cd_\al , \cd_\da] - 4 \si^a_{\al \da} G_a \rp \cl^I
\ = \  - \f{1}{3} \si_{d \al \da} \vep^{dcba} {\ch^I}_{cba}.
\eeq
The difference is of course, that now, as a consequence of the decomposition,
$\ch_{cba}^I$ contains additional terms,
\be {\ch^I}_{cba} \ = \ {H^I}_{cba} - k^I_\Dt {M^\Dt}_{cba}. \eeq

\indent

So far the discussion was quite general, it applied for any particular case
subsumed in the index $\Dt$. In the following we will discuss the various
different cases separately.

The Yang-Mills case, as already discussed in section {\bf 2} is reproduced
in the general formulation presented here with the identifications
\be
S^{\ym} \ = \ -8 \, \tr \lp \cwb_\da \cwb^\da \rp, \cem
T^{\ym} \ = \ -8 \, \tr \lp \cw^\al \cw_\al \rp.
\eeq
and
\be {M^{\ym}}_{cba} \ = \ \tr \lp \cw \si^d \cwb \rp \vep_{dcba}. \eeq

\indent

Before proceeding to the explicit and detailed presentation of the
covariant decomposition for the gravitational curvature-squared
terms $\Psi^{(\pm)}$ introduced above, we would like to draw
attention to another feature of the formulation presented here
which we found quite useful in the analysis of the gravitational
curvature-squared terms.

It corresponds to a certain freedom
in the identification of $\Si^\Dt$ and $M^\Dt$ without changing
$\Psi^\Dt$. In other words, the replacements
$\Si^\Dt \mapsto \Si^\Dt + \si^\Dt$
and $M^\Dt \mapsto M^\Dt + m^\Dt$ do not affect $\Psi^\Dt$ as long as they
satisfy the superspace equation
\be
\si^\Dt + d m^\Dt \ = \ 0.
\label{eq:7.19}
\eeq

A particularly useful solution is given in terms of an arbitrary
unconstrained superfield $\mu^\Dt$ such that the nonvanishing components
of $m^\Dt$ are
\be
{m^\Dt}_\gm {}^\db {}_a \ = \ T_\gm{}^\db{}_a \, \mu^\Dt,
\eeq
and
\bea
{m^\Dt}_{\gm \, ba} &=& 2(\si_{ba})_\gm{}^\vp \cd_\vp \mu^\Dt,
\\
{m^\Dt}^\dg \, {}_{ba} &=& 2(\sib_{ba})^\dg {}_\dv \cd^\dv \mu^\Dt,
\ena
as well as
\be
\vep^{dcba} {m^\Dt}_{cba} \ = \
         \f{3}{2} \sib^{d \da \al} \lp [\cd_\al, \cd_\da]
                - 4 G_{\al \da} \rp \mu^\Dt.
\eeq
In $\si^\Dt$, on the other hand,
the unconstrained superfields $\mu^\Dt$ appear (in obvious notations) as
\bea
s^\Dt &=& \pp \mu^\Dt,
\\
t^\Dt &=& \qq \mu^\Dt.
\label{31L}
\ena
Of course, requiring $\si^\Dt = 0$ would impose the linearity constraints of
curved superspace on $\mu^\Dt$, in accordance with our discussion of the
linear superfields in previous sections. As already mentioned the freedom
in assigning particular values to the arbitrary superfield $\mu^\Dt$ may
turn out to be useful in the gravitational case.

\indent

\subsection{Covariant decomposition of curvature-squared terms}

\indent

We come now back to the covariant decomposition of
curvature-squared terms mentioned above. As we have already
stressed, it is established by an explicit calculation. As a
consequence, the present subsection will be rather technical. In
order to give an impression of the actual procedure employed we
present explicitly the {\footnotesize $(+)$} - sector. The method
consists in successive rearrangements of terms appearing in the
curvature-squared terms
\be
\Psi^\pl \ = \ {R_\bt} \raisebox{.3ex}{${}^\al$} {R_\al}^\bt,
\label{eq:3.1}
\eeq
to arrive at a decomposition
\be
\Psi^\pl \ = \ \Si^\pl + d M^\pl,
\label{eq:3.9}
\eeq
such that the coefficients of the differential forms
$\Si^\pl$ and $M^\pl$,
\bea
M^\pl &=& \f{1}{3!} E^A E^B E^C {M^\pl}_{CBA},
\label{eq:3.10} \\[2mm]
\Si^\pl &=& \f{1}{4!} E^A E^B E^C E^D {\Si^\pl}_{DCBA},
\label{eq:3.11}
\ena
can be completely
expressed in terms of the covariant supergravity superfields and their
covariant derivatives and that
$\Si^\pl$ can be chosen such that (\ref{eq:7.10})
holds\footnote{Note however that $M^\pl$ is only determined
up to the exterior derivative of a 2-form, this means that our decomposition
allows for the replacements
$M^\pl \ \rightarrow \ M^\pl + dm^\pl$ as explained above.}, \ie
the 3-form constraints
\be
{\Si^\pl}_{\undel \, \undgm \, \undbt A} \ = \ 0.
\label{3fcon}
\eeq

To begin with we should have a closer look to the coefficients of
the 4-form
\be
\Psi^\pl \ = \ \f{1}{4!} E^A E^B E^C E^D \, {\Psi^\pl}_{DCBA},
\eeq
which are expressed in terms of the (constrained)
superspace curvatures,
\be
{\Psi^\pl}_{DCBA} \ = \ 2 \oint_{DCB}
         R_{DC} \, {}_\vp {}^\vep R_{BA} \, {}_\vep {}^\vp.
\eeq

From the explicit form of the supergravity curvatures
one finds immediately
\be
\Psi^{\pl \,\dd \dg \db} {}_A \ = \ 0, \cem
{\Psi^\pl}_{\dt \gm \bt \, \undal} \ = \ 0,
\label{pl1}
\eeq
in accordance with the corresponding coefficients in (\ref{3fcon}).
However, a non-trivial contribution arises in
\be
{\Psi^\pl}_{\dt \gm \bt \, a} \ = \
     2 \oint_{\dt \gm \bt} R_{\dt \gm} \, {}_\vp {}^\vep
              R_{\bt a} \, {}_\vep {}^\vp
\ = \ - 16 \rd \oint_{\dt \gm \bt}
           R_{\dt a \,}\raisebox{.2ex}{${}_{\sym{\gm \bt}}$}.
\label{pl3}
\eeq
Replacing $a$ by {\footnotesize $\al \da$} and using supergravity information
leads then to
\be
{\Psi^\pl}_{\dt \gm \bt \, \al \da} \ = \ - 8i \oint_{\dt \gm \bt}
\cd_\dt \lp \eps_{\gm \al} \rd G_{\bt \da}
        + \eps_{\bt \al} \rd G_{\gm \da} \rp.
\eeq
Inspection shows then that the desired decomposition for
the coefficients considered so far can be established by the choice
\be
{M^\pl}_{\undgm \, \undbt \, \undal} \ = \ 0, \cem
{M^{(+) \, \dg \db}}_a \ = \ 0,
\label{eq:3.14}
\eeq
and
\be
{M^\pl}_{\gm \bt \ \al \da} \ = \ -8i \rd \lp
   \eps_{\gm \al} G_{\bt \da} + \eps_{\bt \al} G_{\gm \da} \rp,
\label{eq:3.16} \\
\eeq
with spinor notation,
\be
{M^\pl}_{\gm \bt \ \al \da} \ = \ \si^a_{\al \da} \,
                                            {M^\pl}_{\gm \bt \ a},
\label{eq:3.15}
\eeq
understood. This means that so far we have established
\be
\Si^{\pl \, \dd \dg \db} {}_A \ = \ 0, \cem
{\Si^\pl}_{\dt \gm \bt \, A} \ = \ 0.
\eeq

In the next step we consider
\be
{\Psi^\pl}_{\dt \gm}{}^{\db \da} \ = \
2 \sum _{\dt \gm} R_\dt{}^\db{\,}_\vp{}^\vep R_\gm{}^\da{\,}_\vp{}^\vep
\ = \ 4 \sum _{\dt \gm} G_\dt{}^\db \, G_\gm{}^\da,
\eeq
which serves to identify ${M^\pl}_\gm{}^\db{}_a$ as
\be
{M^\pl}_{\gm \db \ \al \da} \ = \
   -i  \eps_{\gm \al} \, \eps_{\db \da} \, \mu^\pl
   -\i2 \lp G_{\gm \db}\,  G_{\al \da} + G_{\al \db} \, G_{\gm \da} \rp.
\label{eq:3.18}
\eeq
Observe that here we have, in view of the discussion following
eq.(\ref{eq:7.19}), allowed for the appearance of the arbitrary
superfield $\mu^\pl$.

We continue with
\be
{\Psi^\pl}_{\dt \gm}{\,}^\db{}_a \ = \
-16 \rd R^\db{}_a \, \raisebox{.3ex}{${}_{\sym{\dt \gm}}$}
+ 4 \lp R_{\gm a \, \dt}{}^\vp + R_{\dt a \, \gm}{}^\vp \rp G_\vp{}^\db,
\eeq
and
\be
\hs{-2.1cm} \Psi^{\pl \,\dd \dg}{}_{\bt a} \ = \
 4 R^\dg {}_{a \, \bt}{}^\vp G_\vp{}^\dd
+ 4 R^\dd {}_{a \, \bt}{}^\vp G_\vp{}^\dg.
\eeq
At this level the coefficients ${M^\pl}_{\undgm \ ba}$ come in.
Using spinor notation
\be
\hs{-2.2cm}{M^\pl}_{\undgm \ \bt \db \ \al \da} \ = \ \si^b_{\bt \db} \,
\si^a_{\al \da} \, {M^\pl}_{\undgm \ ba},
\label{eq:3.19}
\eeq
with standard tensor decomposition
\be
{M^\pl}_{\undgm \ \bt \db \ \al \da} \ = \
 2\eps_{\db \da} {M^\pl}_{\undgm}
\raisebox{.1ex}{${}_{ \ {\sym{\bt \al}}}$}
-2\eps_{\bt \al} {M^\pl}_{\undgm}
\raisebox{.5ex}{${}_{ \ {\sym{\db \da}}}$},
\label{eq:3.20}
\eeq
and
\bea
{M^\pl}\raisebox{-.2ex}{${}_{\gm}$}{}_{ \ {\sym{\bt \al}}} &=&
{M^\pl}_{\lsym{\gm \bt \al}}
+ \eps_{\gm \bt} {M^\pl}_{\al} + \eps_{\gm \al} {M^\pl}_{\bt},
\label{eq:3.21} \\
{M^\pl} \raisebox{-.2ex}{${}_{\dg}$}
\raisebox{.3ex}{${}_{ \ {\sym{\db \da}}}$} &=&
{M^\pl} \raisebox{.ex}{${}_{\lsym{\dg \db \da}}$}
+ \eps_{\dg \db} {M^\pl}_{\da} + \eps_{\dg \da} {M^\pl}_{\db},
\label{eq:3.22}
\ena
the different irreducible tensors defined here are then
expressed in terms of the basic supergravity fields as follows:
\bea
{M^\pl}\raisebox{-.2ex}{$_{\gm}$}
\raisebox{.2ex}{${}_{ \ {\sym{\db \da}}}$} &=& \f{1}{8} \sum_{\db \da}
\left( 16 \rd \cd_{\db} G_{\gm \da}
      + 4 G_{\gm \db} \cd_{\da} \rd
      - {G^\vp}_\db \cd_\gm G_{\vp \da}
      - 4{G^\vp}_\db \cd_\vp G_{\gm \da} \right),
\label{eq:3.23} \\ [2mm]
{M^\pl}_{\lsym{\gm \bt \al}} &=&
       - 8 \rd W_{\lsym{\gm \bt \al}}
       + \f{1}{24} \oint_{\gm \bt \al}
       {G_\gm}^\dv \lp \cd_\bt G_{\al \dv} + \cd_\al G_{\bt \dv} \rp,
\label{eq:3.24} \\ [3mm]
12 {M^\pl}_\al &=& - 3 \cd_\al \mu^\pl
       - 16 \cd_\al (R\rd)  \nonumber \\[1mm] & &
       - 8 \rd \cd^\dv G_{\al \dv}
       - 18 G_{\al \dv} \cd^\dv \rd
       + 2 G^{\vp \dv} \cd_\al G_{\vp \dv}
       + 5 G^{\vp \dv} \cd_\vp G_{\al \dv},
\label{eq:3.25}
\ena
and
\be
{M^\pl}\raisebox{-.2ex}{${}_{\dg}$}{}_{ \ {\sym{\bt \al}}} \ = \
       - 4 {G^\vp}_\dg W_{\lsym{\vp \bt \al}}
+ \f{1}{8} \sum_{\bt \al}
\lp {G_\bt}^\dv \cd_\dg G_{\al \dv}
      + \f{4}{3} G_{\bt \dg} \lp \cd_\al R - \cd^\dv G_{\al \dv} \rp \rp,
\label{eq:3.26} \\
\eeq
as well as
\bea
{M^\pl}_{\lsym{\dg \db \da}} &=&
        \f{1}{8} \oint_{\dg \db \da}
       {G^\vp}_\dg \lp \cd_\db G_{\vp \da} + \cd_\da G_{\vp \db} \rp,
\label{eq:3.27} \\ [2mm]
4 {M^\pl}_\da &=& \cd_\da \, \mu^\pl
      + 2 {G^\vp}_\da \cd_\vp R
      + G^{\vp \dv} \cd_\dv G_{\vp \da}.
\label{eq:3.28}
\ena
These identifications establish
\be
{\Si^\pl}_{\dt \gm}{}^\db{}_a \ = \ 0, \cem
{\Si^\pl}^{\dd \dg}{}_{\bt a} \ = \ 0,
\eeq
which completes the derivation of the 3-form
constraint structure, eq.(\ref{3fcon}), for $\Si^\pl$.
Recall that this was the crucial goal we wanted
to achieve in this section:
the tensorial structures of the remaining
coefficients of ${\Si^\pl}_{DCBA}$
are now determined from the 3-form geometry,
as for instance (see (\ref{eq:7.11}), (\ref{eq:7.12}))
\bea
{\Si^\pl}_{\dt \gm\ ba} & = &
\f{1}{2} (\si_{ba} \eps)_{\dt \gm} \, S^\pl,
\label{eq:3.29} \\
\Si^{\pl \, \dd \dg}{}_{\ ba} & = &
\f{1}{2} (\bar{\si}_{ba} \eps)^{\dd {\dg}} \, T^\pl,
\label{eq:3.30}
\ena
where the chiral superfields $S^\pl$, $T^\pl$
are now identified in terms of the supergravity superfields
as follows:
\bea
S^\pl &=&
    \pp \left( \mu^\pl + 16 \rd R
                - \f{13}{4} G^{\vp \dv} G_{\vp \dv} \right)
            - 4 \bar{X}_\dv \bar{X}^\dv ,
\label{eq:3.31} \\
T^\pl &=&
     \qq \left( \mu^\pl
                   + \f{3}{4} G^{\vp \dv} G_{\vp \dv} \right)
      +32 W \raisebox{-.5ex}{${}^{\lsym{\gm \bt \al}}$} W_{\lsym{\gm \bt \al}}
      + \f{4}{3} X^\vp X_\vp.
\label{eq:3.32}
\ena
These chiral superfields will be among
the basic ingredients in the construction of the supersymmetric extension of
the various kinds of gravitational curvature-squared terms.

In the ${\Psi^\pl}_\dt{\, }^\db{}_{\, ba}$ - sector, one finds
that the coefficient
\be
{\Si^\pl}_{\dt \dg \ \bt \db \ \al \da} \ = \
\si^b_{\bt \db} \, \si^a_{\al \da} \, {\Si^\pl}_{\dt \dg \, ba},
\eeq
is expressed in terms of just one vector such that
\be
{\Si^\pl}_{\gm \dg \, \bt \db \, \al \da} \ = \
4 \, \eps_{\gm \bt} \, \eps_{\dg \da} \, {\Si^\pl}_{\al \db}
-4 \, \eps_{\gm \al} \, \eps_{\dg \db} \, {\Si^\pl}_{\bt \da}.
\eeq
In the same equation, the component $M^\pl{}_{cba}$,
totally antisymmetric in its indices appears. Written in spinor
notation
\be
{M^\pl}_{\gm \dg \ \bt \db \ \al \da} \ = \
 \si^c_{\gm \dg} \, \si^b_{\bt \db} \, \si^a_{\al \da} \, {M^\pl}_{cba},
\label{eq:3.34}
\eeq
it has the decomposition
\be
{M^\pl}_{\gm \dg \ \bt \db \ \al \da} \ = \
 2i \eps_{\dg \db} \eps_{\gm \al} {M^\pl}_{\bt \da}
-2i \eps_{\dg \da} \eps_{\gm \bt} {M^\pl}_{\al \db},
\label{eq:3.35}
\eeq
reflecting the antisymmetry property in terms of spinor indices.
Following ref.\cite{GG91}, the combination
${M^\pl}_{\al \da}+{\Si^\pl}_{\al \da}$ is then
expressed in terms of the
supergravity superfields as follows:
\bea
\lefteqn{\hs{-1cm}
{M^\pl}_{\al \da} + {\Si^\pl}_{\al \da}
+ \f{1}{8} \lp \left[{\cd}_{\al},{\cd}_{\da} \right]
- 4 G_{\al \da} \rp \mu^\pl \ = }
\nonumber \\[3mm]
&&+\f{1}{16}G^{\vp \dv} \lp
4 \left[{\cd}_{\vp}, {\cd}_{\dv} \right] G_{\al \da}
+ \left[{\cd}_{\al}, {\cd}_{\da} \right] G_{\vp \dv} \rp
- 8 \rd R \, G_{\al \da}
-\f{15}{24} G_{\al \da} G^{\vp \dv} G_{\vp \dv} \nn \\[2mm]
&&-{\cd}_{\al}R \, {\cd}_{\da}\rd
-\f{3}{32} {\cd}^{\dv} G_{\al \dv} \, {\cd}^{\vp}G_{\vp \da}
+\f{3}{2} {T_{\sym{\dv \da}}}^{\vp} \,
    T\raisebox{-.4ex}{${}_{\sym{\vp \al}}$}{}^{\dv}
+8 T\raisebox{-.4ex}{${}^{\sym{\gm \bt}}$}{}_{\da} \,
    W_{\lsym{\gm \bt \al}}
\nonumber \\[1mm]
&&
+ T\raisebox{-.4ex}{${}_{\sym{\vp \al}}$} \, \raisebox{-.2ex}{${}_\da$}
\lp \f{4}{3} {\cd}^{\vp} R + \f{23}{24} {\cd}_{\dv} G^{\vp \dv} \rp
- T_{\sym{\dv \da}} \, \raisebox{-.2ex}{${}_\al$}
\lp 4 {\cd}^{\dv}\rd +\f{3}{8} {\cd}_{\vp} G^{\vp \dv} \rp
\nonumber \\[2mm]
&&+iG^{\vp \dv} \left( {\cd}_{\vp \dv}G_{\al \da}
+\f{1}{2} {\cd}_{\al \da} G_{\vp \dv}
-\f{3}{4} {\cd}_{\vp \da} G_{\al \dv}
-\f{1}{4} {\cd}_{\al \dv} G_{\vp \da} \right)
- 4i\rd \cd_{\al \da} R,
\label{eq:3.38}
\ena
Observe that ${\Si^\pl}_{\al \da}=0$ corresponds to the
conventional constraint (\ref{eq:7.13}).
The remaining coefficients of the 4-form $\Si^\pl$, \ie
\be
{\Si^\pl}_{\undel \ cba}, \cem {\Si^\pl}_{dcba},
\label{eq:3.40}
\eeq
are obtained as spinor derivatives of the superfields $S^\pl$ and
$T^\pl$ as explained in the discussion of the generic properties
of the 3-form geometry in the previous subsection
(eqs.(\ref{eq:7.15}), (\ref{eq:7.16}) and (\ref{eq:7.18})). This
completes our discussion of the {\footnotesize $(+)$} - sector. The
corresponding decomposition in the {\footnotesize $(-)$} - sector
are listed in appendix {\bf D}.

Admittedly, the presentation in this subsection was notationally
quite heavy, the coefficients of the 3-form $M^\pl$ are
rather complicated expressions in terms of the basic supergravity
superfields and their covariant derivatives. But once established,
we then use these coefficients from now on precisely as a shorthand
notation for otherwise complicated expressions, which may be
expanded when necessary.

\indent

\subsection{3-form geometry and modified linearity conditions}

\indent

We have seen that supersymmetric Chern-Simons forms can be
described in the framework of the superspace geometry of
a 3-form gauge potential: on the one hand we have
explicitly etablished the {\em covariant decomposition}
\be
\Psi^\Dt \ = \ \Si^\Dt + d M^\Dt,
\eeq
as anticipated in (\ref{eq:7.6}). On the other hand, from our
starting point (\ref{eq:7.3}) we know
\be
\Psi^\Dt \ = \ d \cq^\Dt,
\eeq
and therefore
\be
\Si^\Dt \ = \ d \lp \cq^\Dt - M^\Dt \rp.
\eeq
Here, the combination $\cq^\Dt - M^\Dt$ corresponds to the
3-form gauge potential, which, under a gauge transformation changes
with the exterior derivative of a 2-form.
$\Si^\Dt$ is subject to the 4-form constraints, which
ensures the existence of the Chern-Simons superfield $\Om^\Dt$,
corresponding to the prepotential of the 3-form, such
that the combinations $S^\Dt$ and $T^\Dt$ as identified
in (\ref{eq:7.11}) and (\ref{eq:7.12}), are given as
\bea
S^\Dt &=& -4 \, \pp \, \Om^\Dt,
\label{prepoS} \\
T^\Dt &=& -4 \ \qq \, \Om^\Dt.
\label{prepoT}
\ena
Under gauge-, Lorentz-, \ka transformations the corresponding
Chern-Simons superfields change as
\be \Om^\Dt \ \mapsto \ \Om^\Dt + \la^\Dt, \label{3fgtr}\eeq
where $\la^\Dt$ are linear superfields, subject to
the conditions
\be \pp \la^\Dt \ = \ 0, \cem \qq \la^\Dt \ = \ 0. \eeq
Of course, this reflects the above-mentioned
fact that the full Chern-Simons
form changes with the exterior derivative of a 2-form.

\indent

This structure is now coupled to the 2-form gauge
potential in the way defined above in the first section of this chapter,
the relevant definitions being (\ref{eq:7.1}), (\ref{eq:7.3}),
(\ref{3.6}), (\ref{3.8}) and (\ref{3.9}).
As we have pointed out there, the presence of Chern-Simons forms
gives rise to modified linearity equations. As these conditions
intervene crucially in the construction of invariant actions, it will be
convenient for our subsequent investigations to summarize them here.

The basic objects of interest are the superfields
$\cl^I$, identified in (\ref{3.22}) and subject to
the modi\-fied linearity conditions
\bea
\pp \cl^I &=& -\f{1}{4} \csi, \label{mlinc1} \\
\qq \cl^I &=& -\f{1}{4} \cti. \label{mlinc2}
\ena
Recall that we have allowed
$\csi$ and $\cti$ to be linear combinations of
terms pertaining to Yang-Mills, gravitational or $U_K(1)$ Chern-Simons forms
as described in the preceding chapters:
\be
\csi \ = \ k^I_\Dt S^\Dt, \cem \cti \ = \ k^I_\Dt T^\Dt,
\eeq
with $k^I_\Dt$ constants, indices $I$ referring to the 2-form
multiplets under consideration and $\Dt$ to the different types of
Chern-Simons forms,
\be
\Dt \ \in \ \left\{ (\cy \cm), (+), (-), U_K(1) \right\}.
\eeq
In the Yang-Mills case we had
\be
S^{\ym} \ = \ -8 \, \tr (\cwb_\da \cwb^\da), \cem
T^{\ym} \ = \ -8 \, \tr (\cw^\al \cw_\al).
\eeq
In the case of gravity, things were slightly more complicated. We found
\bea
S^\pl &=& \pp \lp \mu^\pl +16 R \rd
                      -\f{13}{4} G^{\al \da} G_{\al \da} \rp
                  -4 \bX_\da \bX^\da,  \label{S+} \\
T^\pl &=& \qq \lp \mu^\pl + \f{3}{4} G^{\al \da} G_{\al \da} \rp
               +32 W\raisebox{-.8ex}{${}^{\lsym{\gm \bt \al}}$}
                        W_{\lsym{\gm \bt \al}}
               +\f{4}{3} X^\al X_\al, \label{T+}
\ena
in one chiral sector and
\bea
S^\mn &=& \pp \lp \mu^\mn +\f{3}{4} G^{\al \da} G_{\al \da} \rp
              +32 W_{\lsym{\dg \db \da}}
                 W \raisebox{-.8ex}{${}^{\lsym{\dg \db \da}}$}
              +\f{4}{3} \bX_\da \bX^\da, \label{S-} \\
T^\mn &=& \qq \lp \mu^\mn + 16 R \rd
                       -\f{13}{4} G^{\al \da} G_{\al \da} \rp
               -4 X^\al X_\al, \label{T-}
\ena
in the opposite one. Finally, in the chiral $U_K(1)$ sector we identified
\be
S^\uo \ = \ -2 \bX_\da \bX^\da, \cem
T^\uo \ = \ -2 X^\al X_\al.
\eeq

Clearly, taking into account the relations (\ref{prepoS}) and (\ref{prepoT}),
one may define the truly linear superfields
\be \cl^I - k^I_\Dt \Om^\Dt, \eeq
which are, however, not gauge invariant, in view of (\ref{3fgtr}).
The gauge invariant superfields $\cl^I$, subject to the
modified linearity conditions, will be relevant for
the description of component fields.
In particular, the fieldstrengths
of the antisymmetric tensors are identified in the
covariant superfield expansion of $\cl^I$ as
\be
\lp [\cd_\al , \cd_\da] - 4 \si^a_{\al \da} G_a \rp \cl^I
\ = \  - \f{1}{3} \si_{d \al \da} \vep^{dcba} \ch_{cba}{}^I,
\label{linvec}
\eeq
with
\be \ch_{cba}^I \ = \ H_{cba}^I - k^I_\Dt M_{cba}^\Dt. \eeq


\sect{Dynamics : Green-Schwarz for Gauss-Bonnet}

\indent

The geometric formulation presented in this paper
is quite general and suitable to be employed in
the construction of quite a variety of dynamical
theories involving any kind of gravitational
Chern-Simons forms in the presence of arbitrary
matter and linear multiplet couplings. We shall
leave the discussion of such general constructions
to a separate publication and concentrate here
on the description of a simple, illustrative
example.

This will be obtained from a number of
{\em simplifying assumptions}. First of all
we shall restrict ourselves to
{\em one single antisymmetric tensor gauge field}
which is coupled to the Chern-Simons form relevant
for the {\em Gauss-Bonnet combination} of curvature-squared terms.

Moreover, and in order to exhibit as clearly as possible the
various contributions which arise in the linear
superfield formulation of this Green-Schwarz
coupling we shall {\em neglect} here completely the
{\em matter and Yang-Mills sector}. In technical terms
this means that we may discard the $U(1)$-sector,
\ie work in the framework of traditional
superspace geometry.

The salient features of this dynamical theory will first
be presented in the linear superfield formulation.
A supersymmetric duality transformation, taking into account
the gravitational Chern-Simons superfield, will then be
employed to establish the relation with the dual theory where
the antisymmetric tensor multiplet is replaced by a
chiral multiplet.

\indent

\subsection{From $U(1)$ to traditional superspace}

\indent

The traditional superspace geometry is recovered from
the $U(1)$ superspace, as presented in section {\bf 2},
by simply taking the
kinetic prepotential superfield to vanish. \ie putting
\be
K \ = \ 0.
\label{eq:3.106}
\eeq
As a consequence (see eqs. (\ref{Aal}),(\ref{Ada}) and (\ref{Av}))
one obtains
\be
A_\undal \ = \ 0, \cem A_a \ = \ \frac{3i}{2} \ G_a.
\label{eq:3.105}
\eeq
The equation for the vectorial component of the gauge potential
is a particular artefact of the choice of conventional constraint
for $F_{\al \da}$. On the level of the covariant supergravity
superfields this choice implies
\be X_\alpha \ = \ 0, \cem {\bX}^{\da} \ = \ 0, \eeq
which in turn is tantamount to
\be
\cd_{\! \al} R \ = \ \cd^\da G_{\al \da}, \cem
\cd^\da \rd \ = \  \cd_{\! \al} \, G^{\alpha \da}.
\label{eq:3.109}
\eeq
Moreover, for convenience, we give the additional terms containing
 $G_{\al \da}$ in the basic torsion components,
\bea
\hat{T}_{\ \gamma b}{}^\alpha & = &
   T_{\ \gamma b}{}^\alpha + \delta_\gamma^\alpha \, A_b \ = \
      + \frac{3i}{2} \, \delta_\gamma^\alpha \, G_b
      + \frac{i}{2} \, G^c {(\sigma_c \bar{\sigma}_b)_\gamma}^\alpha,
\label{eq:3.107}
\\
\hat{T}^{\ \dg}{}_{b \da} & = &
   T^{\ \dg}{}_{b \da} - \delta_{\da}^{\dg} \, A_b \ = \
      - \frac{3i}{2} \, \delta_{\da}^{\dg} \, G_b
      - \frac{i}{2} \, G^c {(\bar{\sigma}_c \bar{\sigma}_b)^{\dg}}_{\da},
\label{eq:3.108}
\ena
where now the hatted quantities refer to the traditional superspace
geometry of the so-called old-minimal supergravity multiplet.
Likewise, for the vectorial covariant derivative of a generic
superfield of $U(1)$-weight $w(X)$ we employ the notation
\be \cd_a X \ = \ \hat{\cd}_a \, X + w(X) \, A_a \, X \ = \
\hat{\cd}_a \, X + \frac{3i}{2} \, w(X) \, G_a \, X. \eeq
This last formula will be useful in identifying the additional
$G_a$ contributions arising from the covariant vectorial derivatives
in our equations derived in $U(1)$ superspace.

\indent

\subsection{Identification of Gauss-Bonnet}

\indent

Applying the analysis of chapter {\bf 3} to the case of one
single antisymmetric tensor, and using notations of
appendix {\bf C}, the Gauss-Bonnet combination
is identified in eq.(\ref{3.6}), \ie
\be d H \ = \  k_\Dt \, \Psi^\Dt,
\eeq
in taking $ \ k_\pl \ = \ - \, k_\mn \ = \ k $.
Recall that after taking into account the covariant decomposition
of the curvature-squared terms this becomes
\be
d \ch \ = \ k \lp \Si^\pl - \Si^\mn \rp \ = \ k \, \Si^\gb.
\eeq
In what follows the superscript ${}^\gb$ will be used systematically
to denote quantities referring to the Gauss-Bonnet combination.
In order to write down
the modified linearity conditions for the single superfield $\cl$,
identified in eq.(\ref{3.22}),
\be
\ch_{\gm}{}^\db{}_a \ = \ -2i \, (\si_a \eps)_\gm{}^\db \ \cl,
\eeq
we define (taking $\mu^\pl = \mu^\mn$)
\bea
S^\gb &=& S^\pl - S^\mn \ = \ +8 \, \pp \lp 2 \, \rd R + G^a G_a \rp
    - 32 \, W_{\lsym{\dg \db \da}}
             W \raisebox{-.8ex}{${}^{\lsym{\dg \db \da}}$}, \label{SGB} \\
T^\gb &=& T^\pl - T^\mn \ = \ -8 \, \qq \lp 2 \, \rd R + G^a G_a \rp
    + 32 \, W\raisebox{-.8ex}{${}^{\lsym{\gm \bt \al}}$}
                        W_{\lsym{\gm \bt \al}}. \label{TGB}
\ena
As a consequence, the modified linearity conditions, eqs.(\ref{mlinc1}) to
(\ref{T-}), reduce to
\bea
\pp \cl &=&  - 2k \, \pp \lp 2 \, \rd R + G^a G_a \rp
             +8k \, W_{\lsym{\dg \db \da}}
             W \raisebox{-.8ex}{${}^{\lsym{\dg \db \da}}$},
\label{mlinc1a} \\
\qq \cl &=& + 2k \, \qq \lp 2 \, \rd R + G^a G_a \rp
           -8k \, W\raisebox{-.8ex}{${}^{\lsym{\gm \bt \al}}$}
                        W_{\lsym{\gm \bt \al}}.
\label{mlinc2a}
\ena

\indent

\subsection{Superfield towards component field action}

\indent

Having specified the underlying geometric framework, the action
which describes the coupling of the linear multiplet to supergravity and
gravitational Chern-Simons forms of the Gauss-Bonnet type will
be of the generic form
\be \int E \cf(\cl). \eeq
As matter fields (\ie chiral multiplets) are absent we do not have
to care, for the moment\footnote{beware, however, of eventual
duality transformation from linear superfield formalism to chiral
superfield formalism}, about \ka transformations. Note, however, that
this action will exhibit a field dependent normalization function
of the Einstein term - we will come back to this issue later on.
For the time being we are interested in evaluating explicitly
the component field version and in determining the curvature-squared
contributions.

The starting point for the construction of the component field
action will be the expression for the chiral density as defined
in eq.(\ref{2.36}) with the generic superfield $r$ and its
complex conjugate $\br$ identified as
\be r \ = \ \pp \cf(\cl), \cem \br \ = \ \qq \cf(\cl). \eeq
Inspection of (\ref{2.36}) shows then that the bosonic contributions
to the action will be contained in the projection to lowest
components of the superfield expression
\be
\Box^+ \cf(\cl) \ \equiv \ \left[
  \lp \cd^2 -24 \rd \rp \lp \cdb^2 - 8 R \rp
+ \lp \cdb^2 - 24 R \rp \lp \cd^2 -8 \rd \rp \right] \cf(\cl).
\eeq
In the following we shall restrict ourselves to the discussion
of the purely bosonic terms in the action. To do this
appropriately at the notational level we introduce the
symbol $\ \stackrel{bos}{=} \ $ which means that only bosonic
terms should be retained in the explicit evaluation
(and projection to lowest superfield components is understood).
Applying successively the spinorial covariant derivatives
using explicitly the modified linearity conditions
\be \pp \cl \ = \ - \frac{k}{4} \ S^\gb, \cem
     \qq \cl \ = \ - \frac{k}{4} \ T^\gb, \eeq
relevant for the Gauss-Bonnet combination gives rise to
\bea
\Box^+ \cf(\cl) &\stackrel{bos}{=}&
-4 \, \cf'' \, \cd_{\al \da} \cl \ \cd^{\al\da} \! \cl
+ \cf'' \, [\cd_\al, \cd_\da] \cl \ [\cd^\al, \cd^\da] \cl
\nn \\[2mm]
&& -8\, \lp \cf-\cl \cf'\rp \lp \cd^2 R + \cdb^2 \rd \rp
  +16\, \lp 24 (\cf-\cl \cf')+8\cl^2 \cf'' \rp \rd R
\label{lagbos}
\\[1mm]
&& + \cf' \, \Box^+ \cl
   + \frac{k^2}{8} \, \cf'' \, S^\gb \, T^\gb
   -4k \, R \, \cl \cf'' \, S^\gb
   -4k \, \rd \, \cl \cf'' \, T^\gb, \nn
\ena
where primes denote derivatives of the function $\cf$ with
respect to $\cl$. Note that this equation
contains, in a compact form,
the totality of the bosonic terms in the supersymmetric action.
It remains to work out, in some more detail, the various
contributions of the individual terms.

\indent

\subsection{The basic building blocks and the complete bosonic action}

\indent

We shall now discuss one by one the individual building blocks
for the bosonic part of the action, as they arise in eq.(\ref{lagbos}),
with particular emphasis on the contributions linear and quadratic
in the Gauss-Bonnet coupling $k$. The {\bf first term},
\[ -4 \, \cf'' \, \cd_{\al \da} \cl \ \cd^{\al\da} \! \cl, \]
describes  simply the {\bf kinetic action for the scalar field} which
is identified as the lowest component of the superfield $\cl$.
The {\bf second term},
\[ \cf'' \, [\cd_\al, \cd_\da] \cl \ [\cd^\al, \cd^\da] \cl, \]
is slightly more complicated: among other things it contains the
{\bf kinetic action for the antisymmetric tensor gauge field}
via its covariant field strength $\ch_{cba}$, which appears in
\be
[\cd_\al , \cd_\da] \cl \ = \
\si^a_{\al \da} \left(\tilde{\ch}_a
+ 4 \, \cl \, G_a \right),
\eeq
through its dual defined as
\be \tilde{\ch}^d \ = \ \frac{1}{3!} \, \vep^{dcba}
\ch_{cba}. \eeq
Recall that this field strength is defined as the purely vectorial
component of the 3-form
\be \ch \ = \ d B +  k \lp \cq^\gb - M^\gb \rp, \eeq
with the corresponding component of $M^\gb$ given as
\be
M^\gb_a \ \stackrel{bos}{=} \
-4i \, \cd_a (\rd R) - 2i \, \cd^b (G_b G_a),
\eeq
using obvious notations concerning the Gauss-Bonnet combinations.
The {\bf third term},
\[ -8\, \lp \cf-\cl \cf'\rp \lp \cd^2 R + \cdb^2 \rd \rp, \]
brings in the {\bf curvature scalar with field dependent normalization}
function $\cf-\cl \cf'$, due to the supergravity identity
\be \cd^2 R + \cdb^2 \rd \ = \
        - \frac{2}{3} \, \car + 32 \, \rd R + 4 \, G^a G_a, \eeq
whilst the {\bf fourth term} does not need any comment.
The four {\bf remaining terms} are either linear or quadratic in
the Gauss-Bonnet coupling $k$. Taking into account the relation
\be \cd^2 R - \cdb^2 \rd \ = \ 4i \, \cd_a G^a, \eeq
one finds immediately that the bosonic contribution from
the {\bf last two terms} is simply given as
\be
-4k \, R \, \cl \cf'' \, S^\gb -4k \, \rd \, \cl \cf'' \, T^\gb
\ \stackrel{bos}{=} \
       -256 i \, k \cl \cf'' \ \cd^a \lp \rd R \, G_a \rp. \eeq
In the next step we consider the explicit form of the bosonic
contributions to $S^\gb$ and $T^\gb$,
\bea
S^\gb &\stackrel{bos}{=}& -16 \, \rd
  \lp \f{1}{3} \car -8 \, \rd R +14 \, G^a G_a \rp
   +32i \lp 2 \, G^a \hat{\cd}_a \rd + \rd \hat{\cd}_a G^a \rp, \\
T^\gb &\stackrel{bos}{=}& +16 \, R
  \lp \f{1}{3} \car -8 \, \rd R +14 \, G^a G_a \rp
   +32i \lp 2 \, G^a \hat{\cd}_a R + R \hat{\cd}_a G^a \rp,
\ena
which will be used to establish the explicit expression for the term
\[ \frac{k^2}{8} \, \cf'' \, S^\gb \, T^\gb, \]
at order $k^2$.
Finally we have to substitute for $\Box^+ \cl$ in the last
remaining term. Using the explicit definitions and relations from
appendix {\bf C} we find
\be \Box^+ \cl \ = \ -\frac{k}{4} \lp \cd^2 - 24 \rd \rp T^\gb
       -\frac{k}{4} \lp \cdb^2 - 24 R \rp S^\gb, \eeq
with bosonic part
\be
\Box^+ \cl \ \stackrel{bos}{=} \
  + 2i \, k \ \vep_{fedc} \, \cw^{fe \, , \, ba} {\cw^{dc}}_{\, , \, ba}
  + 48i \, k \, \vep^{dcba} (\hat{\cd}_d G_c) (\hat{\cd}_b G_a)
  +2k \, \Box^- \lp 2 \rd R + G^a G_a \rp.
\eeq
The last term in this equation is basically a space-time divergence
whose explicit form is irrelevant for the present discussion.
If desired, it can be evaluated using the definition of $\Box^-$ in appendix {\bf C}.

Putting all the information concerning the individual terms in
(\ref{lagbos}) together one obtains
\bea
\Box^+ \cf(\cl) &\stackrel{bos}{=}&
\frac{16}{3} \, (\cf-\cl \cf') \, \car
+ 16 \, (\cf-\cl \cf'+ \cl^2 \cf'') \, (8 \rd R -2 G^a G_a)
\nn \\ [2mm] &&
-2 \cf'' \tilde{\ch}^a \tilde{\ch}_a
+8 \cf'' \cd^a \! \cl \, \cd_a \cl
-16 \cl \cf'' G_a \tilde{\ch}^a \\ [2mm] &&
+2i k \, \cf' \, \vep_{fedc} \, \cw^{fe \, , \, ba} {\cw^{dc}}_{\, , \, ba}
+16ik \, \cf' \, \vep^{dcba} (\hat{\cd}_d G_c) (\hat{\cd}_b G_a)
\nn \\ [2mm] &&
+16ik \, \cf' \cd^a \left[
4i (\rd \hat{\cd}_a R - R \hat{\cd}_a \rd)
+2 G^b \tilde{\car}_{ba}
+ \f16 G_a \car -32 G_a \rd R -2 G_a G^b G_b
\right]
\nn \\ [2mm] &&
-\frac{64}{3} \, k^2 \cf'' \, \car \left[
2i G^a (\rd \hat{\cd}_a R - R \hat{\cd}_a \rd)
+(14G^aG_a - 8 \rd R) \rd R + \f16 \car \rd R
\right]
\nn \\ [2mm] &&
-32 k^2 \cf'' \left[
\rd R (14G^aG_a - 8 \rd R)^2
+4 i (14G^b G_b - 8 \rd R) G^a(\rd \hat{\cd}_a R - R \hat{\cd}_a \rd)
\right. \nn \\ [2mm] && \left. \hspace{1.5cm}
+16 G^b G^a (\hat{\cd}_b \rd)(\hat{\cd}_a R)
+8 G^a (\hat{\cd}_b G^b) \hat{\cd}_a (\rd R)
+4 \rd R (\hat{\cd}_b G^b)(\hat{\cd}_a G^a)
\right].
\nn \ena
Taken as the lowest superfield component this expression describes,
up to a factor $e = det e_m{}^a$, the bosonic part of the lagrangian
density for the coupling of the antisymmetric tensor multiplet to
supergravity in the presence of Gauss-Bonnet Chern-Simons forms.
Observe that in the absence of Chern-Simons forms, \ie for $k=0$,
we are just left with the first two lines.

On the other hand, coupling of Chern-Simons form does not just mean
modify the fieldstrength of the antisymmetric tensor: supersymmetry
enforces quite a number of additional couplings, linear and quadratic
in the parameter $k$. This illustrates once more the striking fact that
supersymmetrization of the Green-Schwarz mechanism does not only introduce new
fermionic terms but necessitates genuine new bosonic contributions,
as for instance the square of the Weyl-tensor in the example discussed here.

The Gauss-Bonnet combination itself is not explicitly present in
this action, it appears only after discussion of the equation of motion for
the antisymmetric tensor gauge field.
Instead of discussing the explicit form of the component field
equations of motion, which are quite complex, we shall turn
immediately to the dual theory with the antisymmetric tensor replaced by
a scalar field.

\indent

\subsection{Duality transformation and the dual theory}

\indent

Taking $X$ to be some real, covariant and unconstrained
superfield we consider
\be \int E \left[ \cf(X) + b \lp X - k \, \Om^\gb \rp
         \lp \phi + \phib \rp \right]. \eeq
Variation with respect to $\phi$ and $\phib$, or rather with
respect to their unconstrained prepotentials, entail
linearity conditions
for $X - k \, \Om^\gb$ and thus allow to identify
$X$ with $\cl$, subject to modified linearity conditions
(\ref{mlinc1a}), (\ref{mlinc2a}), thus
getting back the theory already discussed.

On the other hand, variation with respect to $X$
results in
\be \cf' + b \lp \phi + \phib \rp \ = \ 0,
\label{dualrel} \eeq
giving rise to a theory expressed in terms of the
chiral resp. antichiral superfields $\phi$ and $\phib$.
The theory obtained in this way is said to be dual to the one
described previously.

\indent

After taking into account the duality relation (\ref{dualrel})
\ie the (algebraic) equation of motion of the first order action,
one obtains the dual theory described in terms of
one single chiral superfield $\phi$ and its conjugate $\phib$ as
\be \La \ = \ \int E \left[ \cf \lp X [\phi + \phib] \rp
         + b \lp \phi + \phib \rp X [\phi + \phib]
- bk \, \Om^\gb \lp \phi + \phib \rp\right]. \eeq
This action obviously consists of two parts : the usual action
for the kinetic terms of the chiral multiplet and a new
one, containing the Chern-Simons superfield, which will,
at the component field level, give rise to the curvature-squared terms.

In order to match the traditional notations we parametrize
\be -3 e^{-\f{1}{3}K\lp \phi , \phib \rp}
 \ = \ \cf \lp X [\phi + \phib] \rp
         + b \lp \phi + \phib \rp X [\phi + \phib], \eeq
to establish the first part of the dual theory in the
usual notation of standard supergravity
matter coupling with \ka potential $K\lp \phi , \phib \rp$ \ie
\be \La_0 \ = \ -3 \int E \, e^{-\f{1}{3}K\lp \phi , \phib \rp}. \eeq

The explicit supersymmetric component field action may then be obtained from
\be
r_0 \ = \ \frac{3}{8} \qq e^{-\f{1}{3}K\lp \phi , \phib \rp}, \cem
\bar{r}_0 \ = \ \frac{3}{8} \pp e^{-\f{1}{3}K\lp \phi , \phib \rp},
\eeq
in terms of the chiral density construction (\ref{2.36}).
It leads to the usual supergravity-matter action with field dependent
normalization of the curvature scalar term, which we refrain however to
reproduce here in detail.

Instead we concentrate on the second part of the action,
\be \La_1 \ = \ -bk \int E \, \Om^\gb \lp \phi + \phib \rp, \eeq
which reflects the modifications of the standard theory
due to the presence of the Green-Schwarz mechanism
for the Gauss-Bonnet combination. In the following we
will discuss the component field evaluation
of this additional part of the action.

In a first step, still completely in terms of superfields,
we make use of integration by parts in superspace, to write
\be \int E \, \Om^\gb \phi \ = \
    - \frac{1}{8} \int \frac{E}{R} \, \phi \, \qq \Om^\gb
         \ = \ \frac{1}{32} \int \frac{E}{R} \, \phi \, T^\gb, \eeq
where we used (\ref{prepoT}) and $T^\gb$ is given in
eq.(\ref{TGB}). Reasoning in the same manner for the
complex conjugate term in $\La_1$ we then arrive at
\be
\La_1 \ = \ - \frac{bk}{32} \int \frac{E}{R} \, \phi \, T^\gb
            - \frac{bk}{32} \int \frac{E}{\rd} \, \phib \, S^\gb.
\eeq

Employing the language of the chiral density construction
this means that the component field expression is derived
from the chiral, resp. antichiral superfields
\be r_1 \ = \ - \frac{bk}{16} \ \phi \, T^\gb,
      \cem \br_1 \ = \ - \frac{bk}{16} \ \phib \, S^\gb, \eeq
following the usual procedure.
\bea
\La_1 &\stackrel{bos}{=}&
-i\, bk \lp \phi + \phib \rp
\lp \f14 \, \vep^{d c \, d'  c'} \, R_{d c \, , \, b}{}^a
                                     R_{d'  c' \, , \, a}{}^b
+ 4 \, \vep^{d\, c\, b \, a}\, \cd_d G_c \, \cd_b G_a \rp
\nn \\ [1mm] &&
- \frac{bk}{8} \lp \phi - \phib \rp
\vep^{d\, c\,d'  c'} \vep^{b\, a\, b'  a'}R_{dc \, ,\, ba} \, R_{d'  c' \, ,\, b' a'}
\nn \\ [1mm] &&
+\, 8 \, bk \lp \phi - \phib \rp \cd^a \cd_a \! \lp \rd R \rp
+\, 4 \, bk \lp \phi + \phib \rp \cd^a \!
\lp \rd \hat{\cd}_a R - R \hat{\cd}_a \rd \rp
\label{lambda1}
 \\ [1mm] &&
- \frac{bk}{16} \lp \phi + \phib \rp \cd^a
\left\{
G^b \lp \car_{ab} - \f16 \eta_{ab} \car \rp - G_a \lp 2 \rd R + G^b G_b \rp
\right\}
\nn \\ [1mm] &&
+ \, 4 \, bk \lp \phi - \phib \rp \cd^a \cd^b \! \lp G_a G_b \rp
-\, bk \, T^\gb \, \cd^2 \phi -\, bk\, S^\gb \, \cdb^2 \phib.
\nn
\ena

Observe the analogy of this expression with the Yang-Mills case
(\ref{compaf}) discussed in the introductory section:
the curvature-squared term which corresponds to the initial Chern-Simons
combination, in our case Gauss-Bonnet, appears with a factor $\phi - \phib$,
whilst the orthogonal one (in the sense of Hodge duality) acquires a factor
$\phi + \phib$ (in the Yang-Mills case this was just the kinetic term with
field dependent gauge coupling function).

As this example is merely intended as an illustration of the methods of
superspace geometry at work, we will not pursue here this discussion any
further. A more detailed study of this and similar, but more general theories,
is left to forthcoming research, see also the remarks in the concluding
section.


\indent

\sect{Conclusion and outlook}

\indent

The main emphasis of this review was on a concise and complete presentation of
the superspace geometric description of gravitational Chern-Simons forms.
Making use of the structural analogy of Chern-Simons forms in supersymmetric
theories with the geometry of the 3-form multiplet (coupled to supergravity)
makes it possible to cope with otherwise highly complex technicalities.

The second important point we wanted to exhibit is the relevance of this
geometrical description for the construction of supersymmetric dynamical
theories with gravitational Chern-Simons forms.

As an illustration of this point we described in section {\bf 4} the basic
features of a particular and relatively simple example, dealing with
Chern-Simons forms of the Gauss-Bonnet type.
Starting from the linear superfield mechanism and performing then the duality
transformation to the chiral superfield mechanism we displayed the bosonic
parts of the component field action in both versions.

This example was particularly simple in two respects. First of all it was
formulated in the traditional superspace geometry approach (\ie no
explicit $U_K(1)$ factor present in the structure group). Secondly only
one linear, resp. chiral, superfield was taken into account, without any
additional matter or Yang-Mills sectors.

As this specific example was chosen to illustrate the interplay of geometry
and dynamics, we did not analyse the sector of previously auxiliary fields,
which now appear with terms containing space-time derivatives - a subject
which should more conveniently studied in more general situations.

Likewise, the question of the normalization of the curvature scalar term, the
interpretation of its field dependent normalization function as well as
questions of field redefinitions which have the form of Weyl rescalings were
not further pursued. These issues are more conveniently adressed in the
full-fledged $U_K(1)$ superspace geomotry framework.

We wish to emphasize, however, that the geometric superspace formulation of
gravitational Chern-Simons forms, which was the main purpose of this review,
is very well-suited for the discussion of the questions alluded to and which
will be the subject of future investigations.

\appendix
\renewcommand{\sect}[1]{\setcounter{equation}{0}\section{#1}}

\addcontentsline{toc}{section}{Appendices}

\section*{Appendices}

\indent

\sect{3-form gauge potential and Chern-Simons forms}

\indent

\subsection{3-form gauge potential in $U_K(1)$ superspace}

\indent

We consider
\be B^3 \ = \ \f{1}{3!} E^A E^B E^C B^3{}_{CBA}, \eeq
a 3-form gauge potential subject to gauge transformations
\be B^3 \ \mapsto \ {}^\La B^3 \ = \ B^3 + d \La, \eeq
described in terms of a 2-form in superspace,
\be \La \ = \ \f12 E^A E^B \La_{BA}. \eeq
In some more detail
\be
\f{1}{3!} E^A E^B E^C \ {}^\La B^3{}_{CBA}
\ = \ \f{1}{3!} E^A E^B E^C \lp B^3{}_{CBA} + 3 \, \cd_C \La_{BA}
                                    + 3 \, T_{CB}{}^F \La_{FA} \rp,
\eeq
or
\be
{}^\La B^3{}_{CBA} \ = \ B^3{}_{CBA} + \oint_{CBA}
                     \lp \cd_C \La_{BA} + T_{CB}{}^F \La_{FA} \rp,
\eeq
where the graded cyclic sum is defined as
\be \oint_{CBA} \ = \ CBA + (-)^{c(b+a)} BAC + (-)^{a(b+c)} ACB, \eeq
with $a=0$ for vectorial and $a=1$ for spinorial values of the
superspace indices.

The covariant fieldstrength
\be \Si \ = \ d B^3, \eeq
is a 4-form in superspace,
\be \Si \ = \ \f{1}{4!} E^A E^B E^C E^D \, \Si_{DCBA}, \eeq
with coefficients
\be
\f{1}{4!} E^A E^B E^C E^D \, \Si_{DCBA} \ = \
\f{1}{4!} E^A E^B E^C E^D \lp 4 \, \cd_D B^3{}_{CBA}
                            + 6 \, T_{DC}{}^F B^3{}_{FBA} \rp.
\eeq

\indent

\subsection{Explicit solution of the constraints}

\indent

We recall the superspace constraints for the 3-form gauge potential:
\be \Si_{\undel \, \undgm \, \undbt \, A} \ = \ 0. \eeq
In a first step we solve
\be \Si_{\dt \, \gm \, \bt \, A} \ = \ 0, \eeq
by
\be
B^3{}_{\gm \bt A} \ = \ \cd_A U_{\gm \bt} +
    \oint_{\gm \bt} \lp \cd_\gm U_{\bt A} + T_{A \gm}{}^F U_{F \bt} \rp,
\eeq
and the complex conjugate
\be \Si^{\dd \, \dg \, \db}{}_A \ = \ 0, \eeq
    by
\be
B^3{}^{\dg \db}{}_A \ = \ \cd_A V^{\dg \db} +
    \oint^{\dg \db} \lp \cd^\dg V^\db{}_A + T_A{}^{\dg \, F} V_F{}^\db \rp.
\eeq

Since the prepotentials $U_{\bt A}$ and $V^\db{}_A$ should reproduce the gauge
transformations of the gauge potentials $B^3{}_{\gm \bt A}$
and $B^3{}^{\dg \db}{}_A$ we assign
\be U_{\bt A} \ \mapsto \ {}^\La U_{\bt A} \ = \ U_{\bt A} + \La_{\bt A}, \eeq
and
\be V^\db{}_A \ \mapsto \ {}^\La V^\db{}_A \ = \ V^\db{}_A + \La^\db{}_A, \eeq
as gauge transformation laws for the prepotentials. On the other hand,
the so-called {\em pregauge transformations} are defined as the zero-modes
of the gauge potentials themselves, that is transformations which leave
$B^3{}_{\gm \bt A}$ and $B^3{}^{\dg \db}{}_A$ invariant. They are given as
\be U_{\bt A} \ \mapsto \ U_{\bt A}
      + \cd_\bt \chi_A - (-)^a \cd_A \chi_\bt + T_{\bt A}{}^F \chi_F, \eeq
and
\be V^\db{}_A \ \mapsto \ V^\db{}_A
      + \cd^\db \psi_A - (-)^a \cd_A \psi^\db + T^\db{}_A{}^F \psi_F. \eeq

We parametrize the prepotentials now as follows:
\bea
U_\bt{}^\da &=& W_\bt{}^\da + T_\bt{}^{\da \, f} K_f, \\
V^\db{}_\al &=& W_\al{}^\db - T_\al{}^{\db \, f} K_f,
\ena
and
\bea
U_{\bt \, a} &=& W_{\bt \, a} - \cd_\bt K_a, \\
V^\db{}_a &=& W^\db{}_a + \cd^\db K_a.
\ena
Explicit substitution shows that the $K_a$ terms drop out in
$B^3{}_{\gm \bt A}$ and $B^3{}^{\dg \db}{}_A$. Denoting furthermore
\be
U_{\bt \al} \ = \ W_{\bt \al}, \cem {\rm and} \cem V^{\db \da}
\ = \ W^{\db \da},
\eeq
we arrive at
\bea
B^3{}_{\gm \bt A} &=& \cd_A W_{\gm \bt} +
    \oint_{\gm \bt} \lp \cd_\gm W_{\bt A} + T_{A \gm}{}^F W_{F \bt} \rp, \\
B^3{}^{\dg \db}{}_A &=& \cd_A W^{\dg \db} +
    \oint^{\dg \db} \lp \cd^\dg W^\db{}_A + T_A{}^{\dg \, F} W_F{}^\db \rp,
\ena
\ie a pure gauge form for the coefficients
$B^3{}_{\gm \bt A}$ and $B^3{}^{\dg \db}{}_A$
with the 2-form gauge parameter $\La$ replaced by the prepotential
2-form
\be W \ = \ \f12 E^A E^B W_{BA}, \cem {\rm with} \cem W_{ba} \ = \ 0. \eeq
We take advantage of this fact to perform a redefinition of the
3-form gauge potentials, which has the form of a gauge
transformation, in the following way:
\be \hB^3 \ := \ {}^{-W} B^3 \ = \ B^3 - dW. \eeq
This leaves the fieldstrength invariant and leads in particular to
\be
\hB^3{}_{\gm \bt A} \ = \ 0, \cem \mbox{and} \cem
\hB^3{}^{\dg \db}{}_A \ = \ 0,
\eeq
whereas the coefficient $B^3{}_\gm{}^\db{}_a$ is replaced by
\be
\hB^3{}_\gm{}^\db{}_a \ = \ B^3{}_\gm{}^\db{}_a
     - \cd_\gm W^\db{}_a - \cd^\db W_{\gm a} - \cd_a W_\gm{}^\db.
\eeq

We define the tensor decomposition
\be
\hB^3{}_\gm{}^\db{}_a \ = \
     T_\gm{}^\db{}^f \lp \eta_{fa} \, \Om + \hW_{[fa]} + \Omt_{(fa)} \rp,
\eeq
where $\hW_{[fa]}$ is antisymmetric and $\Omt_{(fa)}$ symmetric and traceless,
and perform another redefinition which has again the
form of a gauge transformation,
this time of parameter
\be \hW \ = \ \f12 E^a E^b \hW_{[ba]}, \eeq
such that
\be \Om \ := \ {}^{-\hW} \hB^3 \ = \ \hB^3 - d \hW. \eeq
Note that this reparametrisation leaves $\hB^3{}_{\gm \bt A}$ and
$\hB^3{}^{\dg \db}{}_A$ untouched, they remain zero.

Let us summarize the preceding discussion: we started out with the 3-form
gauge potential $B^3$. The constraints on its fieldstrength led us to
introduce prepotentials. By means of prepotential dependent
redefinitions of $B^3$, which have the form of gauge transformations
(and which, therefore, leave the fieldstrength invariant), we arrived
at the representation of the 3-form gauge potential in terms of $\Om$,
with the particularly nice properties
\be \Om_{\gm \bt A} \ = \ 0, \cem \Om^{\dg \db}{}_A \ = \ 0, \eeq
and
\be
\Om_\gm{}^\db{}_a \ = \
     T_\gm{}^\db{}^f \lp \eta_{fa} \, \Om + \Omt_{(fa)} \rp,
\eeq
Clearly, in this representation, calculations simplify considerably.
We shall, therefore, from now on pursue the
solution of the constraints in terms of $\Om$ and turn to the equation
\be
\Si_{\dt \gm}{}^{\db \da} \ = \ 0 \ = \
      \oint_{\dt \gm}^{\db \da} T_\dt{}^{\dg \, f} \Om_f{}^\db{}_\al,
\eeq
which tells us simply that $\Omt_{(ba)}$ is zero.
Hence,
\be \Om_\gm{}^\db{}_a \ = \ T_\gm{}^\db{}_a \, \Om. \eeq

We turn next to the constraints
\be
\Si_\dt{}^{\dg \db}{}_a \ = \ 0 \ = \ \oint^{\dg \db}
   \lp \cd^\dg \Om_\dt{}^\db{}_a + T_\dt{}^{\dg \, f} \Om_f{}^\db{}_a \rp,
\eeq
and
\be
\Si^\dd{}_{\gm \bt a} \ = \ 0 \ = \ \oint_{\gm \bt}
   \lp \cd_\gm \Om^\dd{}_{\bt a} + T_\gm{}^{\dd \, f} \Om_{f \bt a} \rp,
\eeq
which, after some straightforward spinorial index gymnastics give rise to
\bea
\Om_{\gm \, ba} &=& 2 (\si_{ba})_\gm{}^\vp \, \cd_\vp \Om, \\
\Om^\dg \, {}_{ba} &=& 2 (\sib_{ba})^\dg{}_\dv \, \cd^\dv \Om.
\ena

This completes the discussion of the solution of the constraints,
we discuss next the consequences of this solution for the remaining
components in $\Si$ \ie $\Si_{\undel \, \undgm \, ba}$,
$\Si_{\undel \, cba}$ and $\Si_{dcba}$. As a first step we consider
\be
\Si_{\dt \, \gm \, ba} \ = \
\oint_{\dt \gm} \lp \cd_\dt \Om_{\gm \, ba}
- T_{\dt b \dv} \Om_\gm{}^\dv{}_a + T_{\dt a \dv}\Om_\gm{}^\dv{}_b \rp,
\eeq
and
\be
\Si^{\dd \dg}{}_{ba} \ = \ \oint^{\dd \dg}
   \lp \cd^\dd \Om^\dg{}_{ba}
   - T^\dd{}_b{}^\vp \Om_\vp{}^\dg{}_a + T^\dd{}_a{}^\vp \Om_\vp{}^\dg{}_b \rp.
\eeq

Substituting for the 3-form gauge potentials as determined so far,
and making appropriate use of the supergravity Bianchi identities yields
\be
\Si_{\dt \, \gm \, ba} \ = \ -2 (\si_{ba} \eps)_{\dt \gm} \ \pp \Om,
\eeq
and
\be
\Si^{\dd \dg}{}_{ba} \ = \ -2 (\sib_{ba}\eps)^{\dd \dg} \ \qq \Om,
\eeq
The appearance of the chiral projection operators suggests to define
\bea S &=& -4 \pp \Om, \\ T &=& -4 \qq \Om. \ena
The gauge invariant superfields $S$ and $T$ have chirality properties
\be \cd_\al S \ = \ 0, \cem \cd^\da T \ = \ 0, \eeq
and we obtain
\bea
\Si_{\dt \gm \, ba} &=& \f12 (\si_{ba}\eps)_{\dt \gm} \, S, \\
\Si^{\dd \dg}{\,}_{ba} &=& \f12 (\sib_{ba}\eps)^{\dd \dg} \, T.
\ena

In the next step we observe that, due to the information extracted so far from
the solution of the constraints, the fieldstrength
\be
\Si_\dt{}^\dg{\,}_{ba} \ = \ T_\dt{}^{\dg \, c} \, \Si_{cba},
\eeq
is determined such that $\Si_{cba}$ is totally antisymmetric in its three
vectorial indices. As, in its explicit definition a linear term appears
(due to the constant torsion term), \ie
\be
\Si_\dt{}^\dg{\,}_{ba} \ = \ T_\dt{}^{\dg \, c} \Om_{cba} +
            {\rm \ derivative \ and \ other \ torsion \ terms},
\eeq
we can absorb $\Si_{cba}$ in a modified 3-form gauge potential
\be
\undOm_{cba} \ = \ \Om_{cba} - \Si_{cba},
\eeq
such that the corresponding modified fieldstrength vanishes:
\be
\undSi_\dt{}^\dg{\,}_{ba} \ = \ 0.
\eeq
The outcome  of this equation is then the relation
\be
\lp [\cd_\al, \cd_\da] - 4 G_{\al \da} \rp \Om
             \ = \ \f13 \si_{d\, \al \da} \, \vep^{dcba} \, \undOm_{cba},
\eeq
which identifies $\undOm_{cba}$ in the superfield expansion of the
unconstrained prepotential $\Om$.

Working, from now on, in terms of the modified quantities,
the remaining coefficients, at canonical dimensions
3/2 and 2, i.e. $\undSi_{\unddt \, cba}$ and $\undSi_{dcba}$,
respectively, are quite straightforwardly obtained in terms of
spinorial derivatives of the basic gauge invariant superfields $S$ and $T$.
To be more precise, at dimension 3/2 one obtains
\bea
\undSi_{\dt \, cba} &=&
         -\f{1}{16} \si^d_{\dt \dd} \, \vep_{dcba} \, \cd^\dd \! S, \\
\undSi^\dd {}_{cba} &=&
         +\f{1}{16} \sib^{d \, \dd \dt} \, \vep_{dcba} \, \cd_\dt T,
\ena
and the Bianchi identity at dimension 2 takes the simple form
\be
\lp \cd^2 - 24 \rd \rp T - \lp \cdb^2 - 24 R \rp S \ = \
         \f{8i}{3} \vep^{dcba} \undSi_{dcba}.
\eeq

As to the gauge structure of the 3-form gauge potential we note that
in the transition from $B^3$ to $\Om$, the original 2-form gauge
transformations have disappeared, $\Om$ is invariant under those.
In exchange, however, as already mentioned earlier, $\Om$ transforms
under so-called pregauge transformations (which, in turn, leave $B^3$
unchanged. As a result, the residual pregauge transformations of
the unconstrained prepotential superfield,
\be \Om \ \mapsto \ \Om' \ = \ \Om + \la, \eeq
are parametrized in terms of a linear superfield $\la$ which satisfies
\be \pp \la \ = \ 0, \cem \qq \la \ = \ 0. \eeq
In turn, $\la$ can be expressed in terms of an unconstrained
superfield, as we know from the explicit solution of the superspace
constraints of the 2-form gauge potential, actually defining
the linear superfield geometrically.
In other words, the pregauge transformations should respect the particular
form of the coefficients of the 3-form $\Om$.

\indent

\subsection{Yang-Mills in $U_K(1)$ superspace}

\indent

In section {\bf 2.5} we have presented supersymmetric Yang-Mills theory in
terms of a Lie-algebra valued gauge potential
\be
\ca \ = \ E^A \ca_A^{(r)} T_{(r)} \ = \ \ca^{(r)} T_{(r)},
\eeq
whose fieldstrength
\be
\cf \ = \ d\ca+ \ca\ca,
\eeq
is a 2-form in superspace defined as
\be
\cf \ = \ \f{1}{2} E^A E^B \cf_{BA},
\eeq
with coefficients
\be
\cf_{BA} \ = \ \cd_B \ca_A - (-)^{ab} \cd_A \ca_B - (\ca_B,\ca_A)
                +{T_{BA}}^C \ca_C.
\eeq
The fieldstrength is covariant with respect to superfield
gauge transformations
\be
\ca \ \mapsto \ {}^g \ca \ = \ g^{-1} \ca g - g^{-1} d g.
\eeq

In the present subsection we point out in some detail the solution
of the constraints in terms of prepotentials. In a first step we observe
that the constraints
\be \cf_{\bt \al} \ = \ 0, \cem  \cf^{\db \da} \ = \ 0, \eeq
have the solution
\be
\ca_\al \ = \  -\ct^{-1} D_\al \ct, \cem  \ca^\da \ = \ -\cu^{-1} D^\da \cu.
\eeq
We use here $D_\undal=E_\undal{}^M \prt_M$.
The unconstrained prepotential superfields $\ct$, $\cu$ are related through
\be \ct^{-1} \ = \ \cu^\dagger. \eeq

The gauge transformations of the prepotentials should be defined such
that they reproduce those of the gauge potentials as given above. On the
other hand, there are additional non-trivial gauge transformations which
act on the prepotentials but which do not show up in those of the potentials.
These transformations are called pregauge transformations. Altogether,
gauge and pregauge transformations of the prepotentials are defined as
follows:
\be
\ct \ \mapsto \ \Lab^{-1} \ct g, \cem
\cu \ \mapsto \ \La^{-1} \cu g,
\eeq
the parameters of the pregauge transformations being chiral resp.
antichiral superfields, \ie
\be
D^\da \La \ = \ 0, \cem D_\al \Lab \ = \ 0.
\eeq
We emphasize that the combination
\be \Upsilon \ = \ \ct \cu^{-1}, \eeq
is inert under the $g$-transformations and changes under pregauge
transformations as
\be \Upsilon \ \mapsto\ \Lab^{-1} \Upsilon \La. \eeq
The constraint
\be \cf_\bt{}^\da \ = \ 0, \eeq
is a so-called {\em conventional} one, it expresses the vector
component of the superspace gauge potential in terms of the spinorial
ones:
\be
\ca_{\al \da} \ = \
   \i2 \lp \cd_\al \ca_\da + \cd_\da \ca_\al - \{\ca_\al, \ca_\da\} \rp.
\eeq
As usual in the explicit definition of the fieldstrength,
the derivatives are covariant only
with respect to local Lorentz und $U_K(1)$ transformations.

The component fields of supersymmetric Yang-Mills theory are the
usual gauge potentials as identified in the real representation,
\be
\ca|| \ = \ i \, dx^m \, a_m(x),
\eeq
the covariant "gaugino" fields
\be
\cw^\db| \ = \ i\lab^\db(x), \cem \cw_\bt| \ = \ -i\la_\bt(x),
\eeq
and the auxiliary bosonic field
\be
\cd^\al \cw_\al| \ = \ -2 {\bf D}(x),
\eeq
all of them defined in the real representation as well.
As to the last one, by a slight abuse of notation, we use the same symbol for
the superfield itself and its lowest component.

The appearance of the prepotential superfields and the particular
form of their gauge transformations allows
to dispose completely of the original
$g$ gauge transformations. This is achieved by prepotential
dependent redefinitions of the gauge potential $\ca$ which have the form of a
gauge transformation. In more technical terms, we define
\bea
\vp &=& \cu \ca \cu^{-1} - \cu d \cu^{-1} \ = \ {}^{\cu^{-1}} \ca, \\
\vpb &=& \ct \ca \ct^{-1} - \ct d \ct^{-1} \ = \ {}^{\ct^{-1}} \ca.
\ena
The new gauge potentials are inert under $g$ gauge transformations but
change under pregauge transformations, as induced from the redefinitions, \ie
\bea
\vp \ \mapsto \ {}^\La \vp &=& \La^{-1} \vp \La - \La^{-1} d \La, \\
\vpb \ \mapsto \ {}^{\Lab} \vpb &=& \Lab^{-1} \vpb \Lab - \Lab^{-1} d \Lab,
\ena
We call $\vpb$ the chiral and $\vp$ the antichiral representation because the
corresponding gauge transformations are parametrized in terms of chiral resp.
antichiral superfields. Also, $\ca$ is called the real representation.
The gauge potentials in the chiral and
antichiral representations are related by
\be \vp \ = \ \Upsilon^{-1} \vpb \Upsilon
- \Upsilon^{-1} d \Upsilon \ = \ {}^\Upsilon \vpb. \eeq
The connections $\vp = E^A \vp_A$ and $\vpb = E^A \vpb_A$ depend in a very
simple way on $\Upsilon$:
\be
\vp_\al \ = \ -\Upsilon^{-1} D_\al \Upsilon, \cem \vp^\da \ = \ 0, \cem
\vp_{\al \da} \ = \ \i2 \cd_\da \vp_\al,
\eeq
\be
\vpb_\al \ = \ 0, \cem \vpb^\da \ = \ -\Upsilon D^\da \Upsilon^{-1}, \cem
\vpb_{\al \da} \ = \ \i2 \cd_\al \vpb_\da.
\eeq
Likewise, for the gaugino superfields one finds immediately
\be \cw_\al(\vp) \ = \ -\f{1}{8} \qq \vp_\al, \eeq
\be \cwb^\da(\vpb) \ = \ \f{1}{8} \pp \vpb^\da. \eeq

Clearly, gauge invariant expressions are independent of the the
representation chosen to describe the gauge potentials.

\indent

\subsection{Chern-Simons forms in superspace}

\indent

In this paper we deal with Chern-Simons forms of
the Yang-Mills and gravitational
types. Under gauge transformations these Chern-Simons 3-forms change by the
exterior derivative of a 2-form, which depends on the gauge parameter and the
gauge potential.

Due to this property one may view the Chern-Simons form as a special case of a
generic 3-form gauge potentiel, as
discussed in the first two subsections of
this appendix. This point of view is particularly useful for the supersymmetric
case.

To make this point as clear as possible we first recall, in this subsection,
some general properties of Chern-Simons forms in superspace.

To begin with we consider two gauge potentials $\ca_0$ and $\ca_1$
in superspace. Their field\-strength-squared invariants are related
through
\be
\tr \lp \cf_0 \cf_0 \rp - \tr \lp \cf_1 \cf_1 \rp \ = \
                                        d \cq \lp \ca_0, \ca_1 \rp.
\eeq
This is the superspace version of the Chern-Simons formula, where
\be
\cf_0 \ = \ d \ca_0 + \ca_0 \ca_0, \cem \cf_1 \ = \ d \ca_1 + \ca_1 \ca_1.
\eeq
On the right appears the superspace Chern-Simons
form,
\be
\cq \lp \ca_0, \ca_1 \rp \ = \
          2 \int_0^1 dt \ \tr \left\{ \lp \ca_0 - \ca_1 \rp \cf_t \right\},
\eeq
where
\be \cf_t \ = \ d \ca_t + \ca_t \ca_t, \eeq
is the fieldstrength for the interpolating gauge potential
\be \ca_t \ = \ (1-t) \ca_0 + t \ca_1. \eeq
The Chern-Simons form is antisymmetric in its arguments, \ie
\be \cq \lp \ca_0, \ca_1 \rp \ = \ - \cq \lp \ca_1, \ca_0 \rp. \eeq
In the particular case $\ca_0=\ca$, $\ca_1=0$, one obtains
\be
\cq \lp \ca \rp \ := \ \cq \lp \ca, 0 \rp \ = \
                  \tr \lp \ca \cf- \f{1}{3} \ca \ca \ca \rp,
\eeq
We shall also make use of the identity
\be
\cq \lp \ca_0, \ca_1 \rp + \cq \lp \ca_1, \ca_2 \rp + \cq \lp \ca_2, \ca_0 \rp
\ = \ d \chi \lp \ca_0, \ca_1, \ca_2 \rp,
\eeq
with
\be
\chi \lp \ca_0, \ca_1, \ca_2 \rp
    \ = \ \tr \lp \ca_0 \ca_1 + \ca_1 \ca_2 + \ca_2 \ca_0 \rp.
\eeq

This last relation (the so-called {\em triangular equation}) is particularly
useful for the determination of the gauge transformation of the Chern-Simons
form. The argument goes as follows: first of all, using the definition
given above, one observes
that \be
\cq \lp {}^g \ca, 0 \rp \ = \ \cq \lp \ca, dg \, g^{-1} \rp.
\eeq
Combining this with the triangular equation for the special choices
\be \ca_0 \ = \ 0, \cem \ca_1 \ = \ \ca, \cem \ca_2 \ = \ dg \, g^{-1}, \eeq
one obtains
\be
\cq \lp 0, \ca \rp + \cq \lp {}^g \ca, 0 \rp + \cq \lp dg \, g^{-1}, 0 \rp
   \ = \ d \, \tr \lp \ca \, dg \, g^{-1} \rp,
\eeq
or, using the antisymmetry property
\be
\cq \lp {}^g \ca \rp - \cq \lp \ca \rp \ = \
   d \, \tr \lp \ca \, dg \, g^{-1} \rp - \cq \lp dg \, g^{-1} \rp.
\eeq
The last term in this equation is an exact differential form in superspace
as well, it can be written as
\be \cq \lp dg \, g^{-1} \rp \ = \ d \si, \eeq
where the 2-form $\si$ is defined as
\be
\si \ = \ \int_0^1 dt \ \tr \lp
          \prt_t g_t g_t^{-1} \, dg_t g_t^{-1} \, dg_t g_t^{-1} \rp,
\eeq
with the interpolating group element $g_t$ parametrized such that for
$t \in  [0,1]$
\be g_0 \ = \ 1, \cem g_1 \ = \ g. \eeq

This shows that the gauge transformation of the Chern-Simons form, which
is a 3-form in superspace, is given as the exterior derivative
of a 2-form,
\be \cq \lp {}^g \ca \rp - \cq \lp \ca \rp \ = \ d \Dt(g, \ca), \eeq
with $\Dt = \chi-\si$.

The discussion so far was quite general and valid for some generic gauge
potential. It does not only apply to the Yang-Mills case but to
gravitational Chern-Simons forms as well.

\indent

\subsection{The Chern-Simons superfield}

\indent

We specialize here to the Yang-Mills case, \ie we shall now take into account
the covariant constraints on the fieldstrength, which define supersymmetric
Yang-Mills theory.

It is the purpose of the present subsection to elucidate the relation
between the unconstrained prepotential, which arises
in the constrained 3-form
geometry, and the Chern-Simons superfield.
Moreover, based on this observation and
on the preceeding subsections we present a geometric construction of the
explicit form of the Yang-Mills Chern-Simons superfield in terms of the
unconstrained prepotential of supersymmetric Yang-Mills theory.

In this construction of the Chern-Simons superfield we will combine the
knowledge acquired in the discussion of the 3-form gauge potential
with the special features of Yang-Mills theory in superspace.

Recall that the Chern-Simons superfield $\Om^{\ym}$
is identified in the relations
\bea
\tr \lp \cw_\da \cw^\da \rp &=& \f12 \pp \Om^{\ym}, \\
\tr \lp \cw^\al \cw_\al \rp &=& \f12 \qq \Om^{\ym}.
\ena
The appearance of one and the same superfield under the projectors reflects
the fact that the gaugino superfields $\cw_\undal$ are not only subject
to the chirality constraints (\ref{ymcon1}) but
satisfy the additional condition(\ref{ymcon2}).
It is for this
reason that the Chern-Simons form can be so neatly embedded in the
geometry of the 3-form.

As explained in section {\bf 3} the terms on the left hand side are located
in the superspace 4-form
\be \Si^{\ym} \ = \tr (\cf \cf). \eeq
Of course, the constraints on the Yang-Mills fieldstrength induces special
properties on the 4-form coefficients, in particular
\be \Si_{\undel \, \undgm \, \undbt \, A}{}^{\ym} \ = \ 0, \eeq
which is just the same tensorial structure as the constraints on the
fieldstrength of the 3-form gauge potential. As a consequence the
Chern-Simons geometry can be regarded as a special case of that
of the 3-form gauge potential. Keeping in mind this fact we obtain
\bea
{\Si_{\dt \gm \ ba}}^{\ym} & = &
\f{1}{2} (\si_{ba} \eps)_{\dt \gm} S^{\ym}, \\
{{\Si^{\dd \dg}}_{\ ba}}{}^{\ym} & = &
\f{1}{2} (\bar{\si}_{ba} \eps)^{\dd {\dg}} T^{\ym},
\ena
with
\bea
S^{\ym} &=& - 8 \, \tr (\cwb_\da \cwb^\da),
\\
T^{\ym} &=& - 8 \, \tr (\cw^\al \cw_\al).
\ena

These facts imply the existence and provide a method for the explicit
construction of the Chern-Simons superfield:
comparison of these equations with those obtained earlier in the 3-form
geometry clearly suggests that
the Chern-Simons superfield $\Om^{\ym}$ will be the analogue of the unconstrained
prepotential superfield $\Om$ of the 3-form. In order to establish this
correspondence in full detail we translate the procedure
developped in the case of the 3-form geometry
to the Chern-Simons form
which, as a 3-form in superspace, has the decomposition
\be
\cq \ = \ \f{1}{3!} dz^K dz^L dz^M \cq_{MLK}
        \ = \ \f{1}{3!} E^A E^B E^C \cq_{CBA},
\eeq
with
\be
\f{1}{3!} E^A E^B E^C \cq_{CBA} (\ca) \ = \ \f{1}{3!} E^A E^B E^C \,
      \tr \lp 3 \ca_C \cf_{BA} - 2 \ca_C \ca_B \ca_A \rp.
\eeq

In order to extract the explicit form of the Chern-Simons superfield
we shall now exploit the equation
\be \tr (\cf \cf) \ = \ d \cq(\ca). \eeq

In the 3-form geometry we know unambigiously the exact location of
the prepotential in superspace geometry. Since we have identified
Chern-Simons as a special case of the 3-form, it is now rather
straightforward to identify the Chern-Simons superfield following
the same strategy.

To this end we recall that the prepotential was identified after certain
field dependent redefinitions which had the form of a gauge transformation,
simplifying considerably the form of the potentials. For instance,
the new potentials had the property
\be \Om_{\gm \bt A} \ = \ 0, \cem \Om^{\dg \db}{}_A \ = \ 0, \eeq
Note, en passant, that these redefinitions are not compulsory for the
identification of the unconstrained prepotantial. They make, however,
the derivation a good deal more transparent.

Can these features be reproduced in the Chern-Simons framework?
To answer this question we bring in the
particularity of Yang-Mills in superspace,
namely the existence of different types of gauge potentials corresponding
to the different possible types of gauge resp. pregauge transformations
as described in the previous subsection. These gauge potentials are
superspace one-forms denoted by $\ca$, $\vp$ and $\vpb$ with gauge
transformations parametrized in terms of real, chiral and antichiral
superfields, respectively. Moreover, the chiral and antichiral sectors
are related by a redefinition which has the form of a gauge transformation
involving the prepotential superfield $\Upsilon$:
\be \vp \ = \ \Upsilon^{-1} \vpb \Upsilon
- \Upsilon^{-1} d \Upsilon \ = \ {}^\Upsilon \vpb. \eeq
Writing the superspace Chern-Simons form in terms of $\vp$ shows
immediately that
\be \cq^{\dg \db}{}_A (\vp) \ = \ 0, \eeq
due to $\vp^\da=0$, but
\be \cq_{\gm \bt A} (\vp) \ \neq \ 0. \eeq
Of course, in the antichiral basis, things are just the other way round,
there we have
\be \cq_{\gm \bt A} (\vpb) \ = \ 0. \eeq
On the other hand, due to the relation between $\vp$ and $\vpb$
and the transformation law of the Chern-Simons form given above
we have
\be \cq (\vp) - \cq (\vpb) \ = \ d \Dt (\Upsilon, \vpb), \eeq
where now the group element is replaced by the prepotential superfield
$\Upsilon$. In some more detail, in $\Dt = \chi - \si$, we have
\be \chi \ = \ \chi (0, \vpb, Y) \ = \ \tr (\vpb \, Y), \eeq
where
\be Y \ = \ d \Upsilon \, \Upsilon^{-1} \ = \ E^A Y_A, \eeq
has zero fieldstrength
\be dY + YY \ = \ 0. \eeq
As a 2-form in superspace its coefficients, identified in
\be \chi \ = \ \f12 E^A E^B \chi_{BA}, \eeq
are given as
\be \chi_{BA} \ = \ \tr \lp Y_B \, \vpb_A - (-)^{ab} Y_B \, \vpb_A \rp. \eeq
For the interpolating prepotential $\Upsilon_t$ we define
\be Y_t \ = \ d\Upsilon_t \, \Upsilon_t^{-1} \eeq
to write accordingly
\be \si \ = \ \f12 E^A E^B \si_{BA}, \eeq
with
\be \si_{BA} \ = \
\int_0^1 dt \, \tr \lp \prt_t \Upsilon_t
  \, \Upsilon_t^{-1} (Y_{t \, B}, Y_{t \, A}) \rp.
\eeq
Consider now
\be
\cq_{\gm \bt A} (\vp) \ = \ \cd_A \Dt_{\gm \bt}
  + \oint_{\gm \bt} \lp \cd_\gm \Dt_{\bt A}
  - (-)^a T_{\gm A}{}^F \Dt_{F \bt} \rp.
\eeq
We perform next a redefinition
\be
\hcq \ := \ \cq(\vp) - d \La,
\eeq
where we determine the 2-form $\La$ in terms of the coefficients of
the 2-form $\Dt$ such that
\be \hcq^{\dg \db}{}_A \ = \ 0, \eeq
and maintain, at the same time,
\be \hcq_{\gm \bt A} \ = \ 0. \eeq
This is achieved with the following identification:
\be \La_{\bt A} \ = \ \Dt_{\bt A}, \cem
    \La^\db {}_a \ = \ -\i2 \cd^\db \Dt_a, \cem
    \La_{\db \da} \ = \ 0,
\eeq
and, for later convenience, we put also
\be \La_{ba} \ = \ \i2 \lp \cd_b \Dt_a - \cd_a \Dt_b \rp. \eeq
Here $\Dt_a$ is identified using spinorial notation such that
\be \Dt_\gm {}^\db \ = \ - \i2 T_\gm{} ^{\db \, a} \Dt_a. \eeq
We have, of course, to perform this redefinition on all the other
coefficients, in particular we obtain,
\be
\hcq_\gm{}^\db{}_a \ = \ \cq_\gm{}^\db{}_a (\vp) - \cd^\db \Xi_{\gm a}.
\eeq

In the derivation of this equation one uses the anticommutation relation
of spinorial derivatives und suitable supergravity Bianchi identities
together with the definition
\be \Xi_{\gm a} \ = \ \Dt_{\gm a} + \i2 \cd_\gm \Dt_a \eeq
We parametrize
\be
\hcq_\gm{}^\db{}_a \ = \
       T_\gm{}^\db{}_a \Om^{\ym} + T_\gm{}^{\db \, b} \hcq_{[ba]}{}^{\ym},
\eeq
where we can now identify the explicit form of the Chern-Simons
superfield
\be \Om^{\ym} \ = \  \cq (\vp) - \f{i}{16} \cd^\da      \Xi^\al{}_{\al \da}. \eeq
The first term is obtained from the spinorial contraction of
\be
\cq_\gm{}^\db{}_a (\vp) \ = \ \tr \lp \vp_\gm \cf^\db{}_a(\vp) \rp
   \ = \ -i (\sib_a \eps)^\db{}_\bt \, \tr \lp \vp_\gm \cw^\bt (\vp) \rp
\eeq
\ie
\be
\cq(\vp) \ = \ \f{i}{16} \cq^{\al \da}{}_{\al \da} (\vp)
   \ = \ - \f{1}{4} \tr \lp \vp^\al \cw_\al (\vp) \rp.
\eeq
It remains to read off the explicit form of the
second term from the definitions
above.

In closing we note that a more symmetrical form of the Chern-Simons
superfield may be obtained in exploiting the relation
\be
\cq_\gm{}^\db{}_a (\vp) - \cq_\gm{}^\db{}_a (\vpb) \ = \
       \cd_\gm \Xi^\db{}_a + \cd^\db \Xi_{\gm a}
      + T_\gm{}^{\db \, b} \lp \Dt_{ba} + \i2 (\cd_b \Dt_a - \cd_a \Dt_b) \rp,
\eeq
with
\be \Xi^\db{}_a \ = \ \Dt^\db{}_a + \i2 \cd^\db \Dt_a. \eeq

Observe that different appearances of the Chern-Simons superfields
should be equivalent modulo linear superfields. To establish the
explicit relation of the Chern-Simons superfield presented here
and that given by Ferrara et. al. in \cite{CFV87} is left as an exercise.


\sect{Riemann tensor and its squared - notations}

\indent

This appendix contains some notational information concerning
the Riemann tensor, its tensor decomposition, spinor notation
and curvature squared combinations.

\indent

\subsection{Vector notation}

\indent

The Riemann tensor
\be R_{dc \, , \, ba} \eeq
is separately antisymmetric in the indices $d,c$,
due to the fact that it is a differential 2-form,
and in the indices $b,a$, because it takes its
values in the Lie algebra of the Lorentz group.
As a consequence of these symmetry properties
there are 36 independent components.
But the Riemann tensor is also subject to the
Bianchi identities
\be R_{dc \, , \, ba} + R_{cb \, , \, da} +
    R_{bd \, , \, ca} \ = \ 0, \eeq
which constitute 16 independent equations, thus
reducing to 20 the number of components.
An equivalent way to write the Bianchi identities
is
\be R_{dc \, , \, ba} \ = \ R_{ba \, , \, dc}, \cem
    \vep^{dcba} \, R_{dc \, , \, ba} \ = \ 0. \eeq

The irreducible tensors contained in the Riemann tensor
are the curvature scalar, Ricci tensor and
the Weyl tensor. We define the contractions
\be \car_{ca} \ = \ {R^d}_{c \, , \, da}, \cem
    \car \ = \ {R^{dc}}_{\, , \, dc}. \eeq
The once contracted Riemann tensor gives rise to
a symmetric tensor of 10 components,
\be \car_{ba} \ = \ \car_{ab}, \eeq
whose traceless part
\be
\crt_{ba} \ = \ \car_{ba} - \f{1}{4} \eta_{ba} \car,
\eeq
is called called the Ricci tensor, while its
trace, $\car$, is called the curvature scalar.
The remaining 10 components are arranged in
the Weyl tensor
\be \cw_{dc \, , \, ba} \ = \ R_{dc \, , \, ba}
- \f12 \lp \eta_{db} \car_{ca} - \eta_{da} \car_{cb}
          -\eta_{cb} \car_{da} + \eta_{ca} \car_{db} \rp
+ \f{1}{6}
\lp \eta_{db} \eta_{ca} - \eta_{da} \eta_{cb} \rp \car, \eeq
which is completely traceless. Altogether we have defined
the decomposition of the Riemann tensor
\be R_{dc \, , \, ba} \ = \ \cw_{dc \, , \, ba}
+ \f12 \lp \eta_{db} \crt_{ca} - \eta_{da} \crt_{cb}
          -\eta_{cb} \crt_{da} + \eta_{ca} \crt_{db} \rp
+ \f{1}{12}
\lp \eta_{db} \eta_{ca} - \eta_{da} \eta_{cb} \rp \car, \eeq
in terms of the Weyl tensor, the Ricci tensor and the
curvature scalar. The Weyl tensor can be further
decomposed into self-dual and anti self-dual parts,
\bea
{\cw^\oplus}_{dc \, , \, ba} &=& \f12 \lp \cw_{dc \, , \, ba}
      + \f{i}{2} \vep_{dcfe} \, {\cw^{fe}}_{\, , \, ba} \rp, \\
{\cw^\ominus}_{dc \, , \, ba} &=& \f12 \lp \cw_{dc \, , \, ba}
      - \f{i}{2} \vep_{dcfe} \, {\cw^{fe}}_{\, , \, ba} \rp,
\ena
each consisting of 5 components.

\indent

\subsection{Spinor notation}

\indent

The basic definitions for the Riemann tensor are
\be
R_{dc \, , \, \bt \db \, \al \da} \ = \
\si^b_{\bt \db} \, \si^a_{\al \da} \, R_{dc \, , \, ba},
\eeq
with
\be
R_{dc \, , \, \bt \db \, \al \da} \ = \
2 \eps_{\db \da} \, R_{dc \,}{}_{\sym{\bt \al}}
-2 \eps_{\bt \al} \, R_{dc \,} \raisebox{.6ex}{${}_{\sym{\db \da}}$}.
\eeq

For the 2-form indices $d,c$ an analogous decomposition
holds and one defines altogether
\be
R_{\dt \dd \, \gm \dg \, , \,  \bt \db \, \al \da} \ = \
\si^d_{\dt \dd} \, \si^c_{\gm \dg} \,
\si^b_{\bt \db} \, \si^a_{\al \da} \, R_{dc \, , \, ba},
\eeq
with
\be
R_{\dt \dd \, \gm \dg \, , \, \bt \db \, \al \da} \ = \ 4 \eps_{\dd
\dg} \, \eps_{\db \da} \,
  \chi_{\sym{\dt \gm} \, \sym{\bt \al}}
-4 \eps_{\dd \dg} \, \eps_{\bt \al} \,
   \psi_{\sym{\dt \gm} \, \sym{\db \da}} \nn \\
-4 \eps_{\dt \gm} \, \eps_{\db \da} \,
   \psi_{\sym{\bt \al} \, \sym{\dd \dg}}
+4 \eps_{\dt \gm} \, \eps_{\bt \al} \,
   \chib_{\sym{\dd \dg} \, \sym{\db \da}},
\eeq
and
\bea
\chi_{\sym{\dt \gm} \, \sym{\bt \al}} &=&
  \chi_{\llsym{\dt \gm \bt \al}}
  + (\eps_{\dt \bt} \,\eps_{\gm \al}
    + \eps_{\dt \al} \, \eps_{\gm \bt}) \chi, \\
\chib_{\sym{\dd \dg} \, \sym{\db \da}} &=&
  \chib_{\llsym{\dd \dg \db \da}}
  + (\eps_{\dd \db} \, \eps_{\dg \da}
    + \eps_{\dd \da} \, \eps_{\dg \db}) \chi.
\ena
The relation of these spinor coefficients with curvature scalar,
Ricci and Weyl tensor is easily established. For the curvature
scalar one has
\be \chi \ = \ \f{1}{24} \car. \eeq
For the Ricci tensor we define
\be \car_{\bt \db \, \al \da} \ = \
\si^b_{\bt \db} \, \si^a_{\al \da} \, \car_{ba}, \eeq
and the like for $\crt_{ba}$. The identification is then
\bea
\car_{\bt \db \, \al \da} &=&
  -4\psi_{\sym{\bt \al} \, \sym{\db \da}}
  -\f12 \eps_{\bt \al} \, \eps_{\db \da} \, \car, \\
\crt_{\bt \db \, \al \da} &=&
  -4\psi_{\sym{\bt \al} \, \sym{\db \da}}.
\ena

For the Weyl tensor we define
\be
\cw_{\dt \dd \, \gm \dg \, , \, \bt \db \, \al \da} \ = \
\si^d_{\dt \dd} \, \si^c_{\gm \dg} \,
\si^b_{\bt \db} \, \si^a_{\al \da} \, \cw_{dc \, , \, ba},
\eeq
and the same for the self-dual and anti self-dual parts.
For the Weyl tensor itself on has then
\be
\cw_{\dt \dd \, \gm \dg \, , \, \bt \db \, \al \da} \ = \
4 \eps_{\dd \dg} \, \eps_{\db \da} \,
  \chi_{\llsym{\dt \gm \bt \al}}
+4 \eps_{\dt \gm} \, \eps_{\bt \al} \,
   \chib_{\llsym{\dd \dg \db \da}},
\eeq
whereas for the self-dual and anti self-dual components the
corresponding relations are
\bea
{\cw^\oplus}_{\dt \dd \, \gm \dg \, , \, \bt \db \, \al \da} &=&
4 \eps_{\dd \dg} \, \eps_{\db \da} \,
  \chi_{\llsym{\dt \gm \bt \al}}, \\
{\cw^\ominus}_{\dt \dd \, \gm \dg \, , \, \bt \db \, \al \da} &=&
4 \eps_{\dt \gm} \, \eps_{\bt \al} \,
   \chib_{\llsym{\dd \dg \db \da}},
\ena

\indent

\subsection{Curvature-squared combinations}

\indent

Frequently occuring curvature-squared combinations are
\be
\vep^{dcba} \, R_{dc \, , \, f}{}^e \, R_{ba \, , \, e}{}^f \ = \
2i \, \cw^\oplus{}^{dc \, , \, ba} \, \cw^\oplus{}_{dc \, , \, ba}
-2i \, \cw^\ominus{}^{dc \, , \, ba} \, \cw^\ominus{}_{dc \, , \, ba},
\eeq
\be
\vep^{dcba} \vep^{hgfe} \, R_{hg \, , \, dc} \, R_{fe \, , \, ba} \ = \
-4 \cw^{dc \, , \, ba} \, \cw_{dc \, , \, ba}
+8 \crt^{ba} \, \crt_{ba} - \f{2}{3} \car \, \car,
\eeq
and
\be
R^{dc \, , \, ba} \, R_{dc \, , \, ba} \ = \
\cw^{dc \, , \, ba} \, \cw_{dc \, , \, ba}
+2 \crt^{ba} \, \crt_{ba} + \f{1}{6} \car \, \car.
\eeq
The first two correspond to 4-form coefficients, while the last
one appears in conformal gravity theories. These expressions can be
rewritten in terms of the spinor decomposition using the relations
\bea
\psi^{\sym{\bt \al} \, \sym{\db \da}} \, \psi_{\sym{\bt \al} \, \sym{\db \da}}
&=& \f{1}{4} \crt^{ba} \, \crt_{ba}, \\
\chi \raisebox{-1.2ex}{${}^{\llsym{\dt \gm \bt \al}}$} \,
        \chi_{\llsym{\dt \gm \bt \al}}
&=& \f{1}{4} \cw^{\oplus \, dc \, , \, ba} \, {\cw^\oplus}_{dc \, , \, ba}, \\
\chib_{\llsym{\dd \dg \db \da}} \,
    \chib \raisebox{-1.2ex}{${}^{\llsym{\dd \dg \db \da}}$}
&=& \f{1}{4} \cw^{\ominus \, dc \, , \, ba} \, {\cw^\ominus}_{dc \, , \, ba}.
\ena

In the language of differential forms the Riemann tensor appears as
coefficient of the curvature 2-form
\be R_b{}^a \ = \ \f{1}{2} e^c e^d R_{dc \, , \, b}{}^a, \eeq
which takes its values in the Lie algebra of the Lorentz group
and has the standard decomposition with respect to $SL(2,C)$,
\be
R_\bt{}^\al \ = \ \f{1}{2} e^c e^d R_{dc \, \bt}{}^\al, \cem
R^\db{}_\da \ = \ \f{1}{2} e^c e^d R_{dc \, }{}^\db{}_\da.
\eeq
Correspondingly, for the two curvature-squared combinations one
obtains
\bea
\f{1}{2} R_b{}^a R_a{}^b &=&
R_\bt{}^\al R_\al{}^\bt + R^\db{}_\da R^\da{}_\db, \\
-\f{i}{4} \vep^{dcba} R_{dc} R_{ba} &=&
R_\bt{}^\al R_\al{}^\bt - R^\db{}_\da R^\da{}_\db,
\ena
We use the notations
\be
\Psi^{(+)} \ = \ R_\bt{}^\al R_\al{}^\bt, \cem
\Psi^{(-)} \ = \ R^\db{}_\da R^\da{}_\db.
\eeq
for these 4-forms, their coefficients are defined as
\be
\Psi^{(\pm)}
\ = \ \f{1}{4!} e^a e^b e^c e^d \Psi^{(\pm)}{}_{dcba}
\ = \ - \f{1}{4!} \; \mbox{\bf v} \; \vep^{dcba} \Psi^{(\pm)}{}_{dcba},
\eeq
with $\mbox{\bf v}$ the fundamental 4-form
\be
\mbox{\bf v} \ = \ \f{1}{4!} e^a e^b e^c e^d \vep_{dcba}
    \ = \ e^0 e^1 e^2 e^3.
\eeq
More explictly, using
\bea
\Psi^{(+)} &=&
-\f{1}{4} \; \mbox{\bf v} \;
\vep^{dcba} R_{dc \, \eps}{}^\vp R_{ba \, \vp}{}^\eps, \\
\Psi^{(-)} &=&
-\f{1}{4} \; \mbox{\bf v} \;
\vep^{dcba} R_{dc \,}{}^\dep{}_\dv  R_{ba \,}{}^\dv{}_\dep,
\ena
and taking into account the spinor decomposition gives rise to
\bea
\f{1}{4} \vep^{dcba} R_{dc \, \eps}{}^\vp R_{ba \, \vp}{}^\eps
&=& +i \chi \raisebox{-1.2ex}{${}^{\llsym{\dt \gm \bt \al}}$} \,
        \chi_{\llsym{\dt \gm \bt \al}}
    -i \psi^{\sym{\bt \al} \, \sym{\db \da}} \,
       \psi_{\sym{\bt \al} \, \sym{\db \da}}
    +12i \chi^2, \\
\f{1}{4} \vep^{dcba} R_{dc \,}{}^\dep{}_\dv  R_{ba \,}{}^\dv{}_\dep
&=& -i \chib_{\llsym{\dd \dg \db \da}} \,
    \chib \raisebox{-1.2ex}{${}^{\llsym{\dd \dg \db \da}}$}
    +i \psi^{\sym{\bt \al} \, \sym{\db \da}} \,
       \psi_{\sym{\bt \al} \, \sym{\db \da}}
    -12i \chi^2.
\ena
Another useful relation is (Gauss-Bonnet)
\[ \f{i}{4} \, \vep^{dcba}
  \lp R_{dc \, \eps}{}^\vp R_{ba \, \vp}{}^\eps
      - R_{dc \,}{}^\dep{}_\dv  R_{ba \,}{}^\dv{}_\dep \rp
   \ = \ \f{1}{16} \, \vep^{dcba} \, R_{dc \, , \, d' c'}
           \; R_{ba \, , \, b'a'} \; \vep^{d'c'b'a'}  \]
\[ \hs{-2.3cm}\ = \ - \chi \raisebox{-1.2ex}{${}^{\llsym{\dt \gm \bt \al}}$} \,
        \chi_{\llsym{\dt \gm \bt \al}}
     - \chib_{\llsym{\dd \dg \db \da}} \,
    \chib \raisebox{-1.2ex}{${}^{\llsym{\dd \dg \db \da}}$}
     + 2 \; \psi^{\sym{\bt \al} \, \sym{\db \da}} \,
       \psi_{\sym{\bt \al} \, \sym{\db \da}}
     - 24 \; \chi^2 \]
\be \hs{-3.8cm} \ = \ \f{1}{16}
         \lp - 4 \; \cw^{dc \, , \, ba} \, \cw_{dc \, , \, ba}
      + 8 \; \crt^{ba} \, \crt_{ba} + \f{1}{6} \; \car \car \rp. \eeq


\sect{Supersymmetry and curvature-squared terms}

\indent

Supersymmetric curvature-squared terms arise naturally in
conformal supergravity \cite{FZ78}, \cite{KTN78} and in the study
of supersymmetric extensions of anomalies and topological invariants
\cite{TvN79}, \cite{CD79}, as well as in attempts
at new mechanism of supersymmetry breaking
\cite{HOW96a}, \cite{HOW96b}, \cite{HOW96c}. So far these structures have
been investigated mostly \cite{The86}, \cite{RWZ90}
in the absence of supersymmetric matter and gauge couplings.
As is well known, the general supergravity-matter system
reveals interesting relations between \ka phase transformations,
super-Weyl rescalings and the normalization of the usual Einstein
curvature scalar action, which have a concise geometrical
interpretation in the framework of $U_K(1)$ superspace
(which is also suitable for the description of variant
supergravity theories \cite{MM86}).

It should therefore be useful to discuss the supersymmetric extensions
of curvature-squared terms in a geometrical framework which allows
to take care of the matter sector as well. This will be achieved in this
appendix in working directly in generic $U_K(1)$ superspace, which is
slightly more general: the supergravity-matter system is obtained from it
if one replaces the $U_K(1)$-prepotential with the \ka potential superfield.

But the generic $U_K(1)$ superspace, giving rise to so-called {\em chirally
extended supergravity} \cite{dWvN78}, is also interesting in its own case -
curvature-squared terms in this context have been investigated in
\cite{Led97}.

As is well known, curvature-squared compenent field expressions are
identified in the highest components of the products of the basic
supergravity superfields $\rd R$ and $G^a G_a$ as well as the squared
of the Weyl superfield,
$W\raisebox{-.8ex}{${}^{\lsym{\gm \bt \al}}$} W_{\lsym{\gm \bt \al}}$,
and its complex conjugate. In the presence of the $U_K(1)$ factor in
the structure group, additional terms must be considered, arising from
the square of the superfield $X_\al$ and its complex conjugate.
Different curvature-squared combinations are then obtained from
appropriately chosen linear combinations of these basic superfield products.
In principal, the highest components of these superfield products can be
obtained through a explicit, though somewhat painful, calculation.

In this appendix we will take advantage of the geometric description
in superspace, in particular the covariant decomposition in terms
of the 3-form geometry, to present a more systematic
construction of supersymmetric completions of curvature-squared terms.

We shall start from the Gauss-Bonnet combination
of curvature-squared terms. In our notations it appears
in the purely vectorial coefficient of the superspace
4-form (see (\ref{eq:7.5}))
\be \Psi^{(+)} - \Psi^{(-)}, \eeq
which, in some more detail, is given as (see also appendix {\bf B)} for
the relation
between vector and spinor notation of the curvature-squared terms)
\bea
\lefteqn{\f{i}{24} \, \vep^{dcba}\Psi^{(+)}{}_{dcba}
-\f{i}{24} \, \vep^{dcba}\Psi^{(-)}{}_{dcba}
\ = } \nn \\
&&
- \chi \raisebox{-1.6ex}{${}^{\llsym{\dt \gm \bt \al}}$} \,
        \chi_{\llsym{\dt \gm \bt \al}}
- \chib_{\llsym{\dd \dg \db \da}} \,
    \chib \raisebox{-1.2ex}{${}^{\llsym{\dd \dg \db \da}}$}
+ 2 \, \psi^{\sym{\bt \al} \, \sym{\db \da}} \,
       \psi_{\sym{\bt \al} \, \sym{\db \da}}
- 24 \, \chi^2.
\label{gb1}
\ena
On the other hand, from the covariant decomposition
established in the main text we have
\be
\vep^{dcba} \Si^{(\pm)}{}_{dcba} \ = \ \vep^{dcba}
\lp \Psi^{(\pm)}{}_{dcba} - 4 \, \cd_d M^{(\pm)}{}_{cba}
        -6 \, T_{dc}{}^\undph M^{(\pm)}{}_{\undph ba} \rp,
\eeq
where the expression on the left is related to the
3-form structure by (see \ref{eq:7.18})
\be
\f{8i}{3} \, \vep^{dcba} \Si^{(\pm)}{}_{dcba} \ = \
\lp \cd^2 -24 \rd \rp T^{(\pm)} - \lp \cdb^2 - 24 R \rp S^{(\pm)}.
\label{C.1}
\eeq
This shows how the supersymmetric completion of the
Gauss-Bonnet combination is identified
in the leading term of a supersymmetric chiral
density construction. Using the explicit form of the
superfields $T^{(\pm)}$ and $S^{(\pm)}$
as defined in eqs.(\ref{S+}) - (\ref{T-}) one obtains
\bea
\lefteqn{
8 \ \chi \raisebox{-1.6ex}{${}^{\llsym{\dt \gm \bt \al}}$} \,
        \chi_{\llsym{\dt \gm \bt \al}}
+8 \ \chib_{\llsym{\dd \dg \db \da}} \,
    \chib \raisebox{-1.2ex}{${}^{\llsym{\dd \dg \db \da}}$}
-16 \ \psi^{\sym{\bt \al} \, \sym{\db \da}} \,
       \psi_{\sym{\bt \al} \, \sym{\db \da}}
+ 192 \ \chi^2
+ \mbox{Div}^{(+)} - \mbox{Div}^{(-)} \ = } \nn \\[2mm]
&=& -4 \lp \cd^2 -24 \rd \rp \lp
        W\raisebox{-.8ex}{${}^{\lsym{\gm \bt \al}}$}
          W_{\lsym{\gm \bt \al}} + \f{1}{6} \, X^\al X_\al\rp
    -4 \lp \cdb^2 - 24 R \rp \lp W_{\lsym{\dg \db \da}}
           W \raisebox{-.8ex}{${}^{\lsym{\dg \db \da}}$}
+ \f{1}{6} \, \bX_\da \bX^\da \rp
\nn \\[3mm]
&& - \, \Box^+ \lp  G^a G_a + 2 \, \rd R \rp.
\label{gb2}
\ena
Here we subsumed a number of supercovariant derivative
and nonlinear terms
under the symbols
\be
\mbox{Div}^{(\pm)} \ = \
+ \f{1}{8} \, \Box^- \lp \mu^{(\pm)} +8 \, \rd R + \f{5}{2} \, G^a G_a \rp
+ \f{i}{3} \, \vep^{dcba} \lp 4 \, \cd_d M^{(\pm)}{}_{cba}
      + 6 \, T_{dc}{}^\undph \, M^{(\pm)}{}_{\undph ba} \rp,
\label{c6}
\eeq
and used the notations
\be
\Box^\pm \ = \
\lp \cd^2 -24 \rd \rp \lp \cdb^2 - 8 R \rp
\pm \lp \cdb^2 - 24 R \rp \lp \cd^2 -8 \rd \rp,
\eeq
for the generalized fourth order covariant derivative operators,
understood to act on superfields of vanishing $U_K(1)$ weight.
In particular, the combination $\Box^-$ amounts to a generalized
covariant divergence term. In some more detail, its action on
a generic superfield $X$ of vanishing $U_K(1)$ weight may be written as
\be
\Box^- X \ = \ - 4i \; \cd^{\al \da}
      \lp \left[ \cd_\al,\cd_\da \right] - 4 \; G_{\al \da} \rp X
      + 32 \; S^\al \cd_\al X + 32 \; S_\da \cd^\da X.
\eeq
Using this equation together with the explicit form of the
coefficients of $M^{(\pm)}$ in (\ref{c6}) one verifies directly
that $\mu^{(\pm)}$ drop out in the expression for
$\mbox{Div}^{(\pm)}$. This property allows to cast $\mbox{Div}^{(\pm)}$
in the form
\be
\mbox{Div}^{(\pm)} \ = \
 \f{i}{3} \, \vep^{dcba} \lp 4 \, \cd_d \widehat{M}^{(\pm)}{}_{cba}
      + 6 \, T_{dc}{}^\undph \widehat{M}^{(\pm)}{}_{\undph ba} \rp,
\eeq
with $\widehat{M}^{(\pm)}$ defined as $M^{(\pm)}$ evaluated at the
special values $\mu^{(\pm)} = - 8 \rd R - \f{5}{2} G^a G_a$, \ie
\be
\widehat{M}^{(\pm)} \ = \
M^{(\pm)} \lp \mu^{(\pm)} = - 8 \, \rd R - \f{5}{2} \, G^a G_a \rp.
\eeq
In practical calculations it is sometimes more convenient to
use this same expression in spinor notation :
\be
\mbox{Div}^{(\pm)} \ = \
-4i \, \cd^{\al \da} \widehat{M}^{(\pm)}{}_{\al \da}
+8 \, T \raisebox{-.6ex}{${}^{\sym{\bt \al}}$} \, \raisebox{+.5ex}{${}^\undgm$}
\ \widehat{M}^{(\pm)}\raisebox{-.2ex}{${}_\undgm$} \, {}_{\sym{\bt \al}}
-8 \, T \raisebox{-.6ex}{${}^{\sym{\db \da}}$} \, \raisebox{+.5ex}{${}^\undgm$}
\ \widehat{M}^{(\pm)}\raisebox{-.4ex}{${}_\undgm$} \, {}_{\sym{\db \da}}.
\eeq

Turning back to eq.(\ref{gb2}), we observe that it
identifies the Gauss-Bonnet combination in the
expansion of the basic superfields of $U_K(1)$
superspace\footnote{On the other hand, the description
of the Gauss-Bonnet combination in the traditional superspace
is obtained from this expression by simply turning off the $U_K(1)$
sector, \ie taking $X_\undal$ to be zero.}, that
is a particular combination of the $D$-term of the superfield
$G^a G_a + 2 \, \rd R$ and the $F$-terms of the chiral superfields
$W^2$ and $X^2$ and their conjugates. The term
$\mbox{Div}^{(+)} - \mbox{Div}^{(-)}$, which arises naturally
from the geometric construction, is necessary for the supersymmetric
completion once the projection to component fields of (\ref{gb2})
is employed in the chiral density construction as explained in
section $\bf{2.3}$ of the main text.

Note also that the
particular combination of $W^2$ and $X^2$ occurs in (\ref{gb2})
in order to ensure the absence of the square of the $U_K(1)$
field strength.

Instead of reading eq.(\ref{gb2}) as an expression for the leading
term of the supersymmetric Gauss-Bonnet combination we shall now
turn the argument the other way round and use the same equation to
determine the highest superfield component of $G^a G_a + 2 \, \rd R$.
To do so, we make use of the explicit expressions for the
$F$-terms of the chiral superfields
$W^2$ and $X^2$ and their conjugates.

First of all, the square of the Weyl {\em spinors} provide
the leading terms in the supersymetric
completion of the square of the Weyl {\em tensor}, as can be seen
explicitly from the equations
\bea
\lp \cd^2 -24 \rd \rp W\raisebox{-.8ex}{${}^{\lsym{\gm \bt \al}}$}
          W_{\lsym{\gm \bt \al}} &=&
-2 \, \chi \raisebox{-1.5ex}{${}^{\llsym{\dt \gm \bt \al}}$} \,
        \chi_{\llsym{\dt \gm \bt \al}}
-\f{8}{3} \ f\raisebox{-.6ex}{${}^{\sym{\bt \al}}$} f_{\sym{\bt \al}}
+ 16 \, \rd \, W\raisebox{-.8ex}{${}^{\lsym{\gm \bt \al}}$}
         W_{\lsym{\gm \bt \al}} \nn \\
&&+8i \ W \raisebox{-.8ex}{${}^{\lsym{\gm \bt \al}}$}
\lp \cd_\al{}^\da -i G_\al{}^\da \rp
T \raisebox{-.3ex}{${}_{\sym{\gm \bt}}$} \, \raisebox{-.5ex}{${}_\da$},
\ena
and
\bea
\lp \cdb^2 - 24 R \rp W_{\lsym{\dg \db \da}}
           W \raisebox{-.8ex}{${}^{\lsym{\dg \db \da}}$} &=&
-2 \, \chib_{\llsym{\dd \dg \db \da}} \,
    \chib \raisebox{-1.5ex}{${}^{\llsym{\dd \dg \db \da}}$}
-\f{8}{3} \ f_{\sym{\db \da}} f\raisebox{-.6ex}{${}^{\sym{\db \da}}$}
+16 \, R \, W_{\lsym{\dg \db \da}}
      W \raisebox{-.8ex}{${}^{\lsym{\dg \db \da}}$} \nn \\
&&+8i \ W \raisebox{-.8ex}{${}^{\lsym{\dg \db \da}}$}
\lp \cd^\al{}_\da +i G^\al{}_\da \rp
T_{\sym{\dg \db}} \, \raisebox{-.5ex}{${}_\al$}.
\ena
Note the presence of the squares of the $U_K(1)$ field strength,
\ie the terms $f\raisebox{-.6ex}{${}^{\sym{\bt \al}}$} f_{\sym{\bt \al}}$
and $f_{\sym{\db \da}} f\raisebox{-.6ex}{${}^{\sym{\db \da}}$}$
in these equations.

In turn, the $F$-term of $X^2$ and its conjugate are given as
\be
\lp \cd^2 -24 \rd \rp X^\al X_\al \ = \
    - 32 \ \hat{f}\raisebox{-.6ex}{${}^{\sym{\bt \al}}$}
         \hat{f}_{\sym{\bt \al}}
    + 8i \ X^\al \cd_{\al \da} \bX^\da
    - \lp \cd^\al X_\al \rp^2,
\eeq
\be
\lp \cdb^2 - 24 R \rp \bX_\da \bX^\da \ = \
    - 32 \ \hat{f}_{\sym{\db \da}}
         \hat{f}\raisebox{-.6ex}{${}^{\sym{\db \da}}$}
    + 8i \ \bX^\da \cd_{\al \da} X^\al
    - \lp \cdb_\da X^\da \rp^2.
\eeq
Notations which intervene here are
\be
\hat{f}_{\sym{\bt \al}} \ = \
      f_{\sym{\bt \al}} - \f{3i}{2} \, g_{\sym{\bt \al}}, \cem
\hat{f}_{\sym{\db \da}} \ = \
      f_{\sym{\db \da}} - \f{3i}{2} \, g_{\sym{\db \da}},
\eeq
\be
g_{\sym{\bt \al}} \ = \
\f{1}{4} \lp \cd_\bt{}^\dv G_{\al \dv}
           + \cd_\al{}^\dv G_{\bt \dv} \rp, \cem
g_{\sym{\db \da}} \ = \
- \f{1}{4} \lp \cd^\vp{}_\db G_{\vp \da}
             + \cd^\vp{}_\da G_{\vp \db} \rp,
\eeq \\[1mm]
where $G_{ba} \ = \ \cd_b G_a - \cd_a G_b$ has the same standard spinorial
decomposition as $F_{ba}$, that is
\be G_{\bt \db \ \al \da} \ = \
  2 \, \eps_{\db \da} \, g_{\sym{\bt \al}}
 -2 \, \eps_{\bt \al} \, g_{\sym{\db \da}},
\eeq
and $ \hat{F}_{ba} = F_{ba}- \f{3i}{2} G_{ba}.$
Putting all this information together, one finally obtains
\bea
\lefteqn{\Box^+ \lp G^a G_a + 2 \, \rd R \rp \ =} \nn \\
&&
- 16 \, \psi^{\sym{\bt \al} \, \sym{\db \da}} \,
       \psi_{\sym{\bt \al} \, \sym{\db \da}} + 192 \, \chi^2
+ \lp \mbox{Div}^{(+)} - \mbox{Div}^{(-)} \rp
\nn \\
&& +32i \, W \raisebox{-.8ex}{${}^{\lsym{\gm \bt \al}}$}
      \lp \cd_\al{}^\da -i G_\al{}^\da \rp
      T \raisebox{-.3ex}{${}_{\sym{\gm \bt}}$} \, \raisebox{-.5ex}{${}_\da$}
   +64 \, \rd \, W\raisebox{-.8ex}{${}^{\lsym{\gm \bt \al}}$}
         W_{\lsym{\gm \bt \al}}
\nn \\
&& +32i \, W \raisebox{-.8ex}{${}^{\lsym{\dg \db \da}}$}
      \lp \cd^\al{}_\da +i G^\al{}_\da \rp
      T_{\sym{\dg \db}} \, \raisebox{-.5ex}{${}_\al$}
   +64 \, R \, W_{\lsym{\dg \db \da}}
      W \raisebox{-.8ex}{${}^{\lsym{\dg \db \da}}$}
\nn \\
&&
- \frac{32}{3} \lp f\raisebox{-.6ex}{${}^{\sym{\bt \al}}$} f_{\sym{\bt \al}}
+ 2 \hat{f}\raisebox{-.6ex}{${}^{\sym{\bt \al}}$} \hat{f}_{\sym{\bt \al}} \rp
- \frac{32}{3} \lp f_{\sym{\db \da}} f\raisebox{-.6ex}{${}^{\sym{\db \da}}$}
 + 2 \hat{f}_{\sym{\db \da}} \hat{f}\raisebox{-.6ex}{${}^{\sym{\db \da}}$} \rp
\nn \\[1mm]
&&
+ \frac{16i}{3} \ \bX^\da \cd_{\al \da} X^\al
+ \frac{16i}{3} \ X^\al \cd_{\al \da} \bX^\da
- \frac{4}{3} \, \lp \cd^\al X_\al \rp^2
\ena
One sees that the squares of the
Weyl tensor drop out and we are left with the combination
\be
4 \, \psi^{\sym{\bt \al} \, \sym{\db \da}} \,
       \psi_{\sym{\bt \al} \, \sym{\db \da}} - 48 \, \chi^2
\ = \ \crt^{ba} \, \crt_{ba} - \f{1}{12} \,  \car \car
\ = \ \car^{ba} \, \car_{ba} - \f{1}{3} \, \car \car.
\eeq
Recall from the previous discussion, that
$\mbox{Div}^{(+)} - \mbox{Div}^{(-)}$ hides a number of
derivative and nonlinear terms which are not very illuminating
for the present discussion. Their explicit form may be inferred
from the results presented in appendix {\bf D}, if desired.
Finally, we display the contribution arising from $\rd R$ alone:
\bea
\lefteqn{\Box^+ \rd R \ = \ + \f{2}{9} \, \lp \car + \cd^\al X_\al \rp^2
                            - \f{8}{3} \, \car \lp G^a G_a + 2 \rd R \rp}
\nn \\ [.5mm]
&&  + 16 \, \rd \, \cd^a \cd_a R + 16 \, R \, \cd^a \cd_a \rd
    - 64i \, G^a \lp \rd \cd_a R - R \, \cd_a \rd \rp
    + 8 \, \lp \cd^a G_a \rp^2
\nn \\ [.5mm]
&&- 8i \; \si^a_{\al \da} \lp \cd^\al \! R \; \cd_a \cdb^\da \rd
  + \cdb^\da \! \rd \, \cd_a \cd^\al R
         + 4i \, \cd^\al \! R \, G_a \cdb^\da \rd \rp
\nn \\ [.5mm]
&&
  - \f{8}{3} \lp G^a G_a + 8 \rd R \rp \cd^\al \! X_\al
   +8 \lp G^a G_a - 4 \, \rd R \rp ^2
\nn \\ [1mm]
&& -8 \, R \, \cd_\da \rd \, \cd^\da \rd -8 \, \rd \, \cd^\al R \, \cd_\al R
   -24 \, R \, \bX_\da \cd^\da \rd -24 \ \rd \, X^\al \cd_\al R
\ena

In conclusion, the formulas derived in this appendix provide the
starting point for a constructive procedure to describe the
supersymmetric completion of any combination of curvature-squared
terms by means of the generic chiral density construction.


\sect{The covariant decompositions \
      $\Psi^\Dt \ = \ \Si^\Dt + d \, M^\Dt$}

\indent

An important point in our investigation of
gravitational Chern-Simons forms was the covariant decomposition
$ \ \Psi^\Dt \ = \ \Si^\Dt + d \, M^\Dt \ $,
relating the curvature (resp. fieldstrength)-squared 4-form $\Psi^\Dt$
to the geometrical structure of the 3-form multiplet of supersymmetry
in the four cases
$\Dt \in \{ \, \raisebox{.3ex}{$\pl$} \, , \, \raisebox{.3ex}{$\mn$} \, , \,
\raisebox{.3ex}{$\uo$} \, , \, \raisebox{.3ex}{$\ym$} \, \} $.
As explained in section {\bf 3.1}.
the components of the 4-forms
$\Si^\Dt$ reflect the constraint structure of the geometry of
the 3-form multiplet in their tensor structure and the nonvanishing
components are expressed in terms of the basic covariant superfields
and their (covariant) derivatives. Moreover, the difference between
the original complete curvature-squared 4-form $\Psi^\Dt$ and
the constrained 4-form $\Si^\Dt$
can be cast in the form of an exterior superspace derivative of the
3-form $M^\Dt \, $,  which is expressed in terms of the basic covariant
superfields and their covariant derivatives as well. The resulting expressions,
which are obtained by an explicit calculation in each individual case,
are rather involved in particular in the gravitational
$\raisebox{.3ex}{$\pl$}$ - and $\raisebox{.3ex}{$\mn$}$ - sectors.
It seems therefore preferable to give a compendium of the corresponding
formulae in the four subsections of this appendix.

Although the covariant decomposition have been established by an explicit
calculation in each sector separately, there is a number of features
they have in common.

First of all, resuming the discussion after eq.(\ref{eq:7.6}),
the components of $\Si^\Dt$ reflect the 3-form constraints, \ie
\be
{\Si^\Dt}_{\undel \undgm \undbt A} \ = \ 0.
\eeq
Furthermore, given these restrictions, the Bianchi identities
$ \, d \Si^\Dt \ = \ 0 \, $ imply a number of consequences for
the remaining components. Most importantly it turns out that
all the components of $\Si^\Dt$ are completely described in
terms of two superfields $S^\Dt$ and $T^\Dt$, appearing in
\be
{\Si^\Dt}_{\dt \gm \ ba} \ = \
\f{1}{2} (\si_{ba} \eps)_{\dt \gm} \, S^\Dt,
\cem
\Si^{\Dt \, \dd \dg}{}_{\ ba} \ = \
\f{1}{2} (\bar{\si}_{ba} \eps)^{\dd {\dg}} \, T^\Dt,
\eeq
which are subject to the chirality conditions
\be \cd_\al S^\Dt \ = \ 0, \cem \cd^\da T^\Dt \ = \ 0. \eeq
Furthermore one finds
(cf. eqs.(\ref{eq:7.15}), (\ref{eq:7.16}) and (\ref{eq:7.18})).
\be
\Si^\Dt{}_{\ \dt \ cba} \ = \ - \f{1}{16} \, \si^d_{\dt \dd} \, \vep_{dcba}
                          \, \cd^\dd \! S^\Dt,
\cem
\Si^{\Dt \ \dd \ }{}_{cba} \ = \  + \f{1}{16} \bar{\si}^{d \, \dd \dt}
                          \, \vep_{dcba} \, \cd_\dt T^\Dt,
\eeq
and
\be 2i {\bf \Si}^\Dt \ = \ - \f{1}{32} \pp T^\Dt + \f{1}{32} \qq S^\Dt, \eeq
where the boldscript scalar superfields ${\bf \Si}^\Dt$ is defined as
\be {\Si^\Dt}_{dcba} \ = \ \vep_{dcba} \, {\bf \Si}^\Dt. \eeq
As already explained above the superfields $S^\Dt$ and $T^\Dt$ have
a different form in each sector, but once they are known (and they
will be given explicitly below), the 4-form $\Si^\Dt$ is
completely determined.

Recall also that the component ${\Si^\Dt}_{\dt \dg \; ba}$ is found to have
the tensor structure
\be
{\Si^\Dt}_{\dt \dg \ \bt \db \ \al \da} \ = \
\si^b_{\bt \db} \, \si^a_{\al \da} \, {\Si^\Dt}_{\dt \dg \; ba}
\ = \
4 \, \eps_{\gm \bt} \, \eps_{\dg \da} \, {\Si^\Dt}_{\al \db}
-4 \, \eps_{\gm \al} \, \eps_{\dg \db} \, {\Si^\Dt}_{\bt \da}.
\eeq
In other words this means that
$ \ \sib^{\dg \gm}_c \ {\Si^\Dt}_{\gm \dg \; ba} \ $ is completely
antisymmetric in the three vector indices $c$, $b$ and $a$. A closer
look at the decomposition
$ \ \Psi^\Dt \ = \ \Si^\Dt + d \, M^\Dt \ $
shows then that this component appears always in a particular linear
combination with ${M^\Dt}_{cba}$. More precisely, using the
decomposition
\be
{M^\Dt}_{\gm \dg \ \bt \db \ \al \da} \ = \
 \si^c_{\gm \dg} \, \si^b_{\bt \db} \, \si^a_{\al \da} \, {M^\Dt}_{cba}
\ = \
2i \eps_{\dg \db} \eps_{\gm \al} {M^\Dt}_{\bt \da}
-2i \eps_{\dg \da} \eps_{\gm \bt} {M^\Dt}_{\al \db} \; ,
\eeq
this combination is $ \ {M^\Dt}_{\al \da} + {\Si^\Dt}_{\al \da} \ $.
As a consequence, one has the freedom to redefine individually
$ \ {M^\Dt}_{\al \da} \ $ and $ \ {\Si^\Dt}_{\al \da} \ $, provided
their sum remains unchanged. Such special assignements are called
{\em conventional constraints} and one might, for instance absorb
$ \ {\Si^\Dt}_{\al \da} \ $ completely in a redefinition of
$ \ {M^\Dt}_{\al \da} \ $, such establishing the conventional constraint
mentioned in eq. (\ref{eq:7.13}).

In the remaining part of this preamble we give the definitions of the
tensor decompositions of the components of the 3-form $M^\Dt$ which are the
same in the four cases.
One has
\be {M^\Dt}_{\undgm \, \undbt \ \al \da} \ = \
    \si^a_{\al \da} \, {M^\Dt}_{\undgm \, \undbt \ a},
\eeq
and
\be \hs{-1.5cm}
{M^\Dt}_{\undgm \ \bt \db \ \al \da}  \ = \
 2\eps_{\db \da} {M^\Dt}_{\undgm}
\raisebox{.1ex}{${}_{ \ {\sym{\bt \al}}}$}
-2\eps_{\bt \al} {M^\Dt}_{\undgm}
\raisebox{.5ex}{${}_{ \ {\sym{\db \da}}}$} \ ,
\eeq
with
\bea \hs{-1cm}
{M^\Dt}\raisebox{-.2ex}{${}_{\gm}$}{}_{ \ {\sym{\bt \al}}} &=&
{M^\Dt}_{\lsym{\gm \bt \al}}
+ \eps_{\gm \bt} {M^\Dt}_{\al} + \eps_{\gm \al} {M^\Dt}_{\bt},
\\[1mm] \hs{-1cm}
{M^\Dt} \raisebox{-.2ex}{${}_{\dg}$}
\raisebox{.3ex}{${}_{ \ {\sym{\db \da}}}$} &=&
{M^\Dt} \raisebox{.ex}{${}_{\lsym{\dg \db \da}}$}
+ \eps_{\dg \db} {M^\Dt}_{\da} + \eps_{\dg \da} {M^\Dt}_{\db}.
\ena
In the following the explicit expressions
for these superfields in the different sectors will be given.

\indent

\subsection{$\Psi^\pl \ = \ \Si^\pl + d \, M^\pl$}

\indent

Although part of these results have already been exposed in section {\bf 3},
the complete set of expressions is displayed here.
For the components of the 3-form $M^\pl$ one obtains
\be \hs{-3.3cm}
{M^\pl}_{\undgm \, \undbt \, \undal} \ = \ 0, \cem
{M^{(+) \, \dg \db}}_a \ = \ 0, \\[1.3mm]
\eeq
\be \hs{-2.4cm}
{M^\pl}_{\gm \bt \ \al \da} \ = \ -8i \rd \lp
   \eps_{\gm \al} G_{\bt \da} + \eps_{\bt \al} G_{\gm \da} \rp, \\[2mm]
\eeq
\be
{M^\pl}_{\gm \db \ \al \da} \ = \
   -i  \eps_{\gm \al} \, \eps_{\db \da} \, \mu^\pl
   -\i2 \lp G_{\gm \db}\,  G_{\al \da} + G_{\al \db} \, G_{\gm \da} \rp,
\eeq
at dimension 3/2 and 2, whereas at dimension 5/2
the various irreducible components are given as
\bea
{M^\pl}\raisebox{-.2ex}{$_{\gm}$}
\raisebox{.2ex}{${}_{ \ {\sym{\db \da}}}$} &=& \f{1}{8} \sum_{\db \da}
\left( 16 \rd \cd_{\db} G_{\gm \da}
      + 4 G_{\gm \db} \cd_{\da} \rd
      - {G^\vp}_\db \cd_\gm G_{\vp \da}
      - 4{G^\vp}_\db \cd_\vp G_{\gm \da} \right),
 \\ [2mm]
{M^\pl}_{\lsym{\gm \bt \al}} \  &=&
       - 8 \rd W_{\lsym{\gm \bt \al}}
       + \f{1}{24} \oint_{\gm \bt \al}
       {G_\gm}^\dv \lp \cd_\bt G_{\al \dv} + \cd_\al G_{\bt \dv} \rp,
 \\[2mm]
12 \  {M^\pl}_\al  &=& - 3 \cd_\al \mu^\pl
       - 16 \cd_\al (R\rd)  \nonumber \\[1mm] & &
       - 8 \rd \cd^\dv G_{\al \dv}
       - 18 G_{\al \dv} \cd^\dv \rd
       + 2 G^{\vp \dv} \cd_\al G_{\vp \dv}
       + 5 G^{\vp \dv} \cd_\vp G_{\al \dv} \, ,
\ena
and
\be
{M^\pl}\raisebox{-.2ex}{${}_{\dg}$}{}_{ \ {\sym{\bt \al}}} \ = \
       - 4 {G^\vp}_\dg W_{\lsym{\vp \bt \al}}
+ \f{1}{8} \sum_{\bt \al}
\lp {G_\bt}^\dv \cd_\dg G_{\al \dv}
      + \f{4}{3} G_{\bt \dg} \lp \cd_\al R - \cd^\dv G_{\al \dv} \rp
\rp,
\eeq \vs{-.6cm}
\bea
{M^\pl}_{\lsym{\dg \db \da}} &=&
        \f{1}{8} \oint_{\dg \db \da}
       {G^\vp}_\dg \lp \cd_\db G_{\vp \da} + \cd_\da G_{\vp \db} \rp,
 \\[4mm]
4 \  {M^\pl}_\da &=& \cd_\da \, \mu^\pl
      + 2 {G^\vp}_\da \cd_\vp R
      + G^{\vp \dv} \cd_\dv G_{\vp \da}.
\ena

The superfields $S^\pl$ and $T^\pl$ are defined as
\bea
S^\pl &=& \pp \left( \mu^\pl + 16 \rd R
                   - \f{13}{4} G^{\vp \dv} G_{\vp \dv} \right)
            - 4 \bar{X}_\dv \bar{X}^\dv ,
\\[2mm]
T^\pl &=& \qq \left( \mu^\pl
                   + \f{3}{4} G^{\vp \dv} G_{\vp \dv} \right)
       +32 W \raisebox{-.5ex}{${}^{\lsym{\gm \bt \al}}$} W_{\lsym{\gm \bt \al}}
       + \f{4}{3} X^\vp X_\vp.
\ena
For the remaining components at dimension 3 one finds
\bea
\lefteqn{\hs{-1.2cm}
{M^\pl}_{\al \da} + {\Si^\pl}_{\al \da}
+ \f{1}{8} \lp \left[{\cd}_{\al},{\cd}_{\da} \right]
- 4 G_{\al \da} \rp \mu^\pl \ = }
\nonumber \\[2.5mm]
&&+\f{1}{16}G^{\vp \dv} \lp
4 \left[{\cd}_{\vp}, {\cd}_{\dv} \right] G_{\al \da}
+ \left[{\cd}_{\al}, {\cd}_{\da} \right] G_{\vp \dv} \rp
- 8 \rd R \, G_{\al \da}
-\f{15}{24} G_{\al \da} G^{\vp \dv} G_{\vp \dv} \nn \\[2mm]
&&-{\cd}_{\al}R \, {\cd}_{\da}\rd
-\f{3}{32} {\cd}^{\dv} G_{\al \dv} \, {\cd}^{\vp}G_{\vp \da}
+\f{3}{2} {T_{\sym{\dv \da}}}^{\vp} \,
    T\raisebox{-.4ex}{${}_{\sym{\vp \al}}$}{}^{\dv}
+8 T\raisebox{-.4ex}{${}^{\sym{\gm \bt}}$}{}_{\da} \,
    W_{\lsym{\gm \bt \al}}
\nonumber \\[1mm]
&&
+ T\raisebox{-.4ex}{${}_{\sym{\vp \al}}$} \, \raisebox{-.2ex}{${}_\da$}
\lp \f{4}{3} {\cd}^{\vp} R + \f{23}{24} {\cd}_{\dv} G^{\vp \dv} \rp
- T_{\sym{\dv \da}} \, \raisebox{-.2ex}{${}_\al$}
\lp 4 {\cd}^{\dv}\rd +\f{3}{8} {\cd}_{\vp} G^{\vp \dv} \rp
\nonumber \\[2mm]
&&+iG^{\vp \dv} \left( {\cd}_{\vp \dv}G_{\al \da}
+\f{1}{2} {\cd}_{\al \da} G_{\vp \dv}
-\f{3}{4} {\cd}_{\vp \da} G_{\al \dv}
-\f{1}{4} {\cd}_{\al \dv} G_{\vp \da} \right)
- 4i\rd \cd_{\al \da} R.
\ena
The remaining coefficients of the 4-form $\Si^\pl$, \ie
\be
{\Si^\pl}_{\undel \ cba}, \cem {\Si^\pl}_{dcba},
\eeq
are obtained as spinor derivatives of the superfields $S^\pl$ and
$T^\pl$ as explained in the preamble to this appendix.

\indent

\subsection{$\Psi^\mn \ = \ \Si^\mn + d \, M^\mn$}

\indent

Although the basic structure in this sector follows the same pattern as
in the previous subsection (it is basically the complex conjugate),
we will be slightly more explicit and try to give a flavour of the
sequence of arguments used to establish the 3-form constraints
\be
{\Si^\mn}_{\undel \, \undgm \, \undbt A} \ = \ 0,
\label{-3fcon} \\[2mm]
\eeq
starting from the curvature-squared 4-form
\be \hs{-1.2cm}
\Psi^\mn \ = \ {R^\db} {}_\da R\raisebox{.3ex}{${\,}^\da {}_\db$}.
\label{eq:c.1}
\eeq
In more detail, and in analogy to eqs.(\ref{pl1}) and (\ref{pl3}), one has
\be \hs{-1.5cm}
{\Psi^\mn}_{\dt \gm \bt \, A} \ = \ 0, \cem \cem
\Psi^{(-) \, \dd \dg \db} {}_\undal \ = \ 0,
\eeq
\be
{\Psi^\mn}_{\dd \dg \db \ \al \da} \ = \ - 8i \oint_{\dd \dg \db}
\cd_\dd \lp \eps_{\dg \da} R \, G_{\al \db}
+ \eps_{\db \da} R \, G_{\al \dg} \rp,
\eeq
due to the constraints on the curvatures themselves.
Identifying (modulo the discussion at the end of section {\bf 3.1},
eqs.(\ref{eq:7.19}) - (\ref{31L}))
\be \hs{-1.0cm}
{M^\mn}_{\undgm \, \undbt \, \undal} \ = \ 0, \cem
{M^\mn}_{\gm \bt a} \ = \ 0,
\label{eq:c.14}
\eeq
and
\be
{M^\mn}_{\dg \db \ \al \da} \ = \ -8i R \lp
   \eps_{\dg \da} G_{\al \db} + \eps_{\db \da} G_{\al \dg} \rp,
\label{eq:c.16} \\
\eeq
establishes
\be
{\Si^\mn}_{\dt \gm \bt \, A} \ = \ 0, \cem
\Si^{(-) \, \dd \dg \db} {}_A \ = \ 0.
\eeq
In the next step, taking into account
\be
\Psi^\mn{}_{\dt \gm}{}^{\db \da} \ = \
2 \sum _{\dt \gm} R_\dt{}^\db{\,}_\vp{}^\vep R_\gm{}^\da{\,}_\vp{}^\vep
\ = \ 4 \sum _{\dt \gm} G_\dt{}^\db \, G_\gm{}^\da,
\eeq
one is lead to parametrize
\be
{M^\mn}_{\gm \db \ \al \da} \ = \
   -i  \eps_{\gm \al} \, \eps_{\db \da} \, \mu^\mn
   -\i2 \lp G_{\gm \db}\,  G_{\al \da} + G_{\al \db} \, G_{\gm \da} \rp,
\label{eq:c.18}
\eeq
where we note the appearance of the arbitrary superfield $\mu^\mn$.

At dimension 5/2 the components of the curvature-squared 4-form are given as
\be
{\Psi^\mn}_\dt{}^{\dg \db}{}_a \ = \
- 16 R \, R_{\dt a \, }\raisebox{-.8ex}{${}^{\sym{\dg \db}}$}
- 4 \lp R^\db{}_a{\,}^\dg{}_\dv + R^\dg{}_a{\,}^\db{}_\dv \rp
      G_\dt{}^\dv,
\eeq
and
\be \hs{-2cm}
{\Psi^\mn}_{\dt \gm}{}^\db{}_a \ = \
-4 R_{\dt a \, }{}^\db{}_\dv G_\gm{}^\dv
-4 R_{\gm a \, }{}^\db{}_\dv G_\dt{}^\dv. \\[2mm]
\eeq
This in turn determines the components ${M^\mn}_{\undgm \, ba}$.
Employing the decompositions defined in the preamble to this appendix,
the different irreducible tensors appearing here are given as
\bea \hs{-1cm}
{M^\mn}\raisebox{-.2ex}{$_{\dg}$}
\raisebox{.2ex}{${}_{ \ {\sym{\bt \al}}}$} &=& -\f{1}{8} \sum_{\bt \al}
\left( 16 R \, \cd_\bt G_{\al \dg}
      + 4 G_{\bt \dg} \cd_\al R
      - {G_\bt}^\dv \cd_\dg G_{\al \dv}
      - 4 {G_\bt}^\dv \cd_\dv G_{\al \dg} \right),
\label{eq:c.23} \\ [1mm]
{M^\mn}_{\lsym{\dg \db \da}}  &=&
       - 8 R \, W_{\lsym{\dg \db \da}}
       - \f{1}{24} \oint_{\dg \db \da}
       {G^\vp}_\dg \lp \cd_\db G_{\vp \da} + \cd_\da G_{\vp \db} \rp,
\label{eq:c.24} \\ [2.5mm]
12 \ {M^\mn}_\da &=&  3 \cd_\da \mu^\mn
       + 16 \cd_\da (R\rd)  \nonumber \\[2mm] & &
       + 8 R \, \cd^\vp G_{\vp \da}
       + 18 G_{\vp \da} \cd^\vp R
       - 2 G^{\vp \dv} \cd_\da G_{\vp \dv}
       - 5 G^{\vp \dv} \cd_\dv G_{\vp \da},
\label{eq:c.25} \\[4mm]
{M^\mn}\raisebox{-.5ex}{${}_\gm$}{}_{ \ {\sym{\db \da}}} &=&
       - 4 G\raisebox{-.3ex}{${}_\gm$}{}^\dv
           W\raisebox{.3ex}{${}_{\lsym{\dv \db \da}}$}
- \f{1}{8} \sum_{\db \da}
\lp {G^\vp}_\db \cd_\gm G_{\vp \da}
      + \f{4}{3} G_{\gm \db} \lp \cd_\da R - \cd^\vp G_{\vp \da} \rp \rp,
\label{eq:c.26} \\[2mm]
{M^\mn}_{\lsym{\gm \bt \al}} &=&
       - \f{1}{8} \oint_{\gm \bt \al}
       {G_\gm}^\dv \lp \cd_\bt G_{\al \dv} + \cd_\al G_{\bt \dv} \rp,
\label{eq:c.27} \\ [3mm]
4 {M^\mn}_\al &=& -\cd_\al \mu^\mn
      - 2 {G_\al}^\dv \cd_\dv \rd
      - G^{\vp \dv} \cd_\vp G_{\al \dv}.
\label{eq:c.28}
\ena
In this way one ensures that
\be
{\Si^\mn}_{\dt \gm}{}^\db{}_a \ = \ 0, \cem
\Si^{\mn \, \dd \dg}{}_{\bt a} \ = \ 0.
\eeq
The two chiral superfields  $S^\mn$ and $T^\mn$ are then determined to be
\bea
S^\mn &=& \pp \lp \mu^\mn +\f{3}{4} G^{\al \da} G_{\al \da} \rp
              +32 W_{\lsym{\dg \db \da}}
       W\raisebox{-.7ex}{${}^{\lsym{\dg \db \da}}$}
              +\f{4}{3} \bX_\da \bX^\da, \label{eq:c.31} \\[2mm]
T^\mn &=& \qq \lp \mu^\mn + 16 R \rd
                       -\f{13}{4} G^{\al \da} G_{\al \da} \rp
               -4 X^\al X_\al.
\ena
The analysis of the ${\Psi^\mn}{}_{\, ba}$ - sector
shows then that the component ${\Si^\mn}_{\dt \dg \: ba}$,
is expressed in terms of one single vector ${\Si^\mn}_{\al \da}$
(see preamble to this appendix again)
which in turn combines with the purely vectorial component
of ${M^\mn}_{\al \da}$ such that
\bea
\lefteqn{
{M^\mn}_{\al \da} + {\Si^\mn}_{\al \da}
+ \f{1}{8} \lp \left[{\cd}_{\al},{\cd}_{\da} \right]
- 4 G_{\al \da} \rp \mu^\mn \ = }
\nonumber \\[3mm]
&&+\f{1}{16}G^{\vp \dv} \lp
4 \left[{\cd}_{\vp}, {\cd}_{\dv} \right] G_{\al \da}
+ \left[{\cd}_{\al}, {\cd}_{\da} \right] G_{\vp \dv} \rp
- 8 \rd R \, G_{\al \da}
-\f{15}{24} G_{\al \da} G^{\vp \dv} G_{\vp \dv} \nn \\[2mm]
&&-{\cd}_{\al}R \, {\cd}_{\da}\rd
-\f{3}{32} {\cd}^{\dv} G_{\al \dv} \, {\cd}^{\vp}G_{\vp \da}
+\f{3}{2} {T_{\sym{\dv \da}}}^{\vp} \,
    T\raisebox{-.4ex}{${}_{\sym{\vp \al}}$}{}^{\dv}
-8 T\raisebox{-.4ex}{${}^{\sym{\dg \db}}$}{}_{\al} \,
    W_{\lsym{\dg \db \da}}
\nonumber \\[1mm]
&&
-T \raisebox{-.4ex}{${}_{\sym{\vp \al}}$} \, \raisebox{-.2ex}{${}_\da$}
\lp 4 {\cd}^{\vp} R + \f{3}{8} {\cd}_{\dv} G^{\vp \dv} \rp
+T_{\sym{\dv \da}} \, \raisebox{-.2ex}{${}_\al$}
\lp \f{4}{3} {\cd}^{\dv}\rd + \f{23}{24}{\cd}_{\vp} G^{\vp \dv} \rp
\nonumber \\[2mm]
&&-iG^{\vp \dv} \left( {\cd}_{\vp \dv}G_{\al \da}
+\f{1}{2} {\cd}_{\al \da} G_{\vp \dv}
-\f{1}{4} {\cd}_{\vp \da} G_{\al \dv}
-\f{3}{4} {\cd}_{\al \dv} G_{\vp \da} \right)
+ 4iR \cd_{\al \da} \rd.
\ena
Again, the remaining coefficients of the 4-form $\Si^\mn$, i.e.
${\Si^\mn}_{\undel \ cba}$ and ${\Si^\mn}_{dcba}$,
are obtained as spinor derivatives of the superfields $S^\mn$ and $T^\mn$
as explained in the preamble.
\indent

\subsection{$\Psi^\uo \ = \ \Si^\uo + d \, M^\uo$}

\indent

The $U_K(1)$ - sector with $\Psi^\uo = FF \ $ is slightly less involved than
the preceding gravitational sectors. The components of the 3-form $M^\uo$
at dimension 3/2 and 2 are given as
\be
{M^\uo}_{\undgm \, \undbt \, \undal} \ = \ 0, \cem
{M^\uo}_{\gm \bt a} \ = \ 0, \cem
{M^\uo}_{\dg \db \ a} \ = 0,
\eeq
and
\be
{M^\uo}_{\gm \db \ \al \da} \ = \
   -i  \eps_{\gm \al} \, \eps_{\db \da} \, \mu^\uo
   -\f{9i}{4} \lp G_{\gm \db}\,  G_{\al \da} + G_{\al \db} \, G_{\gm \da} \rp.
\eeq
At dimension 5/2 the components ${M^\uo}_{\undgm \ ba}$have irreducible components given as
\bea
{M^\uo}\raisebox{-.2ex}{$_{\dg}$}
\raisebox{.2ex}{${}_{ \ {\sym{\bt \al}}}$}
&=& -\f{3}{8} \sum_{\bt \al} \ {G_\bt}^\dv
\lp \  3 \, T_{\sym{\dv \dg}} \, \raisebox{-.2ex}{${}_\al$} +
\eps_{\dv \dg} (X_\al -S_\al) \rp, \\ [2mm]
{M^\uo}_{\lsym{\dg \db \da}} &=&
        \f{3}{4} \oint_{\dg \db \da}
       {G^\al}_\dg \ T_{\sym{\db \da}} \, \raisebox{-.2ex}{${}_\al$} \; ,
 \\ [2mm]
4 {M^\uo}_\da &=&  \cd_\da \mu^\uo
       +\f{9}{2} G^b \, \cd_\da G_b
- \f{3}{2} G^{ \vp \dv}
\lp  T_{\sym{\dv \da}} \, \raisebox{-.2ex}{${}_\vp$} +
\eps_{\dv \da} (S_\vp -X_\vp) \rp,
\\[2mm]
{M^\uo}\raisebox{-.5ex}{${}_\gm$}{}_{ \ {\sym{\db \da}}} &=&
-\f{3}{8} \sum_{\db\da} \ {G^\vp}_\db
\lp  3 \,
T \raisebox{-.2ex}{${}_{\sym{\vp \gm}}$} \,\raisebox{-.2ex}{${}_\da$}
- \eps_{\vp \gm} (\bX_\da -\bS_\da) \rp, \\ [2mm]
{M^\uo}_{\lsym{\gm \bt \al}} &=&
       \f{3}{4} \oint_{\gm \bt \al}
       {G_\gm}^\dv \
T \raisebox{-.4ex}{${}_{\sym{\bt \al}}$} \,
\raisebox{-.4ex}{${}_\dv$}  \; , \\ [2mm]
4 {M^\uo}_\al &=& -\cd_\al \mu^\uo - \f{9}{2} G^b \, \cd_\al G_b -\f{3}{2}
G^{\vp \dv}
\lp T \raisebox{-.4ex}{${}_{\sym{\vp \al}}$} \,\raisebox{-.2ex}{${}_\dv$} +
\eps_{\vp \al} (\bX_\dv +\bS_\dv) \rp.
\ena
The chiral superfields $S^\uo$ and $T^\uo$ are given as
\bea
S^\uo &=& \pp \lp \mu^\uo +\f{9}{4} G^b G_b \rp
             - 2 \bX_\da \bX^\da,\\[2mm]
T^\uo &=& \qq \lp \mu^\uo + \f{9}{4} G^b G_b \rp
               - 2X^\al X_\al.
\ena
Finally, one obtains
\bea
\lefteqn{
{M^\uo}_{\al \da} + {\Si^\uo}_{\al \da}
+ \f{1}{8} \lp \left[{\cd}_{\al},{\cd}_{\da} \right]
- 4 G_{\al \da} \rp \lp\mu^\uo + \f{9}{4}  G^{b} G_{b} \rp \ = }
\nonumber \\[3mm]
&& - \f{1}{2} X_\al \bX_\da
 - 3 ( {G_ \al }^\dv \ f_{ \sym{ \dv \da}}
+ {G^\vp} _\da \  f_{\sym{ \vp \al}} )
+ \f{9i}{8}G^{\vp \dv} \left(
 {\cd}_{\vp \da} G_{\al \dv}
-{\cd}_{\al \dv} G_{\vp \da} \right).
\ena

\indent

\subsection{$\Psi^{\ym} \ = \ \Si^{\ym} + d \, M^{\ym}$}

\indent

In the Yang-Mills sector the components of $M^{\ym}$ and $\Si^{\ym}$ are
given as
\be
{M^{\ym}}_{\undgm \, \undbt \, \undal} \ = \ 0, \cem
{M^{\ym}}_{\gm \bt a} \ = \ 0, \cem
{M^{\ym}}_{\dg \db \ a} \ = 0, \\[1mm]
\eeq
\be
{M^{\ym}}_{\gm \db \ \al \da} \ = \
   -i  \eps_{\gm \al} \, \eps_{\db \da} \, \mu^{\ym},
\eeq
\bea
{M^{\ym}}_{\gm \ \bt \db \ \al \da} &=&
- \f{1}{2} \eps_{\db \da} \lp \eps_{\gm \bt} \cd_\al \mu^{\ym}
 +\eps_{\gm \al} \cd_\bt \mu^{\ym} \rp \\[2mm]
{M^{\ym}}_{\dg \ \bt \db \ \al \da} &=&
- \f{1}{2} \eps_{\bt \al} \lp \eps_{\dg \db} \cdb_\da \mu^{\ym}
 +\eps_{\dg \da} \cdb_\db \mu^{\ym} \rp,
\ena
as well as
\bea
S^{\ym} &=& \pp \ \mu^{\ym}
             - 8 \bW_\da \bW^\da, \\[2mm]
T^{\ym} &=& \qq \ \mu^{\ym}
               - 8W^\al W_\al,
\ena
where $S^{\ym}$ and $T^{\ym}$ are as usual related to the
components $\Sigma^{\ym}{}_{ \dt \gm \, ba}$
and $ \Sigma^{{\ym} \ \dd \dg} {}_{\; ba}$, and
\be
{M^{\ym}}_{\al \da} + {\Sigma^{\ym}}_{\al \da}
+ \f{1}{8} \lp \left[{\cd}_{\al},{\cd}_{\da} \right]
- 4 G_{\al \da} \rp \mu^{\ym}  = - 2 W_\al  \bW_\da.
\eeq
It is clear, that in this case the decomposition is trivial in the sense
that one can take $\mu^{\ym} = 0$ and $M^{\ym} = 0$ as a superspace 3-form.


\end{document}